\documentclass[12pt]{article}
\usepackage{epsfig,cite}
\usepackage{amsmath}
\usepackage{amssymb}
\usepackage{color}
\usepackage{graphicx}
\usepackage{verbatim,mathrsfs,subfig,feynmf}
\include{epsf}
\newcount\timecount
\newcount\hours \newcount\minutes  \newcount\temp \newcount\pmhours
\hours = \time
\divide\hours by 60
\temp = \hours
\multiply\temp by 60
\minutes = \time
\advance\minutes by -\temp
\def\hour{\the\hours}
\def\minute{\ifnum\minutes<10 0\the\minutes
            \else\the\minutes\fi}
\def\clock{
\ifnum\hours=0 12:\minute\ AM
\else\ifnum\hours<12 \hour:\minute\ AM
      \else\ifnum\hours=12 12:\minute\ PM
            \else\ifnum\hours>12
                 \pmhours=\hours
                 \advance\pmhours by -12
                 \the\pmhours:\minute\ PM
                 \fi
            \fi
      \fi
\fi
}

\def\monthname{\relax\ifcase\month 0/\or January\or February\or
   March\or April\or May\or June\or July\or August\or September\or
   October\or November\or December\else\number\month/\fi}

\def\bold#1{\setbox0=\hbox{$#1$}%
     \kern-.025em\copy0\kern-\wd0
     \kern.05em\copy0\kern-\wd0
     \kern-.025em\raise.0433em\box0 }

\def\beq{\begin{equation}}
\def\eeq{\end{equation}}

\def\ga{\mathrel{\raise.3ex\hbox{$>$\kern-.75em\lower1ex\hbox{$\sim$}}}}
\def\la{\mathrel{\raise.3ex\hbox{$<$\kern-.75em\lower1ex\hbox{$\sim$}}}}
\def\gev{{\rm \, Ge\kern-0.125em V}}
\def\tev{{\rm \, Te\kern-0.125em V}}
\def\gyr{{\rm \, G\kern-0.125em yr}}

\def\gappeq{\mathrel{\rlap {\raise.5ex\hbox{$>$}}
{\lower.5ex\hbox{$\sim$}}}}
\def\lappeq{\mathrel{\rlap{\raise.5ex\hbox{$<$}}
{\lower.5ex\hbox{$\sim$}}}}
\def\Toprel#1\over#2{\mathrel{\mathop{#2}\limits^{#1}}}

\def\m12{m_{1\!/2}}

\def\bea{\begin{eqnarray}}
\def\eea{\end{eqnarray}}

\def\beqar{\begin{eqnarray}}
\def\eeqar{\end{eqnarray}}

\usepackage{lscape}

\def\beq{\begin{equation}}
\def\eeq{\end{equation}}

\begin{document}

\begin{titlepage}
\pagestyle{empty}
\rightline{KCL-PH-TH/2015-42, LCTS/2015-29, CERN-PH-TH/2015-229,}
\rightline{UMN--TH--3502/15, FTPI--MINN--15/41, IPMU15--0174, CETUP2015-021}
\vspace{0.5cm}
\begin{center}
{\bf {\large Beyond the CMSSM without an Accelerator:} \\
\vspace{0.1cm}
Proton Decay and Direct Dark Matter Detection}

\end{center}
\vspace{0.5cm}
\begin{center}
{\bf John~Ellis}$^{1,2}$,
{\bf Jason L. Evans}$^3$,
{\bf Feng Luo}$^{2}$, {\bf Natsumi Nagata}$^{3,4}$, \\
\vspace{0.1cm}
{\bf Keith~A.~Olive}$^{3}$ and {\bf Pearl Sandick}$^{5}$\\
\vskip 0.2in
{\small {\it
$^1${Theoretical Physics and Cosmology Group, Department of Physics, \\ King's College London, Strand,
London~WC2R 2LS, UK}\\
$^2${TH Division, Physics Department, CERN, CH-1211 Geneva 23, Switzerland}\\
$^3${William I. Fine Theoretical Physics Institute, School of Physics and Astronomy,\\
University of Minnesota, Minneapolis, MN 55455,\,USA}\\
$^4${Kavli IPMU (WPI), UTIAS, University of Tokyo, Kashiwa, Chiba
 277-8583, Japan}\\
$^5${Department of Physics and Astronomy, University of Utah, Salt Lake City, Utah 84112, USA}} \\
}
\vspace{1cm}
{\bf Abstract}
\end{center}
{\small
We consider two potential non-accelerator signatures of generalizations of the well-studied
constrained minimal supersymmetric standard model (CMSSM). In one generalization, the universality
constraints on soft supersymmetry-breaking parameters are applied at some input
scale $M_{in}$ {\it below} the grand unification (GUT) scale $M_{GUT}$, a scenario referred to as `sub-GUT'.
The other generalization we consider is to retain GUT-scale universality for the squark and slepton masses,
but to relax universality for the soft supersymmetry-breaking contributions to the masses of the Higgs doublets.
As with other CMSSM-like models, the measured Higgs mass requires supersymmetric particle masses near or beyond the TeV
scale.  Because of these rather heavy sparticle masses, the embedding of these CMSSM-like models
in a minimal SU(5) model of grand unification can yield a proton lifetime consistent with current experimental limits, and may be
accessible in existing and future proton decay experiments. Another possible signature of these CMSSM-like models is
direct detection of supersymmetric dark matter. The direct dark
matter scattering rate is typically below the reach of the LUX-ZEPLIN (LZ) experiment if $M_{in}$
is close to $M_{GUT}$, but may lie within its reach if $M_{in} \lesssim 10^{11}$~GeV.
Likewise, generalizing the CMSSM to allow non-universal supersymmetry-breaking contributions to the Higgs
offers extensive possibilities for models within reach of the LZ experiment that have long proton lifetimes.}


\vfill
\leftline{September 2015}
\end{titlepage}

\section{Introduction}

Supersymmetry remains a favored extension of the Standard Model (SM),
despite its non-appearance during Run~1 of the LHC\cite{ATLAS20,CMS20}.  Indeed, the
discovery of a 125-GeV Higgs boson at the LHC \cite{lhch}
has supplemented the traditional arguments for supersymmetry, including the
naturalness of the electroweak scale, the unification of the fundamental
interactions and the existence of a cold dark matter candidate (if $R$-parity is conserved).
The minimal supersymmetric extension of the SM (MSSM)
predicts the existence of a Higgs boson with mass $m_h \lesssim 130$~GeV, and
is a prime example of new physics capable of stabilizing the
electroweak vacuum for $m_h \sim 125$~GeV \cite{125}. Moreover, global fits
in the framework of simple supersymmetric models suggest that the couplings of
the lightest supersymmetric Higgs boson should be very similar to those of the
Higgs boson in the SM, as is indicated by the ATLAS and CMS
experiments \cite{ATLASmu,CMSmu}. When the supersymmetric particle masses are large,
which is the case we consider, the Higgs couplings become even more like the SM couplings.

If these arguments are valid, the following questions must be answered:
which supersymmetric model is found in Nature, and how may it be tested?
To begin to answer these questions, we focus here on the
MSSM, and more specifically on constrained versions
in which the soft supersymmetry-breaking scalar masses $m_0$, gaugino
masses $m_{1/2}$ and trilinear terms $A_0$ are assumed to have universal values at some high
input mass scale $M_{in}$.  Typically, $M_{in}$ is chosen to be at the
grand-unified-theory (GUT) scale, a scenario called the
constrained MSSM (CMSSM)~\cite{funnel,cmssm,efgo,cmssmwmap,eo6,ehow+},
in which the ratio of Higgs vacuum expectation values, $\tan \beta$, is a free parameter.

In order to find models with less-constrained dark matter scenarios and
simultaneously a sufficiently long lifetime for the proton,
we focus here on two one-parameter extensions of the CMSSM: `sub-GUT' models~\cite{subGUT}
in which $M_{in} < M_{GUT}$ is free, and the NUHM1~\cite{nuhm1,eosknuhm}, in which
the two Higgs soft masses are equal at the input scale, $m_1=m_2$, but are allowed to differ from $m_0$.
We will also discuss `sub-GUT' models obtained from minimal supergravity (mSUGRA),
which are more constrained than the CMSSM, since the
gravitino mass $m_{3/2} = m_0$ and the
trilinear and bilinear soft supersymmetry-breaking terms are related: $A_0 - B_0 = m_0$~\cite{bfs,vcmssm}.
Since mSUGRA models  have one fewer parameter, $\tan\beta$ is no longer free.
Sub-GUT mSUGRA models have the same number of free parameters as in the CMSSM,
but viable models can readily be found.

Although the standard CMSSM with $M_{in}=M_{GUT}$ is still viable,
there remain only restricted regions of the parameter space of the CMSSM
(and, {\it a fortiori}, of the more restrictive mSUGRA model) in which a
successful prediction for $m_h$ can be reconciled with the measured cold
dark matter
density~\cite{eo6,ehow+,mc75,mc8,mc9,ELOS,post-mh,fp,eoz,elo,yal}. The parameter spaces of these
models become more restricted when they are embedded in an SU(5) GUT,
because they tend to have a
proton lifetime which is shorter than the current experimental
limits~\cite{Abe:2014mwa,Shiozawa, Babu:2013jba}, even if the
supersymmetric sparticle masses are rather heavy.

These problems can be avoided, however, if $M_{in}$ is identified with
some scale lower than the typical GUT scale.  An effective scale
of supersymmetry breaking significantly below the GUT scale,
$M_{in}<M_{GUT}$, is not without theoretical motivation.  For example,
mirage unification models~\cite{mixed} and other scenarios such
as~\cite{Monaco:2011fe} give exactly such boundary conditions for the soft supersymmetry-breaking
parameters. Phenomenologically, these sub-GUT models with
$M_{in}<M_{GUT}$ have been shown to have an appropriate cold dark matter density
in a considerably larger parameter space~\cite{ELOS}. As
could be expected, a significant part of this parameter space contains points
that are compatible with the LHC measurement of $m_h$ and other
phenomenological constraints, such as the non-detection of
supersymmetric particles at the LHC~\cite{ATLAS20,CMS20}~\footnote{Note
that we do not impose any constraint from the anomalous magnetic moment
of the muon, $g_\mu - 2$~\cite{newBNL}.}. This reduced tension in
sub-GUT models is due to the reduced running of the soft
masses, which leads to a sparticle spectrum that is, in
general, more compressed than models with $M_{in} =
M_{GUT}$. Moreover, this compression of the spectrum leads to more avenues for
coannihilation~\cite{Griest:1990kh}, which is effective in reducing the relic neutralino density into the range allowed by
cosmology.

Similarly, if the Higgs soft masses are allowed to differ from the soft masses of the matter scalars, as in the NUHM,
there are more viable options for dark matter. In both the CMSSM and in the NUHM,
the Higgs mixing mass, $\mu$, and the pseudoscalar mass, $m_A$, are determined
through the minimization of the Higgs potential.
However, either $\mu$ and/or $m_A$ can be traded for the Higgs soft mass,
which can be calculated using the minimization of the Higgs potential.

Here we examine two potential non-accelerator observables in the contexts of these less-constrained models:
the proton lifetime and the elastic scattering cross section for the direct detection of dark matter.

In \cite{mp}, the proton lifetime was computed by renormalization group (RG) running the gauge couplings
up to the GUT scale, defined to be where the two electroweak couplings are equal.
The imperfection in the unification of the electroweak couplings with the strong coupling was then used to determine the size of the
color-triplet Higgs threshold, which then determined the color-triplet Higgs
mass \cite{Hisano:1992mh, Hisano:1992jj, Hisano:2013cqa}.
Using this procedure with sub-TeV stops and Higgsinos and decoupled first- and second-generation sfermions,
it was shown that the lifetime of the proton was shorter than the experimental constraints. Since the
experimental constraints are now stronger, this problem has become even worse.

This problem can be avoided in many
ways. One particularly simple way is to include an additional pair of ${\bf 5}$ and ${\bf \bar 5}$
Higgs boson supersymmetric multiplets
that do not couple to any of the SM fields. Below the SU(5) breaking scale,
the colored and flavored Higgs mass become free parameters.  If the portion of the Higgs
supersymmetric multiplet that has SU(3) charges is lighter than the portion with SU(2) charges,
the thresholds in the couplings will be different from those of minimal SU(5), and the colored
Higgs masses can be made sufficiently heavy that the proton decay constraints can be met~\cite{mp}.
Other possibilities for alleviating this problem include forbidding the dimension-five operator leading to proton decay
using extra dimensions~\cite{Kawamura:2000ev}, more complicated Higgs sectors~\cite{Babu:1993we},
flipped SU(5)~\cite{flipped}, or a Peccei-Quinn symmetry~\cite{PQ}. The problem is
 also alleviated in models with scalar masses that are
 $\mathcal{O}(100)$ TeV \cite{Hisano:2013exa,McKeen:2013dma, Nagata:2013sba},
 as in pure gravity mediation \cite{evno}.

The LHC constraints on sfermion masses and the observed Higgs mass of $125$ GeV
motivate us to consider once again the decoupling limit as an explanation for the long lifetime of the proton.
The decoupling limit was unsuccessful in~\cite{mp} due to the assumption of a light third generation.
However, if the third generation is also decoupled, the proton lifetime is extended.
Since a heavier third generation is favoured by the $125$ GeV mass of the Higgs,
we find this to be a reasonable scenario for suppressing proton decay.
However, the real challenge in this scenario is to find regions of parameter space
that combine a viable dark matter candidate with an acceptably long proton lifetime.

We will find that the minimal supersymmetric grand unified theory based
on SU(5) \cite{Dimopoulos:1981zb} with a CMSSM spectrum
does have a very small region of parameter space that combines a $125$ GeV Higgs,
a sufficiently long proton lifetime, and a viable dark matter candidate. This may occur either in the focus-point region \cite{fp2}
or in a region where the dark matter density is obtained by stop coannihilation with the bino \cite{stop,eoz}.
However, as we show below, sub-GUT and NUHM1 models are less restricted by proton decay constraints.
The proton lifetime is longer in sub-GUT models, in general, because the stop masses are larger due to reduced RG running.
Since the lifetime of the proton scales as a power of the stop mass, this enhances the proton lifetime.
Moreover, these sub-GUT models have an acceptable dark matter density in regions where the bino and the
lighter stau coannihilate \cite{stau}.
In the NUHM1, $\sim$ TeV Higgsinos are possible for any value of $m_{0}$ and $m_{1/2}$ and,
if the Higgsino mass is $\mathcal{O}$(TeV), the Higgsino can be a thermal relic dark matter candidate \cite{osi}.
If $m_0$ and $m_{1/2}$ are large, the proton lifetime is greatly enhanced because of the large stop mass.
In all cases, compatibility with minimal SU(5) requires relatively low values of $\tan \beta \lesssim 5$.

We also examine whether such models are compatible with present
experimental constraints on the direct detection of dark matter through
spin-independent elastic scattering, as provided, e.g., by the LUX experiment \cite{lux},
and whether they can be probed by the next generation of such experiments, e.g., XENON1T~\cite{xenon1t}
and LUX-ZEPLIN (LZ)~\cite{Malling:2011va}.

For the purpose of our study, we use {\tt FeynHiggs}~\cite{FH} to calculate $m_h$ as a
function of the model input parameters. Since one expects an uncertainty $\sim \pm 1.5$~GeV in this
calculation, we assume that any model yielding a prediction $m_h \in [124, 127]$~GeV may be
acceptable. Even with this theoretical uncertainty,
we find that the $m_h$ measurement generally gives a stronger constraint on the model parameters than do the
direct LHC searches for supersymmetric particles published so far. As we also discuss,
another important constraint is provided by the experimental search for $B_s \to \mu^+ \mu^-$ decay~\cite{bmm},
particularly at large $\tan \beta$. Since we do not impose any $g_\mu - 2$ constraint,
and the Higgs and other LHC constraints exclude small values of $(m_0, m_{1/2})$, the impact
of the $b \to s \gamma$ constraint \cite{bsgex} is reduced.
We use SSARD \cite{SSARD} to calculate the particle spectrum, proton lifetimes, and elastic scattering cross sections.

The layout of this paper is as follows. In Section~2 we summarize the features of the CMSSM
and mSUGRA models that are relevant for our analysis, and introduce their extensions to sub-GUT
and NUHM1 models. In Section 3, we discuss the basics of our calculations of the proton
life-time and elastic scattering cross sections in  CMSSM-like models. Section~4 displays our results.  Finally,
Section~5 summarizes our conclusions.

\section{The Models}

\subsection{mSUGRA}

Minimal supergravity  (mSUGRA) models have
a quadratic K\"ahler potential for the chiral superfields, and the effective scalar potential is~\cite{Fetal,acn,bfs}
 \begin{eqnarray}
V  & =  &  \left|\frac{\partial W}{\partial \phi^i}\right|^2 +
\left( A_0 W^{(3)} + B_0 W^{(2)} + \text{h.c.}\right)  + m_{3/2}^2 \phi^i \phi_i^*  \, ,
\label{pot}
\end{eqnarray}
where $W$ is the superpotential for the matter superfields $\phi_i$~\footnote{We use the same notation
for the chiral superfields and their spin-zero components.}, which takes the following form in the MSSM:
\beq
W =  \bigl( y_e H_1 L \overline{e} + y_d H_1 Q \overline{d} + y_u H_2
Q \overline{u} \bigr) +  \mu H_1 H_2  \, .
\label{WMSSM}
\eeq
We denote by $L$ and $Q$ ($e$, $u$, and $d$) the left- (right-)handed matter superfields,
the Yukawa couplings are denoted by the $y_\alpha$, and $H_{1,2}$ are the pair of MSSM
Higgs doublets with superpotential mixing coefficient $\mu$.
The $W^{(3)}$ in Eq.~\eqref{pot} are the trilinear terms in the superpotential,
$W^{(2)}$ is the bilinear part corresponding to the $\mu$ term, and $m_{3/2}$ is the gravitino mass.
In mSUGRA one finds scalar mass universality with $m_0 = m_{3/2}$, and there is a relation
between the tri- and bilinear supersymmetry-breaking terms:
\begin{equation}
A_0 = B_0 + m_0 \, .
\label{mSUGRAcondition}
\end{equation}
These conditions apply at an input renormalization scale, $M_{in}$, which may or may not
be identified with the grand unification scale $M_{GUT}$. If the gauge kinetic function is minimal,
there is also gaugino mass universality, with a common mass $m_{1/2}$ that we assume to
apply at the same input scale $M_{in}$.

The two electroweak vacuum conditions are
\beq
\mu^2=\frac{m_1^2-m_2^2\tan^2\beta+\frac{1}{2}m_Z^2(1-\tan^2\beta)+\Delta_{\mu}^{(1)}}{\tan^2\beta-1+\Delta_{\mu}^{(2)}} \,
\label{eq:mu}
\eeq
and
\beq
B \mu = - \frac{1}{2}(m_1^2+m_2^2+2\mu^2)\sin 2\beta +\Delta_B \, ,
\label{eq:muB}
\eeq
where the soft supersymmetry-breaking Higgs masses denoted by $m_{1,2}$ are here evaluated at the weak scale,  and $\Delta_B$ and $\Delta_\mu^{(1,2)}$ are loop
corrections~\cite{Barger:1993gh}. An mSUGRA model has just three continuous parameters:
$m_{1/2}$, $m_0$ and $A_0$. The conditions (\ref{eq:mu}, \ref{eq:muB}) can be used to determine $\tan \beta$
as well as the magnitude of $\mu$, but the sign of $\mu$ is undetermined.
We consider in this paper both signs of $\mu$ in selected cases.

\subsection{The CMSSM}

The CMSSM is effectively a one-parameter generalization of mSUGRA, in which
the relation (\ref{mSUGRAcondition}) between $A_0$ and $B_0$ is dropped,
which allows $\tan \beta$ to be taken as an extra free parameter. In addition,
$m_0 \ne m_{3/2}$ in general, which is possible in SUGRA models only if the
supergravity K\"ahler potential has non-minimal kinetic terms.
Thus the CMSSM
is specified by four parameters, $m_{1/2}$, $m_0$, $A_0$, $\tan \beta$,  the sign of $\mu$.
Here we assume that $m_{3/2}$ is sufficiently large to be irrelevant.

\subsection{Sub-GUT versions of the CMSSM and mSUGRA}

Generalizations of both mSUGRA and the CMSSM are possible if
the input scale for universality of the supersymmetry-breaking terms differs
from $M_{GUT}$. We concentrate here on `sub-GUT' models with $M_{in} < M_{GUT}$~\cite{subGUT}:
a value of $M_{in}$ above the GUT scale~\cite{superGUT,dlmmo}
would introduce many more GUT parameters, requiring a separate in-depth study.
Sub-GUT versions of
mSUGRA have four parameters: $m_{1/2}, m_0 = m_{3/2}, A_0 = B_0 + m_0$, and $M_{in}$,
whereas sub-GUT versions of the CMSSM have $\tan \beta$ as an extra parameter (assuming again that $m_{3/2}$
is irrelevantly large).
We found in~\cite{ELOS} that sub-GUT mSUGRA models are phenomenologically viable in a relatively
restricted range of $A_0$ that straddles the Polonyi value $A_0 = (3 -\sqrt{3}) \times
m_{3/2}$~\cite{Polonyi,bfs}.

\subsection{The NUHM1}

Another one- or two-parameter generalization of the CMSSM is the NUHM~\cite{nonu,efgo},
in which the values of the soft supersymmetry-breaking contributions to the Higgs masses
$m_1$ and $m_2$ at the input scale are allowed to differ from
the universal scalar mass $m_0$. In the NUHM1 considered here, it is assumed that
$m_1 = m_2$ at the input scale~\cite{nuhm1,eosknuhm}.
One may choose either $\mu$ or
$m_A$ (through its relation to $B\mu$) as a free parameter, and use the minimization conditions
(\ref{eq:mu} and \ref{eq:muB}) to solve for $m_1 = m_2$. The examples shown here treat $\mu$ as a free parameter,
since this displays more readily the interesting results in this scenario.

The two-parameter extension known as the NUHM2 \cite{nuhm2,eosknuhm} drops the requirement
that $m_1 = m_2$ at the input scale. In this case, both $m_1$ and $m_2$ are allowed to be
free input parameters.  Alternatively, one can choose both $\mu$ and $m_A$ at the weak
scale as free input parameters and use the minimization conditions (\ref{eq:mu} and \ref{eq:muB}) to solve for $m_1$ and $m_2$.
We do not study the NUHM2 in this paper, as most of  the interesting aspects of the NUHM2 are contained in NUHM1 scans.

\section{Calculations}

\subsection{Proton Decay Lifetimes}
\label{pdecay}

In this subsection we describe how we calculate proton decay
rates in the minimal supersymmetric SU(5) GUT model~\cite{Dimopoulos:1981zb}: for further
discussions and detailed formulae, see~\cite{Hisano:2013exa,Nagata:2013sba, evno, Goto:1998qg}.
This model is the simplest supersymmetric extension of the original Georgi-Glashow
model \cite{Georgi:1974sy}, in which the MSSM matter superfields
are embedded into a $\overline{\bf 5}\oplus {\bf 10}$ representation of
SU(5) for each generation, and the MSSM Higgs superfields $H_1$ and $H_2$
are incorporated in a pair of $\overline{\bf 5}$ and {\bf 5}
superfields, respectively. The SU(3)$_C$ components of the $\overline{\bf 5}$ and {\bf 5} are called the
color-triplet Higgs fields. The dominant contribution to proton decay in this model is given by the
exchange of
these color-triplet Higgs fields~\cite{Sakai:1981pk}, which
induce dimension-five baryon-number violating operators, whereas the exchanges of SU(5)
gauge bosons yield dimension-six operators. In this case, the
dominant proton decay channel is the $p\to K^+\bar{\nu}$ mode, and we focus on the
partial decay rate for this channel in the following.

We obtain the low-energy effective Lagrangian below the GUT scale
by first integrating out the color-triplet Higgs fields. In superfield
notation, the effective Lagrangian is given by
\begin{equation}
{\cal L}_5^{\rm eff}= C^{ijkl}_{5L}{\cal O}^{5L}_{ijkl}
+C^{ijkl}_{5R}{\cal O}^{5R}_{ijkl}
+{\rm h.c.}~,
\label{eq:efflaggut}
\end{equation}
with the effective operators ${\cal O}^{5L}_{ijkl}$ and ${\cal
O}^{5R}_{ijkl}$ defined by
\begin{align}
 {\cal O}^{5L}_{ijkl}&\equiv\int d^2\theta~ \frac{1}{2}\epsilon_{abc}
(Q^a_i\cdot Q^b_j)(Q_k^c\cdot L_l)~,\nonumber \\
{\cal O}^{5R}_{ijkl}&\equiv\int d^2\theta~
\epsilon^{abc}\overline{u}_{ia}\overline{e}_j\overline{u}_{kb}
\overline{d}_{lc}~,
\end{align}
where $a,b,c$ are SU(3)$_C$ color indices, $i,j,k,l$ are
generation indices, and $\epsilon_{abc}$ is the totally antisymmetric tensor.
The Wilson coefficients of the above operators, $C^{ijkl}_{5L}$
and $C^{ijkl}_{5R}$ at the GUT scale are
evaluated from the tree-level color-triplet Higgs exchange diagrams, with
the results given in Appendix~\ref{sec:protondecayapp}. As shown in
Eq.~\eqref{eq:wilson5} of the appendix, these Wilson coefficients include up-type
quark Yukawa couplings $y_{u_i}$ and down-type quark/lepton Yukawa
couplings $y_{d_l}$, which should be unified at the GUT scale in the
minimal SU(5) GUT. Note, however, that although the minimal SU(5) GUT
relation between the bottom and $\tau$ masses~\cite{SU5mbmtau} is
approximately consistent with the experimental values, this is not the
case for the strange and $\mu$ masses, nor for the down and $e$
masses. The GUT Higgs couplings must therefore be more complicated than
in the minimal SU(5) GUT, e.g., with additional higher-dimensional Higgs
representations~\cite{GM} and/or contributions to the fermion masses
from higher-dimensional superpotential terms~\cite{EG}.  In practice, the
ambiguity in choosing whether the down-type quark or lepton Yukawa couplings sets the scale of proton decay results in about a factor of 20 uncertainty in the proton decay
calculation \cite{evno}, which represents our ignorance of GUT-scale physics in the
Yukawa sector. In the following calculation, we take down-type quark
Yukawa couplings as the GUT-scale Yukawa couplings, which leads to
longer proton decay lifetimes and thus gives rather conservative bounds
on the model parameter space.

In addition, the GUT-scale Yukawa couplings introduce two extra phase
factors \cite{Ellis:1979hy}, which give rise to additional
uncertainty \cite{Goto:1998qg}. It turns out, however, that the effects
of these unknown phases are actually negligible, as shown in
Appendix~\ref{sec:protondecayapp}. Thus, we neglect these effects in our
analysis.

\begin{figure}[t]
\begin{center}
\subfloat[Wino contribution]{
\includegraphics[height=45mm]{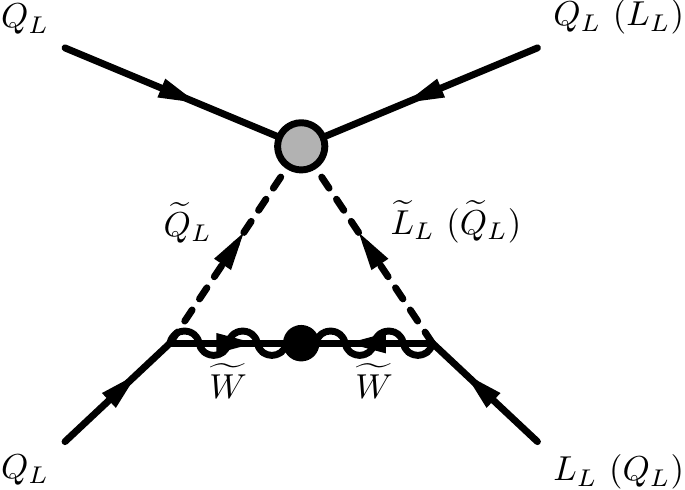}
}
\hspace{10mm}
\subfloat[Higgsino contribution]{
\includegraphics[height=45mm]{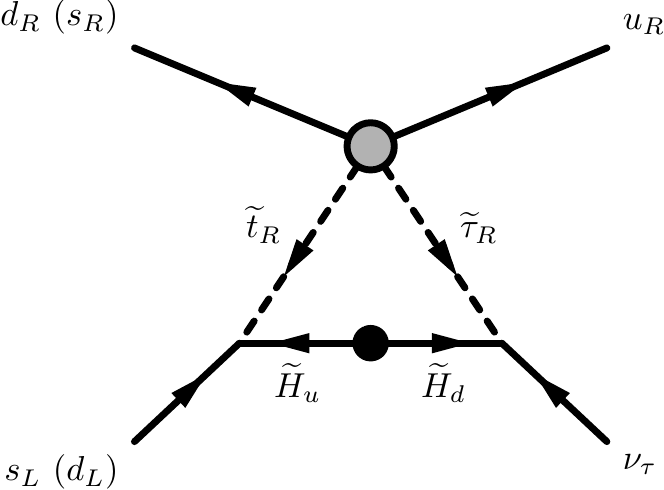}
}
\caption{\it One-loop diagrams that yield dimension-six
 four-fermion operators. Diagrams (a) and (b) are generated by the charged wino
 and higgsino exchange processes, respectively. The gray dots indicate the
 dimension-five effective interactions (\protect\ref{eq:efflaggut}), and the black dots represent
 the wino and Higgsino mass terms.}
\label{fig:1loop}
\end{center}
\end{figure}

After integrating out the color triplet Higgs boson, the GUT-scale Wilson coefficients are then evolved down to
the sfermion mass scale according to the renormalization group equations
(RGEs), which are also presented in
Appendix~\ref{sec:protondecayapp}. At the sfermion mass threshold,
sfermions in the external lines of the dimension-five effective
operators are integrated out via the one-loop diagrams illustrated in
Fig.~\ref{fig:1loop}, yielding
dimension-six four-fermion operators~\cite{Weinberg:1979sa,
Abbott:1980zj}. In the absence of flavor violation in
the sfermion sector~\footnote{The effects on proton decay of possible flavor violation in the
sfermion sector are discussed in~\cite{Nagata:2013sba}.}, only the operators, ${\cal
O}^{5L}_{ii1j}$ and ${\cal O}^{5R}_{331k}$ with $i=2,3$, $j=1,2,3$, and
$k =1,2$ give sizable contributions to proton decay; the contributions
of the other operators are suppressed by small Yukawa couplings and/or the
off-diagonal CKM matrix elements. The one-loop diagrams in
Fig.~\ref{fig:1loop} yield the following effective Lagrangian below the
sfermion mass scale:
\begin{align}
 {\cal L}^{\text{eff}}_6&=C^{\widetilde{H}}_i {\cal O}_{1i33}
+ C^{\widetilde{W}}_{jk}\widetilde{\cal O}_{1jjk}
+ C^{\widetilde{W}}_{jk}\widetilde{\cal O}_{j1jk}
+ \overline{C}^{\widetilde{W}}_{jk}\widetilde{\cal O}_{jj1k}
~,
\label{eq:effdim6}
\end{align}
with the operators composed of SM fermion fields,
\begin{align}
 {\cal O}_{ijkl} &\equiv \epsilon_{abc}(u^a_{Ri}d^b_{Rj})
(Q_{Lk}^c \cdot L_{Ll}) ~, \nonumber \\
 \widetilde{\cal O}_{ijkl} &\equiv \epsilon_{abc} \epsilon^{\alpha\beta}
\epsilon^{\gamma\delta} (Q^a_{Li\alpha}Q^b_{Lj\gamma})
(Q_{Lk\delta}^c L_{Ll\beta}) ~,
\end{align}
corresponding, respectively, to the operators $O^{(1)}$ and $\widetilde{O}^{(4)}$ in~\cite{Abbott:1980zj}, where
$\alpha,\beta,\gamma,\delta$ are SU(2)$_L$ indices,
$\epsilon^{\alpha\beta}$ is the totally antisymmetric tensor, and
$i =1,2$, $j=2,3$, and $k=1,2,3$. The coefficients of the operators in (\ref{eq:effdim6}) are also given in
Appendix~\ref{sec:protondecayapp}. Note that, since a chirality
flip in the internal wino/Higgsino propagator is required in the
processes shown in Fig.~\ref{fig:1loop}, the operator coefficients $C^{\widetilde{H}}_i, C^{\widetilde{W}}_{jk}$
contain factors of $M/M_{\widetilde{f}}^2$ if $M\lesssim
M_{\widetilde{f}}$, where $M$ is the wino or Higgsino mass and
$M_{\widetilde{f}}$ is the mass of a sfermion running in the loop. As a
result, if the magnitude of the Higgsino mixing term $|\mu|$ is much smaller than that of the wino mass
$|M_2|$, the wino contribution dominates the Higgsino contribution, and
vice versa if $|\mu| \gg |M_2|$.

The coefficients in (\ref{eq:effdim6}) are then run down to the
electroweak scale using the one-loop RGEs given in~\cite{Alonso:2014zka}.
At the electroweak scale, we transform to an
operator basis in the low-energy $\text{SU}(3)_C\otimes \text{U}(1)_{\text{em}}$
theory, and the operator coefficients are evolved to the hadron scale,
$Q_{\text{had}} =2$~GeV, using the two-loop RGEs obtained in~\cite{Nihei:1994tx}.
Finally, using the hadron matrix elements
of the operators at $Q_{\text{had}} =2$~GeV, we obtain the partial decay
width of the $p\to K^+ \bar{\nu}$ channel. These matrix elements are
evaluated using the QCD lattice simulation performed in~\cite{Aoki:2013yxa}.
This procedure, as well as the relevant
formulae, is also summarized in Appendix~\ref{sec:protondecayapp}.

As can be seen from \eqref{eq:wilson5}, the proton decay rate
depends on the mass of the color-triplet Higgs field $M_{H_C}$. Thus,
to evaluate the proton lifetime, we need to determine the size of
$M_{H_C}$. To that end, we use the method discussed
in~\cite{Hisano:1992mh, Hisano:1992jj, Hisano:2013cqa}.
In this method, the GUT-scale threshold corrections to the gauge
coupling constants are used to estimate the masses of the GUT
particles. The GUT-scale matching conditions for the gauge
coupling constants at one-loop level in the $\overline{\rm DR}$ scheme
\cite{Siegel:1979wq} in the minimal SU(5) GUT are given as follows
\cite{Weinberg:1980wa, Hall:1980kf, Ellis:2015jwa},
assuming no additional GUT-scale physics:
\begin{align}
 \frac{1}{g_1^2(M_{GUT})}&=\frac{1}{g_5^2(M_{GUT})}
+\frac{1}{8\pi^2}\biggl[
\frac{2}{5}
\ln \frac{M_{GUT}}{M_{H_C}}-10\ln\frac{M_{GUT}}{M_X}
\biggr]~,\nonumber \\
 \frac{1}{g_2^2(M_{GUT})}&=\frac{1}{g_5^2(M_{GUT})}
+\frac{1}{8\pi^2}\biggl[
2\ln \frac{M_{GUT}}{M_\Sigma}-6\ln\frac{M_{GUT}}{M_X}
\biggr]~,\nonumber \\
 \frac{1}{g_3^2(M_{GUT})}&=\frac{1}{g_5^2(M_{GUT})}
+\frac{1}{8\pi^2}\biggl[
\ln \frac{M_{GUT}}{M_{H_C}}+3\ln \frac{M_{GUT}}
{M_\Sigma}-4\ln\frac{M_{GUT}}{M_X}
\biggr]~,
\end{align}
where $g_1$, $g_2$, $g_3$, and $g_5$ are the gauge coupling constants of
U(1), SU(2)$_L$, SU(3)$_C$, and SU(5), respectively, with $g_1$ related
to the hypercharge gauge coupling $g^\prime$ through $g_1 = g^\prime
\sqrt{5/3}$, and $M_X$ and $M_\Sigma$ are the masses of the heavy gauge
bosons and the adjoint Higgs fields, respectively.
Note that these conditions do not include scale-independent terms
since we use the $\overline{\rm DR}$ scheme for the
renormalization. These equations then yield
\begin{align}
 \frac{3}{g_2^2(M_{GUT})}- \frac{2}{g_3^2(M_{GUT})}
- \frac{1}{g_1^2(M_{GUT})}
&=-\frac{3}{10\pi^2}\ln \biggl(\frac{M_{GUT}}{M_{H_C}}\biggr)
~, \nonumber \\
 \frac{5}{g_1^2(M_{GUT})}- \frac{3}{g_2^2(M_{GUT})}
- \frac{2}{g_3^2(M_{GUT})}
&=-\frac{3}{2\pi^2}\ln\biggl(
 \frac{M_{GUT}^3}{M_X^2M_{\Sigma}}\biggr)~,
\label{conditions}
\end{align}
and the upper relation allows one to evaluate $M_{H_C}$ from the coupling
constants of the SM gauge interactions at the GUT scale determined
using the RGEs~\cite{Hisano:1992mh, Hisano:1992jj, Hisano:2013cqa}.

Before concluding this subsection, we discuss the qualitative dependence of the proton
decay lifetime on the MSSM parameters. As already
mentioned above, the loop functions for the diagrams in
Fig.~\ref{fig:1loop} give rise to a factor of $\sim
M/M_{\widetilde{f}}^2$. Therefore, the proton lifetime becomes longer if
the sfermion masses are taken to be larger. In addition, as can be seen
from \eqref{eq:wilson5}, the decay amplitude contains both the
up- and down-type Yukawa couplings, which leads to a factor of $1/\sin
2\beta$. Moreover, the Higgsino-exchange contribution also has an extra
factor of $1/\sin 2\beta$. As a result, the proton decay rate is
strongly enhanced for moderate/large values of $\tan
\beta$. For these reasons, large sfermion masses and small $\tan\beta$
are favorable for evading the proton decay constraints.

\subsection{Elastic Scattering Cross Sections}
\label{sec:elsccross}

Next, we review the calculation of the neutralino-nucleus elastic scattering
cross sections that we use in the following analysis. There are two
types of interactions that induce dark matter-nuclei scattering:
spin-independent (SI) and spin-dependent (SD). Since
there is no interference between these two interactions, we can evaluate
the SD and SI scattering cross sections separately.

We first consider SI scattering. The SI elastic scattering cross
section of the neutralino LSP with a nucleus is expressed in terms of
the SI neutralino-nucleon effective coupling $f_N$ ($N=p, n$) as follows:
\begin{equation}
 \sigma_{\text{SI}} = \frac{4}{\pi}\left(\frac{m_\chi m_T}
{m_\chi+ m_T}\right)^2 \left[Z f_p +(A-Z) f_n\right]^2 ~,
\label{eq:sinuclei}
\end{equation}
where $m_\chi$ and $m_T$ are the masses of the neutralino LSP and
the target nucleus, respectively, and $Z$ ($A$) denotes the atomic (mass)
number of the target nucleus.

The SI neutralino-nucleon scattering matrix elements are induced by the exchange of
squarks and neutral Higgs bosons. To evaluate the effective coupling
$f_N$, we first obtain the neutralino-quark/gluon effective operators by
integrating out the squarks and Higgs bosons. Then, using the
nucleon matrix elements of these effective operators, we can calculate
the effective coupling $f_N$. For more details, see \cite{Drees:1993bu,
Jungman:1995df, Falk:1998xj, Ellis:2008hf, Hisano:2010ct,
Hisano:2015bma}. As a result, $f_N$ is expressed in terms of the
coefficients of the neutralino-quark effective scalar interactions,
$\alpha_{3q} \overline{\chi}\chi \overline{q}q$ (where $\chi$ and $q$
denote the neutralino LSP and quarks, respectively), as
\cite{Falk:1998xj, Ellis:2008hf}
\begin{equation}
 \frac{f_N}{m_N} = \sum_{q=u,d,s} f^{(N)}_{T_q}
\frac{\alpha_{3q}}{m_q}
+\frac{2}{27}f^{(N)}_{TG} \sum_{q=c,b,t}\frac{\alpha_{3q}}{m_q} ~.
\label{eq:sifn}
\end{equation}
Here the $m_q$ are the quark masses,
$m_N$ is the nucleon mass, the $f_{T_q}^{(N)} \equiv \langle N| m_q
\overline{q} q |N\rangle / m_N$ are the nucleon matrix elements of the
light-quark mass operators, and $f_{TG}^{(N)} \equiv 1- \sum_{q=u,d,s}
f_{T_q}^{(N)}$ denotes the gluon contribution to the nucleon mass.
We extract the values of $f_{T_q}^{(N)}$ from the pion-nucleon
$\sigma$-term $\Sigma_{\pi N}=50$~MeV and $\sigma_0 = 36$~MeV
\cite{Borasoy:1996bx}~\footnote{This choice corresponds to $y\equiv
2\langle N|\overline{s}s|N\rangle /\langle N| \overline{u}u +
\overline{d}d |N\rangle = 0.28$, which is larger than the values obtained in
lattice QCD simulations \cite{Ohki:2008ff}. We note that the
uncertainties in these quantities significantly affect the resultant
scattering cross sections \cite{Ellis:2008hf}.}. Analytic
expressions for the $\alpha_{3q}$ are presented in
\cite{Falk:1998xj,Ellis:2008hf}. The second term of the right-hand side in
Eq.~\eqref{eq:sifn} represents the (long-distance) contribution of heavy
quarks to the neutralino-gluon interactions, which can be related to the
quark couplings $\alpha_{3q}$ via the triangle diagrams associated with
the trace anomaly of the energy-momentum tensor
\cite{Shifman:1978zn}. In Eq.~\eqref{eq:sifn}, we neglect the effects of the
twist-2 operators \cite{Drees:1993bu} as well as the short-distance
contribution of quarks to the gluon operators \cite{Hisano:2010ct,
Hisano:2015bma}, since they are numerically small when squarks are
rather heavy, which is the case we discuss below.

In the models we study, the dominant contribution to $\alpha_{3q}$ is given by the
exchange of neutral Higgs bosons, since the squarks tend to be
heavy. Moreover, in a wide range of parameter space, the Higgs sector is
close to the decoupling limit, and the LSP is a bino-Higgsino mixed state
with $|M_1|$, $|\mu|$, $|M_1-|\mu|| \gg m_Z$. In this case, the
expression for $\alpha_{3q}$ is approximated by
\begin{equation}
 \alpha_{3q} \simeq - \frac{g_2^2  m_q \tan^2 \theta_W}{4(\mu^2-M_1^2)}
\left(\frac{M_1+\mu \sin 2\beta}{m_h^2} +\frac{\mu \cos 2\beta }{m_H^2}
 c_q \right) ~,
\label{eq:siappbino}
\end{equation}
for the bino LSP case, while
\begin{equation}
 \alpha_{3q} \simeq -\frac{\text{sgn}(\mu)g_2^2m_q}{8}\left(\frac{\tan^2 \theta_W}
{M_1-|\mu|}+\frac{1}{M_2-|\mu|}\right)
\left[\frac{1+\text{sgn}(\mu)\sin 2\beta}{m_h^2}+
\frac{\text{sgn}(\mu)\cos 2\beta}{m_H^2}c_q\right] ~,
\label{eq:siapphiggsino}
\end{equation}
for the Higgsino LSP case. Here, $\theta_W$ is the weak mixing angle,
$m_h$ ($m_H$) is the mass of the
lighter (heavier) neutral Higgs boson, and $c_q = \cot \beta $ and
$-\tan \beta$ for up- and down-type quarks, respectively. As can be seen
from these expressions, the neutralino-nucleon scattering cross sections
decrease when the difference between $M_1$ and $|\mu|$ gets large. In
addition, it is found that the SI effective coupling depends on the sign
of $\mu$ and, in particular, when $\mu$ is negative the coupling can be
significantly suppressed due to cancellations (this feature is sometimes
called the ``blind spot'' \cite{Falk:1998xj,cancel,Cheung:2012qy}).

We next discuss SD scattering, for which the
neutralino-nucleus scattering cross section is given by
\begin{equation}
\sigma_{\text{SD}}= \frac{32}{\pi}G_F^2\Lambda^2 J (J+1)\left(\frac{m_\chi m_T}
{m_\chi+ m_T}\right)^2 ~,
\end{equation}
where $G_F$ is the Fermi constant, $J$ is the total spin of the target
nucleus, and
\begin{equation}
 \Lambda = \frac{1}{J}(a_p \langle S_p \rangle + a_n \langle S_n
  \rangle) ~,
\end{equation}
with $\langle S_p \rangle $ ($\langle S_n \rangle $) being the expectation
value of the total spin of protons (neutrons) in the target nucleus. The
SD neutralino-nucleon effective coupling, $\alpha_N$, is expressed as
\begin{equation}
 a_N =\sum_{q=u,d,s} \frac{\alpha_{2q}}{\sqrt{2} G_F} \Delta^{(N)}_q ~,
\end{equation}
where we use $\Delta^{(N)}_q$ given in
\cite{Ellis:2008hf}, and $\alpha_{2q}$ denotes the SD neutralino-quark
couplings, which are induced by the exchange of $Z$-boson and
squarks. The analytic formula for $\alpha_{2q}$ is again given in
\cite{Falk:1998xj, Ellis:2008hf}. As in the SI case, the squark contribution
is suppressed compared with the $Z$ boson contribution in the parameter
region we are interested in. Furthermore, when $|M_1|$, $|\mu|$,
$|M_1-|\mu|| \gg m_Z$, $a_{2q}$ is approximated by
\begin{equation}
 a_{2q} \simeq  \frac{g_2^2\tan^2 \theta_W}{8 (\mu^2-M_1^2)} \cos
  2\beta T_q^3 ~,
\end{equation}
for a bino-like LSP, whilst for a Higgsino-like LSP we have
\begin{equation}
 \alpha_{2q}\simeq  \frac{g_2^2}{16|\mu|} \cos  2\beta T_q^3
\left(
\frac{\tan^2\theta_W}{M_1}+\frac{1}{M_2}
\right) ~,
\end{equation}
where $T_q^3$ denotes the third component of the SU(2)$_L$
generators.

As seen above, the neutralino-nucleus scattering cross sections are
suppressed when gauginos/Higgsinos are heavy. In such cases,
electroweak-loop contributions may dominate the tree-level Higgs and $Z$
boson contributions \cite{Hisano:2010ct, Hisano:2004pv}.
It turns out, however, that in the case of a bino-Higgsino
LSP, the electroweak loop contributions are quite small
\cite{Hisano:2015rsa}, and thus we neglect them in our calculation.

Because of the coherent nature of the SI neutralino-nucleus scattering
as shown in Eq.~\eqref{eq:sinuclei}, the current and future
direct detection experiments are much more sensitive to the SI scattering
compared to SD scattering. For this reason, we mainly discuss
SI scattering in the following.

\section{Results}

\subsection{CMSSM}

\begin{figure}[htb!]
\begin{minipage}{8in}
\includegraphics[height=3.3in]{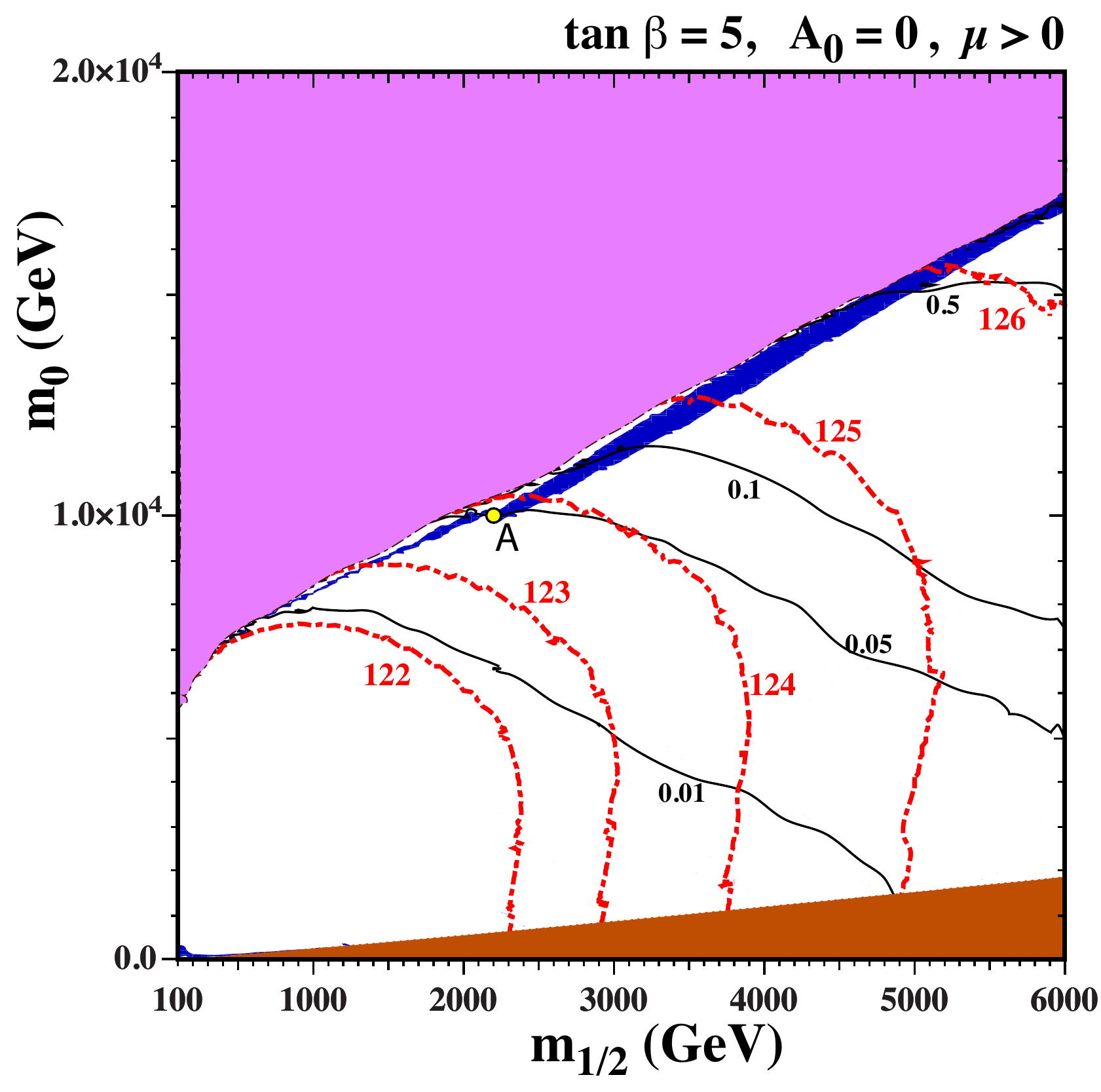}
\hspace*{-0.17in}
\includegraphics[height=3.3in]{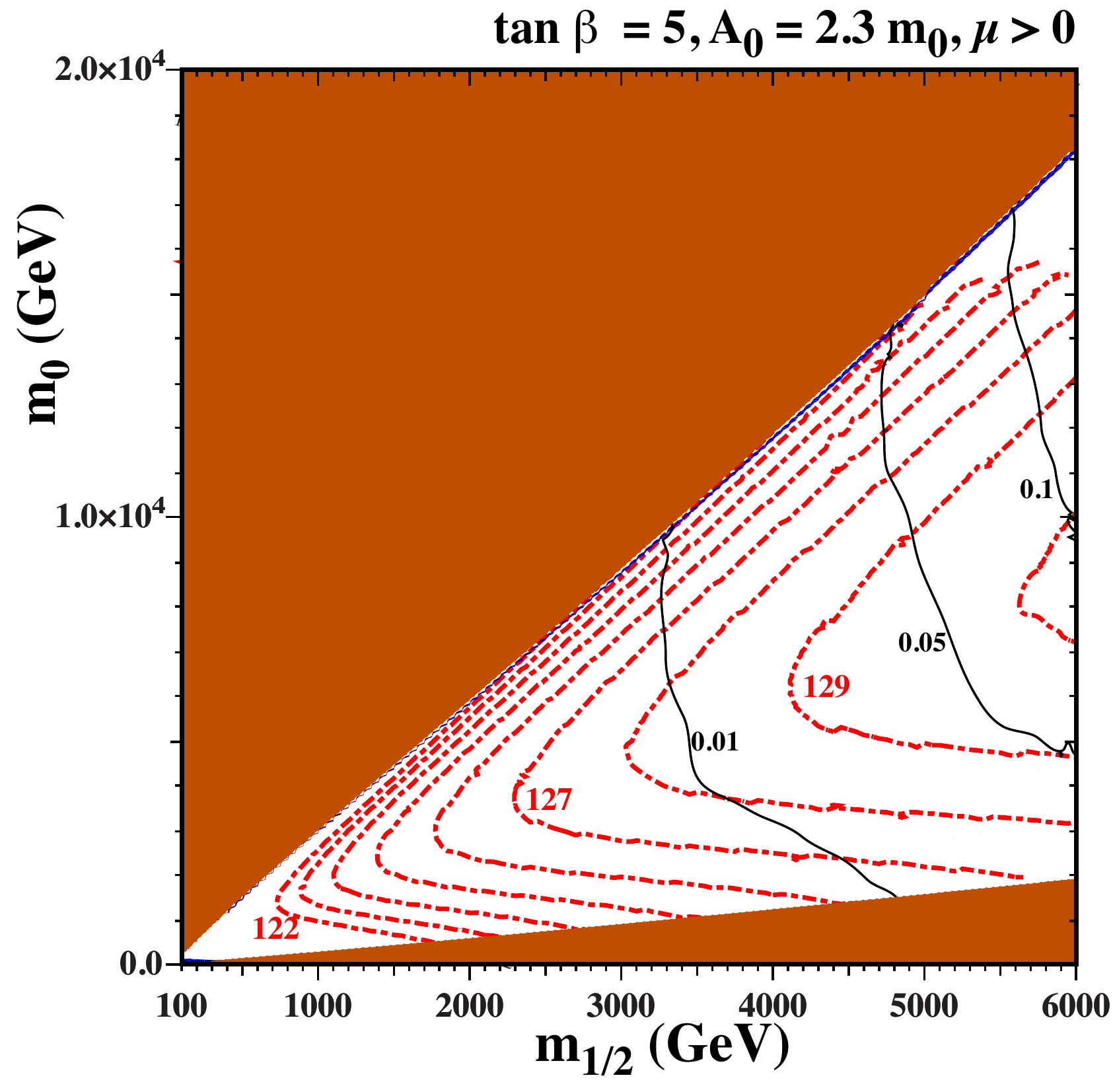}
\hfill
\end{minipage}
\begin{minipage}{8in}
\includegraphics[height=3.3in]{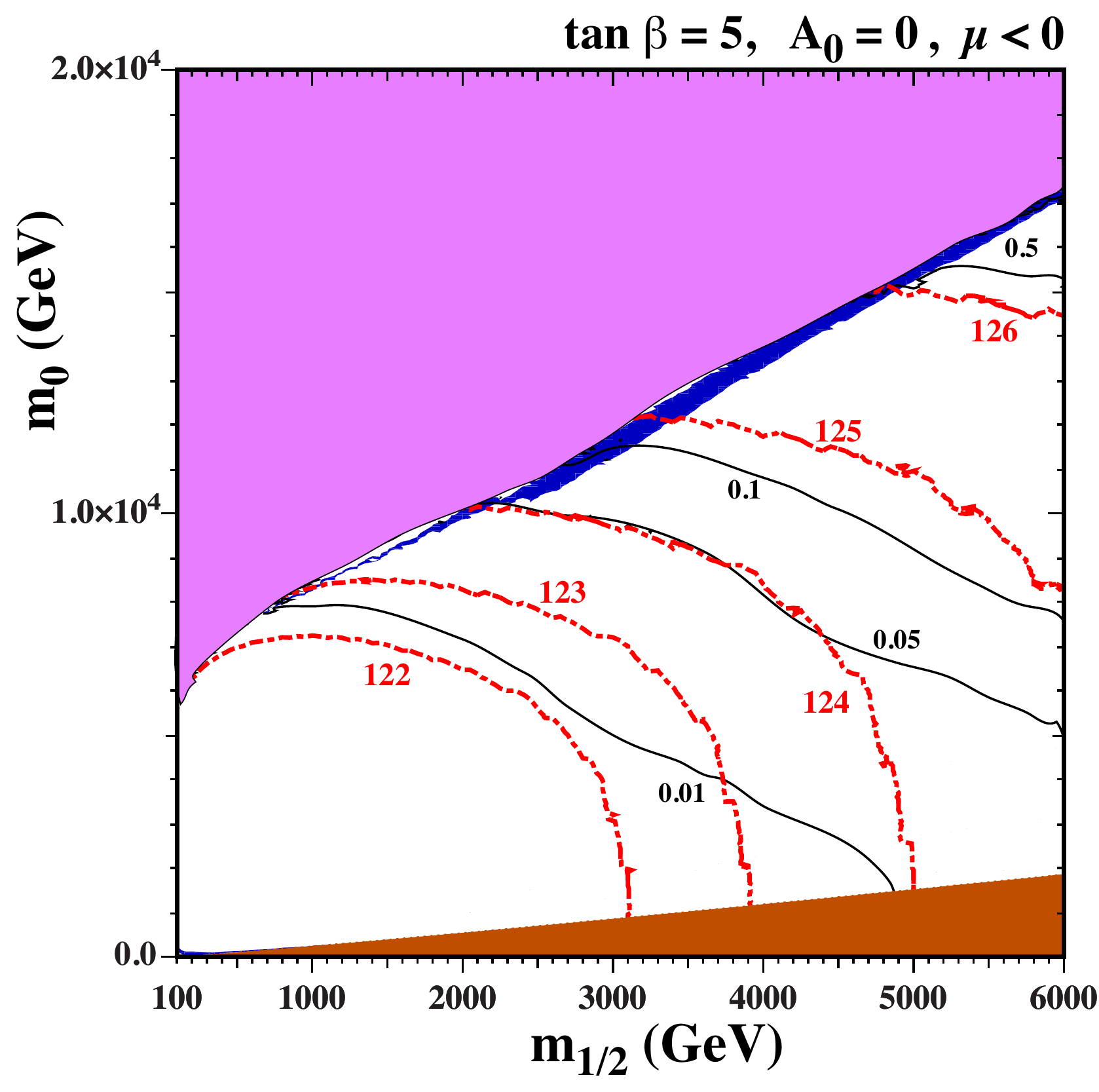}
\hspace*{-0.17in}
\includegraphics[height=3.3in]{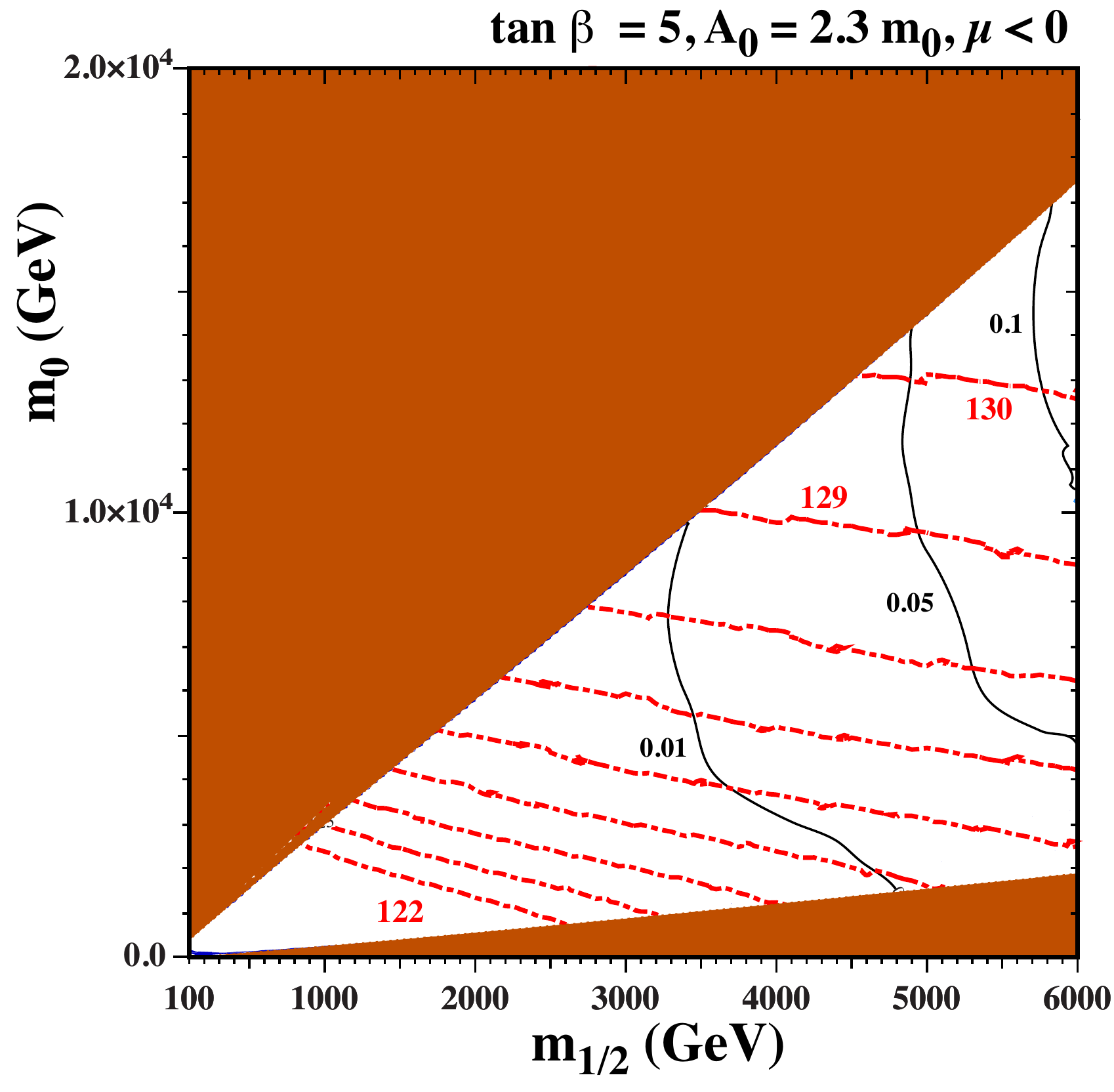}
\end{minipage}
\caption{
{\it
The CMSSM $(m_{1/2}, m_0)$ planes for $\tan \beta = 5$ with $\mu > 0$ (upper) and $\mu < 0$ (lower), and with
 and $A_0 = 0$ (left) and $A_0 = 2.3 \, m_0$ (right).
In the light mauve shaded regions, it is not possible to satisfy the electroweak symmetry breaking (EWSB) conditions.
In the brown shaded regions, the LSP is charged and/or colored. The dark blue shaded regions
shows the areas where $0.06 < \Omega_\chi h^2 < 0.2$.  The red dot-dashed contours
indicate the Higgs mass, labelled in GeV, and the solid black contours indicate the proton lifetime
in units of $10^{35}$~yrs. The point labelled A refers to the point tested for phase dependence in the
Appendix.
}}
\label{fig:CMSSM}
\end{figure}

In view of the discussion in Section \ref{pdecay}, in our study of the proton
lifetime we focus on relatively small values of $\tan \beta$, and have chosen
$\tan \beta = 5$ in Fig.~\ref{fig:CMSSM}.  For larger values of $\tan \beta$,
the proton lifetime becomes smaller than the current experimental bound,
and minimal SU(5) is not viable.

We show in Fig.~\ref{fig:CMSSM} four examples of $(m_{1/2}, m_0)$ planes in the CMSSM
with $\tan \beta = 5$.  In the left panels we choose $A_0 = 0$, whereas in the right panels
we choose $A_0 = 2.3 \, m_0$. We take $\mu > 0$ in the upper panels and $\mu < 0$ in the lower panels.
Higgs mass contours are shown as red dot-dashed curves labelled
by $m_h$ in GeV in 1 GeV intervals starting at 122 GeV. We recall that to calculate $m_h$
we use {\tt FeynHiggs}~\cite{FH}, which carries a roughly $\pm$1.5 GeV uncertainty.
In the left panels, the light mauve shaded region in the upper part of the figure is excluded
because there are no solutions to the Higgs minimization conditions: along this boundary $\mu^2 = 0$.
Electroweak symmetry breaking (EWSB) fails here because the Higgs soft masses at the GUT scale are large
and the RG running to the weak scale does not suppress the Higgs soft masses
sufficiently for EWSB to occur.  Because large gluino masses can assist electroweak symmetry breaking
effectively at two loops, the value of $m_0$ that is allowed increases for increasing $m_{1/2}$.
Just below the region where EWSB fails, there is a dark blue shaded region where the relic density falls within the
range determined by CMB experiments~\cite{Planck15}. Since the relic density of dark matter is now
determined quite accurately  ($\Omega_\chi h^2  = 0.1193 \pm 0.0014$), for the purpose of visibility we have
shown the strip for which the relic density lies in the range $[0.06, 0.20]$.
This strip is in the focus-point region~\cite{fp,fp2} where the Higgsinos are much lighter than the stops.
The correct dark matter density is realized either by coannihilation of the Higgsino with the bino when $m_{1/2}$ is smaller,
or by Higgsino annihilations when the Higgsino mass is of order a TeV for larger values of $m_{1/2}$.
The TeV-scale Higgsino dark matter region continues well beyond the bounds of the figure.
We note also that the brown shaded regions at the bottoms of the panels are excluded because there the LSP is the lighter
charged stau lepton. The planes also feature stau-coannihilation strips close to the boundary
of the brown shaded region. They extend to about $m_{1/2} \simeq 1$ TeV, but are very difficult to see
on the scale of this plot, even with our enhancement of the relic density range.
We note that for this value of $\tan \beta$ there
are no relevant constraints from rare B decays.

Contours of the proton lifetime using down-type Yukawa couplings (see
the discussion given in Sec.~\ref{pdecay}) are shown as solid black curves that are labelled in units of
$10^{35}$ yrs.  Thus the limit of $\tau_p > 5 \times 10^{33}$~yrs would exclude everything below the
curve labelled 0.05.  For the nominal value of $m_h = 125$ GeV, neglecting the
theoretical uncertainties in the calculation of $m_h$, we see that
in the upper left plane of Fig.~\ref{fig:CMSSM} the Higgs contour intersects the focus-point
region where $\tau_p \approx 2 \times 10^{34}$~yrs. Much of the focus-point strip in this figure
may be probed by future proton decay experiments. Changing the sign of $\mu$ has almost no effect
on the proton lifetime, as seen in the lower left panel of Fig.~\ref{fig:CMSSM},
but the calculated Higgs mass is smaller by $\sim 1$ GeV,
which is less than the uncertainty in the {\tt FeynHiggs} calculation of $m_h$.

In the right panels of Fig.~\ref{fig:CMSSM}, since increasing $A_0$ drives a larger splitting between
the two stops, there are excluded regions shaded in brown in the upper halves of the panels,
where the lighter
stop becomes the LSP.  Close to this boundary (but again difficult to see) there is a narrow blue strip
where the dark matter density is brought into the allowed range by coannihilation with
the lighter stop. The relatively large value of $A_0 = 2.3 \, m_0$
leads to large Higgs masses in most of the plane, but the Higgs mass is somewhat smaller
along this strip for $\mu > 0$. We recall that the Higgs mass is sensitive to the off-diagonal element in the
stop mass matrix, which is proportional to $X_t = A_t + \mu \cot \beta$ and peaks when $X_t$ is roughly
2.5 times the geometric mean of the two stop masses. For positive $\mu$, $X_t$ is relatively large
and we are past the peak where $m_h$ is maximized.  In contrast, for $\mu < 0$, $X_t$ is smaller
(as there is some cancellation between the two terms) and we are closer to the peak and $m_h$ is larger. This effect is pronounced along the upper left edge because the stop is much lighter in this region.
For $\mu > 0$, the $m_h = 124$~GeV contour, which is consistent with the experimental value when uncertainties in the theoretical calculation are considered, intersects the strip at $m_{1/2} \sim 4.7$ TeV where
$\tau_p = 5 \times 10^{33}$~yrs. Note that we terminate the larger $m_h$ contours where the calculated value
becomes unreliable: near the endpoints of these curves, the uncertainty in the {\tt FeynHiggs}
calculation of $m_h$ is $\gtrsim 5$~GeV.
For $\mu < 0$, the $m_h = 125$ GeV contour intersects the stop coannihilation strip when $m_{1/2} \sim 1.4$
TeV and the proton lifetime is significantly smaller ($< .001$ in these units).
For slightly lower $A_0$ than the value $2.3 \, m_0$ shown in these panels
of Fig.~\ref{fig:CMSSM}, large uncertainties in $m_h$ from {\tt FeynHiggs}
appear when $\tau_p < 5 \times 10^{33}$~yrs. When $A_0/m_0 \lesssim 2.0$ the stop coannihilation strip
is no longer present.  On the other hand, when $A_0/m_0 \gtrsim 2.4$ the central value of the Higgs mass along
the stop strip drops below 122 GeV when $\mu > 0$, which is unacceptably small.

\begin{figure}[htb!]
\begin{minipage}{8in}
\includegraphics[height=2.14in]{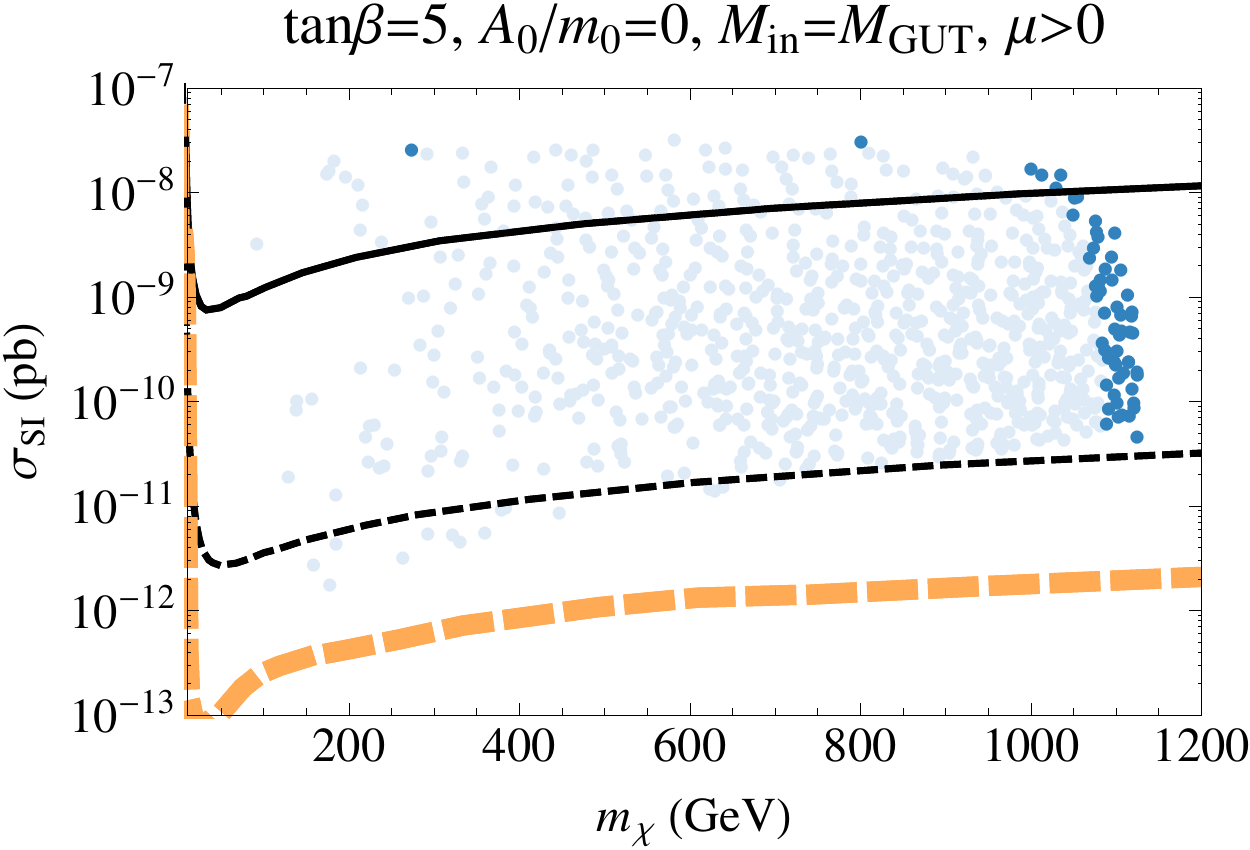}
\includegraphics[height=2.14in]{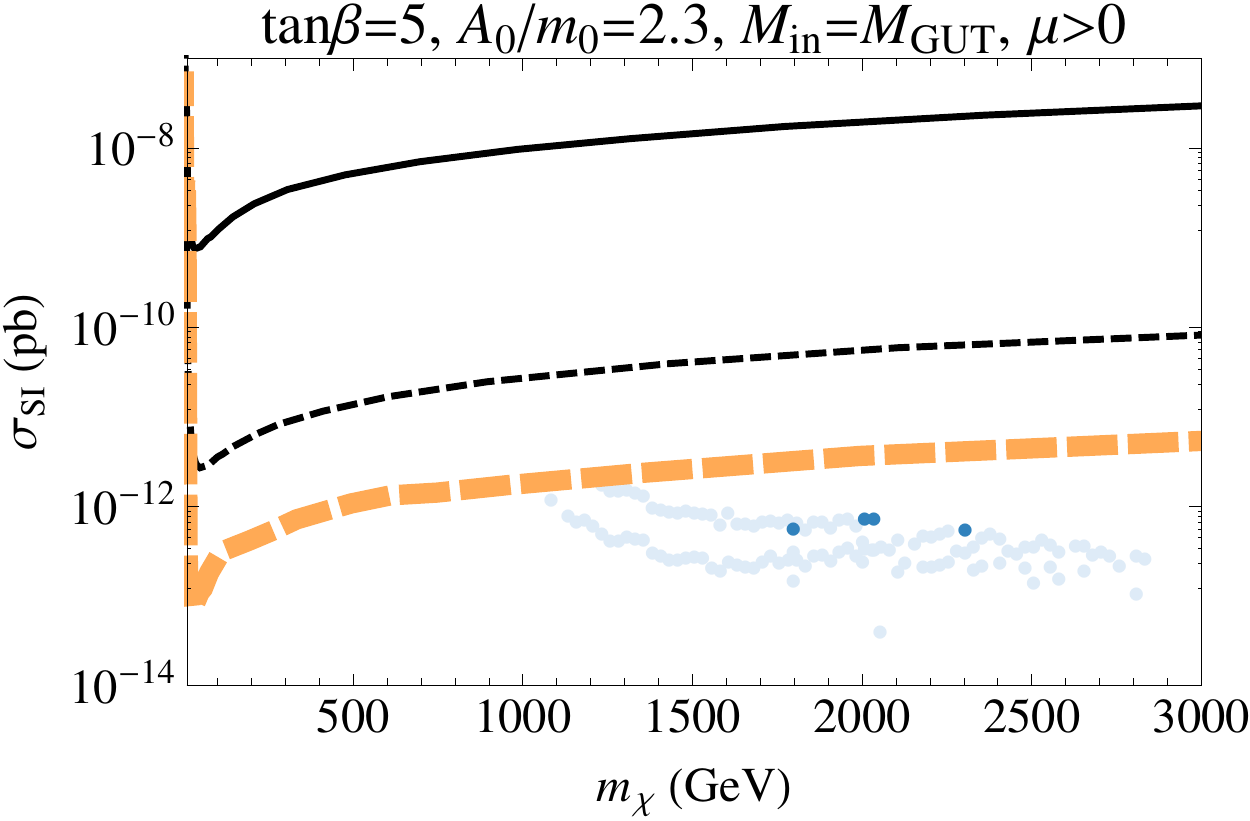}
\hfill
\end{minipage}
\begin{minipage}{8in}
\includegraphics[height=2.14in]{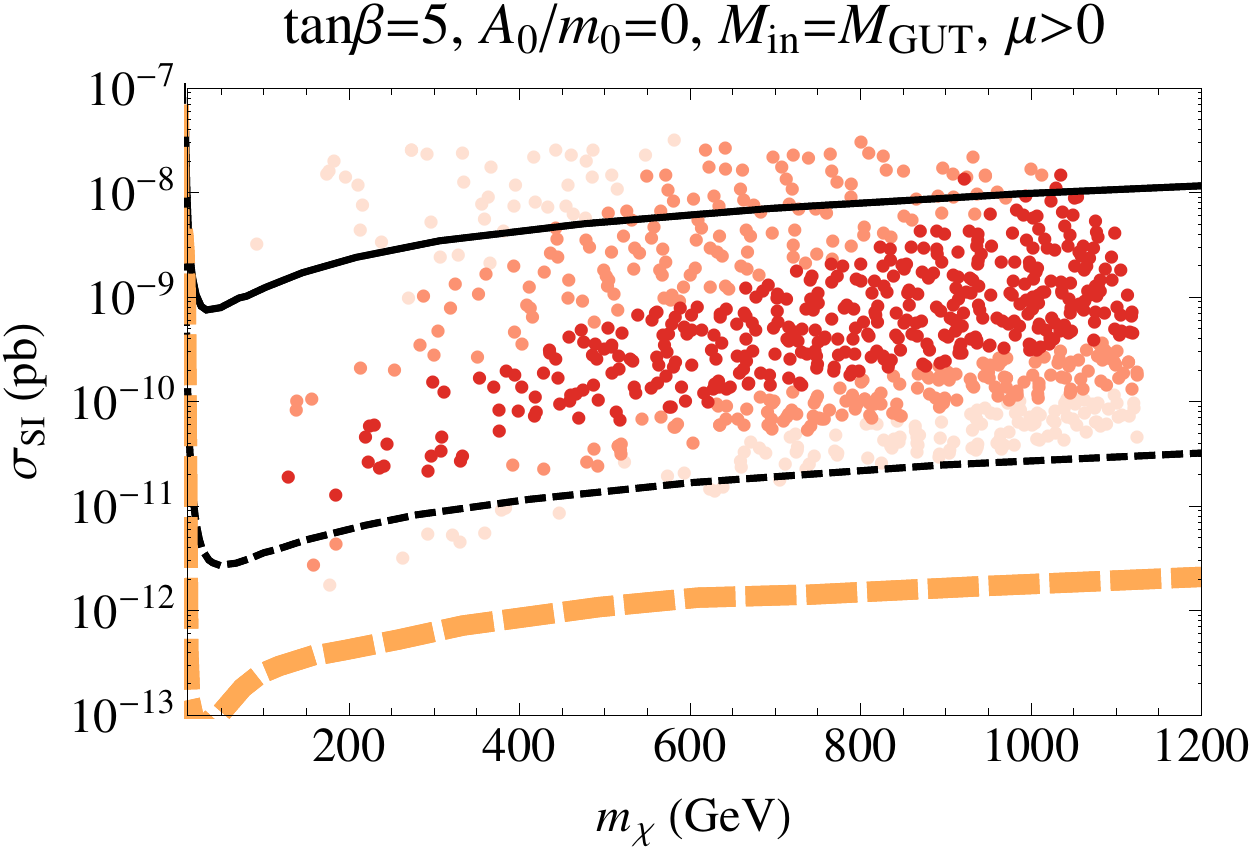}
\includegraphics[height=2.14in]{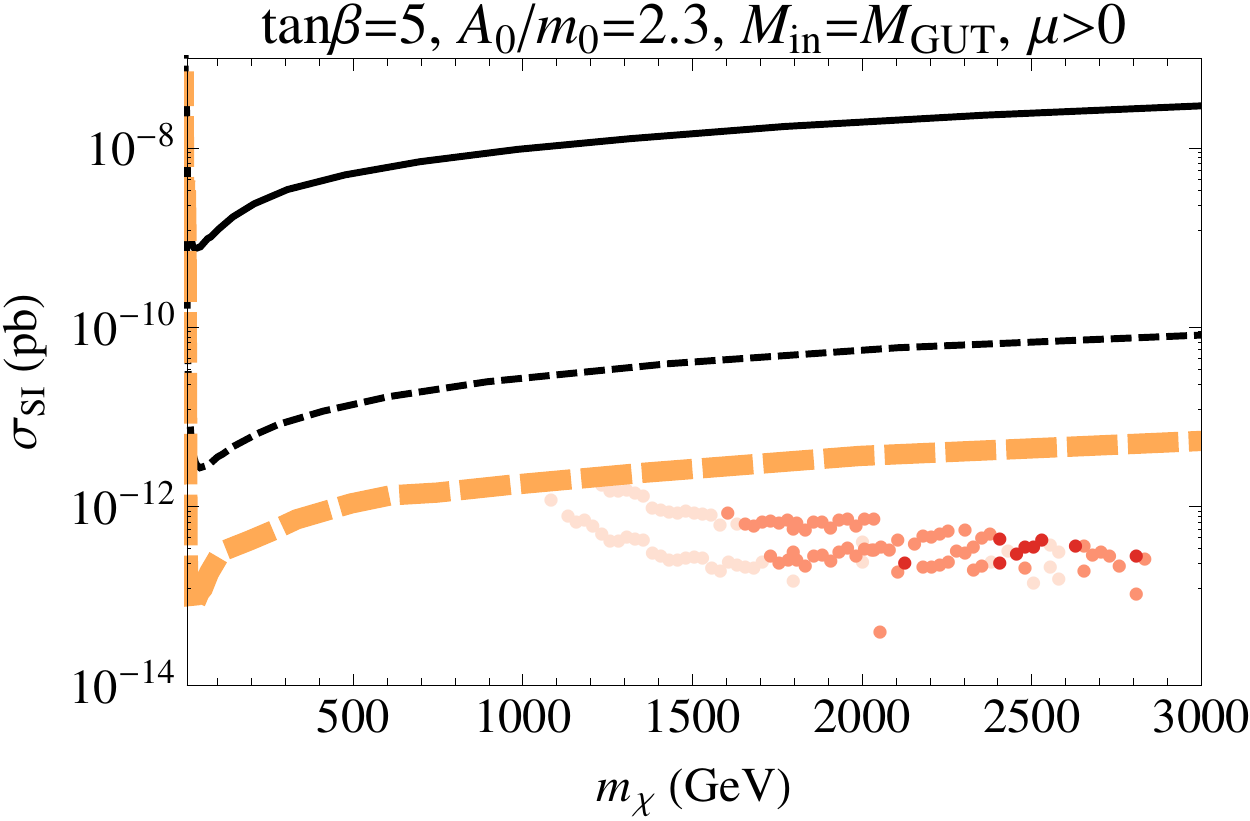}
\hfill
\end{minipage}
\caption{
{\it
The spin-independent elastic scattering cross section in the
CMSSM as a function of the neutralino mass for $\mu > 0$, with
$\tan \beta = 5$ and $A_0 = 0$ (left) and $A_0 = 2.3 \, m_0$ (right).
The upper panels show points where the relic density is within 3$\sigma$ of
the central Planck value colored darker blue, and those where the relic
density is below the Planck value as lighter blue points.  The lower panels show
the same set of points colored according to the calculation of the Higgs mass:
124--126 GeV (darkest), 123--124 and 126--127 GeV (lighter), 122--123 and 127--128 GeV (lightest).
The black solid curve is the current LUX bound. The black dashed curve is the projected LZ sensitivity and
the dashed orange curve is the neutrino background level. }}
\label{fig:CMSSMSI}
\end{figure}

We show in Fig.~\ref{fig:CMSSMSI} the spin-independent cross section,
$\sigma_{\rm SI}$, as a
function of the neutralino mass for the two upper panels in Fig.~\ref{fig:CMSSM} with $\mu > 0$.
The points in each panel represent results of a scan of the parameter space.
In the upper panels, darker points fall within 3$\sigma$ of the dark matter relic density that fits best the Planck data.
Lighter points have smaller relic densities and should not be excluded. However,
whenever the relic density is below the central value determined by Planck, we
scale the cross section downwards by the ratio of the calculated density
to the Planck density. From these panels, we find that the $A_0 = 0$
cases give relatively large SI scattering cross sections, while those
for $A_0 = 2.3m_0$ are significantly suppressed. In the case of $A_0 =
0$, the values of $\mu$ and $M_1$ are close to each other, and thus the LSP
is a well-mixed bino-Higgsino state. This leads to a large SI scattering
cross section, as can be seen from Eqs.~\eqref{eq:siappbino} and
\eqref{eq:siapphiggsino}. The set of darkly shaded points with good relic density
are found mostly at $m_\chi \simeq 1100$ GeV due to the fact that these points are
mainly Higgsino LSPs. As the bino mass is increased, the scattering cross section
decreases. However, the points sampled here all have $m_h < 128$ GeV which
produces the lower boundary of the points displayed. Because of the constraints coming from the Higgs mass, the scattering cross sections for all dark matter candidates in these models are accessible at LZ.
Along the focus point, the LSP mass varies downward as $\mu$ is decreased and the cross
section is maximal at around 3 $\times 10^{-8}$ pb.   Due to the small uncertainty in the Planck
relic density, we find very few darkly shaded points in this region. Points with
smaller cross section are found between the focus point strip and the no-EWSB boundary
where the relic density is below the Planck density.
On the other hand, for the $A_0 = 2.3m_0$
cases, the LSP is almost pure bino and squarks are quite heavy; for
these reasons, we obtain very small SI cross sections.
The solid curve in Fig.~\ref{fig:CMSSMSI} corresponds to
the current LUX limit \cite{lux} and thus some models
(including the focus point models) are excluded by this limit when $A_0 = 0$.
The thin black dashed curve corresponds to the projected LZ  sensitivity \cite{Malling:2011va,cushman}
and almost all of the points sampled when $A_0 = 0$ are therefore testable.
The thick orange dashed line corresponds
to the irreducible neutrino background \cite{neutrino,cushman}. All of the points sampled
when $A_0 = 2.3 m_0$ fall below the neutrino background and probing them would require a
directional recoil detector~\cite{Fairbairn}.

In the lower panels of Fig.~\ref{fig:CMSSMSI}, we see the same points,
now colored to show the Higgs mass ranges. The darkest points have the
calculated Higgs mass in the range 124--126~GeV, medium shaded points have $m_h$ in the ranges 123--124~GeV or 126--127~GeV, and the lightest points have
$m_h$ in the ranges 122--123~GeV or 127--128~GeV. All of these are
compatible with the experimental measurement, within twice the {\tt
FeynHiggs} uncertainty.

\begin{figure}[htb!]
\begin{minipage}{8in}
\includegraphics[height=2.14in]{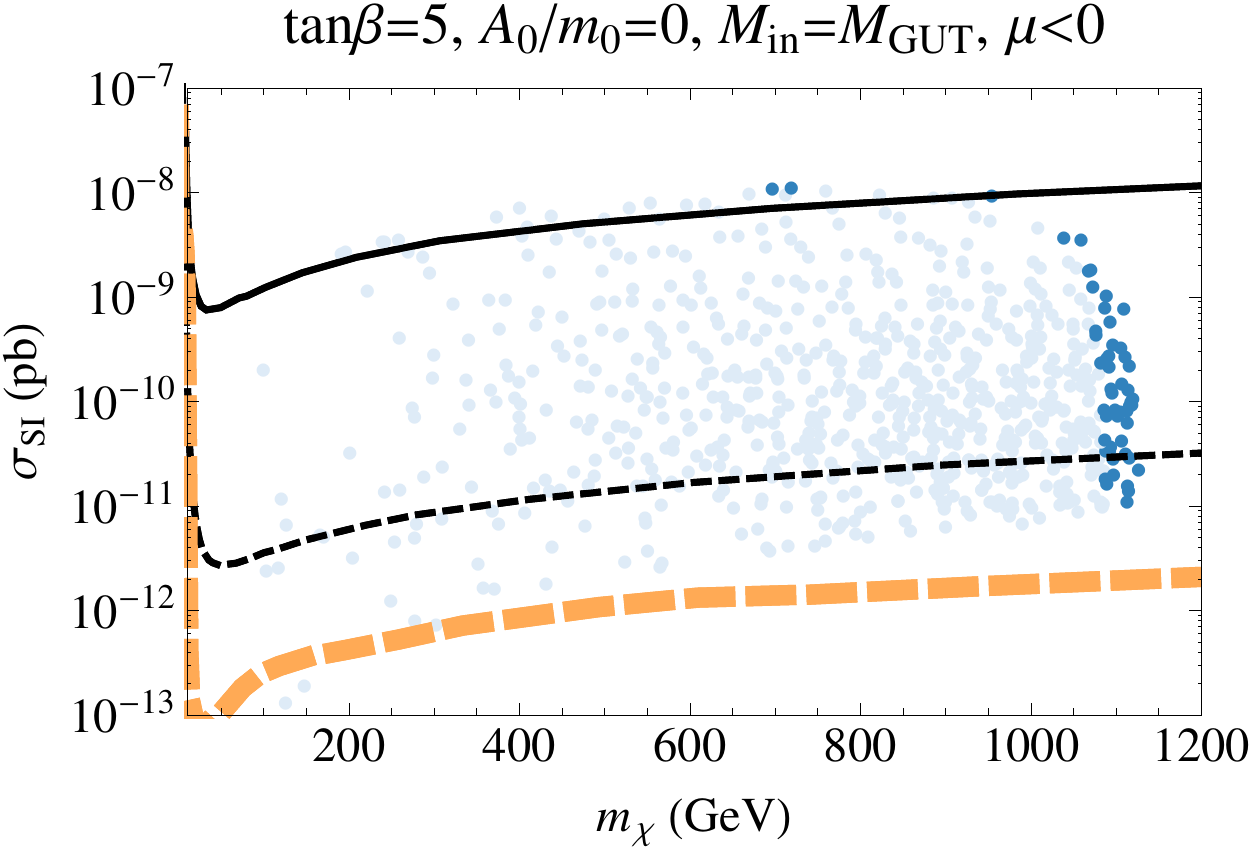}
\includegraphics[height=2.14in]{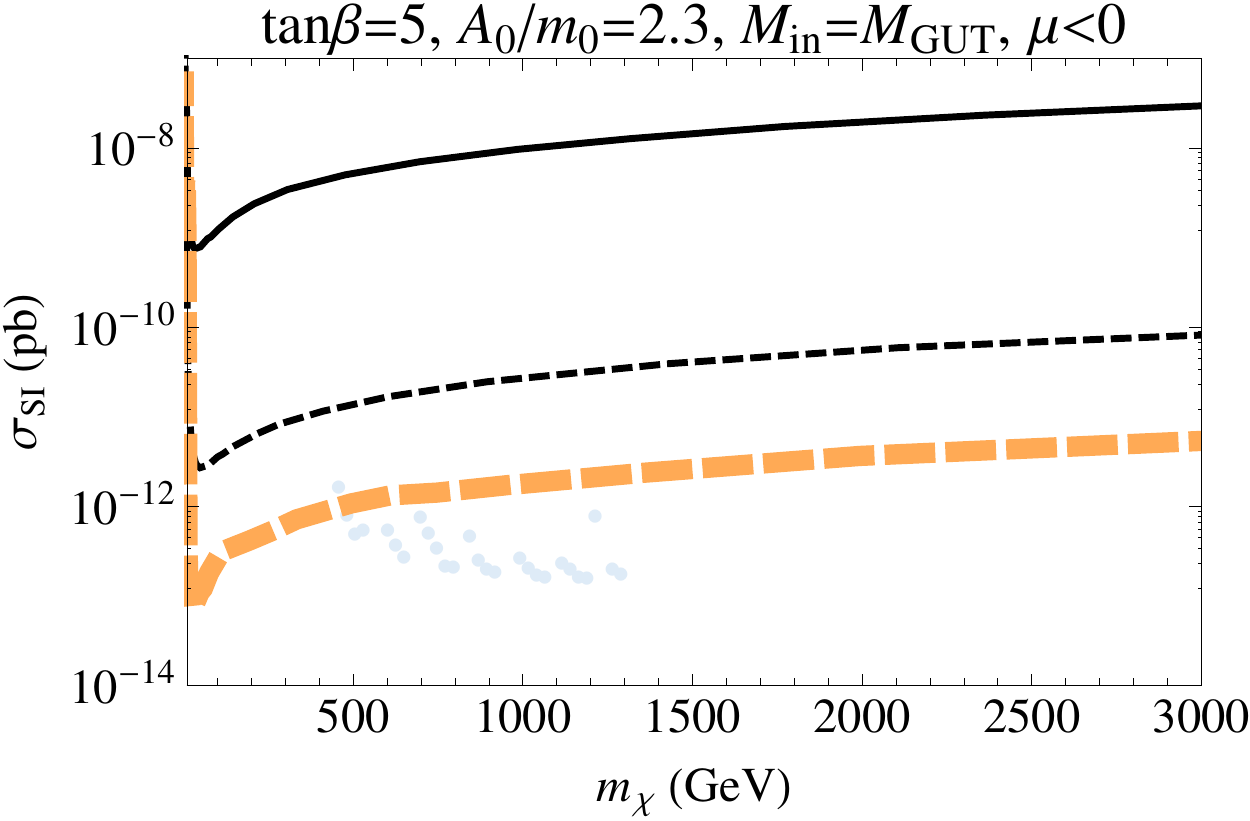}
\hfill
\end{minipage}
\begin{minipage}{8in}
\includegraphics[height=2.14in]{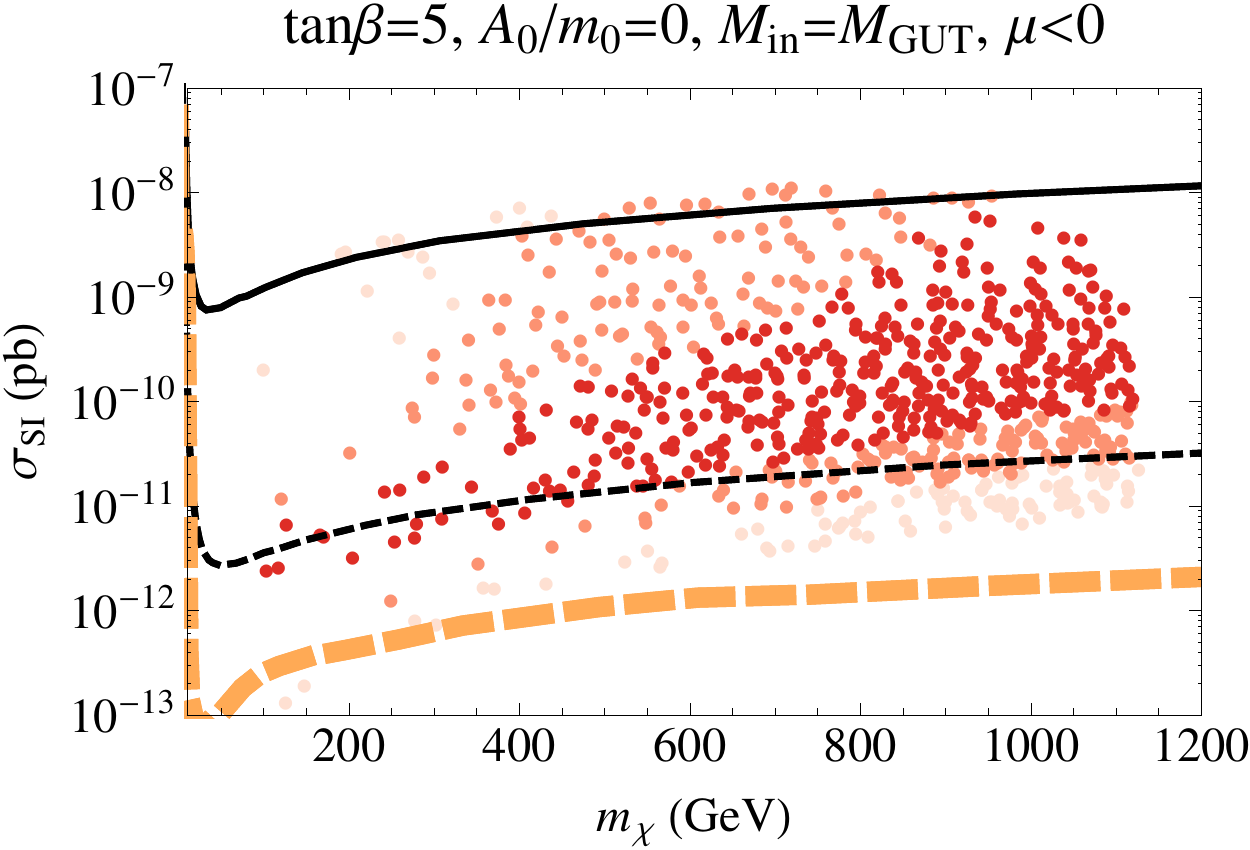}
\includegraphics[height=2.14in]{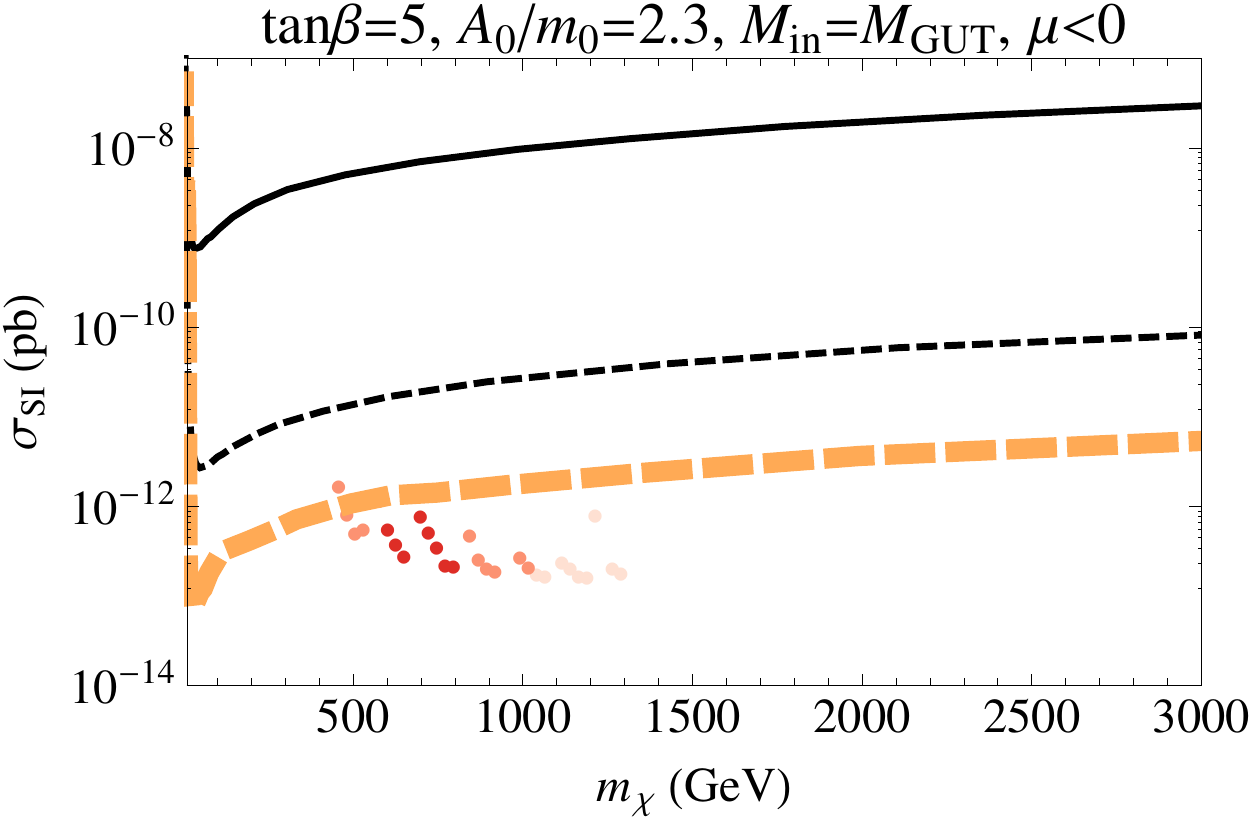}
\hfill
\end{minipage}
\caption{
{\it
As in Fig.~\protect\ref{fig:CMSSMSI}, but for $\mu < 0$. }}
\label{fig:CMSSMSIn}
\end{figure}

In Fig.~\ref{fig:CMSSMSIn}, we show the points corresponding to the
lower panels of  Fig.~\ref{fig:CMSSM} with $\mu < 0$. There is
relatively little change in the scattering cross sections for $\mu
<0$. When $A_0 = 0$, the cross sections are in general somewhat lower
due to the cancellation mentioned in Sec.~\ref{sec:elsccross}. For $A_0
= 2.3m_0$ the points have moved to lower $m_\chi$, but remain for the
most part unobservable. We again see that the Higgs mass constraint puts the bulk of the points within reach of LZ.  

\begin{figure}[htb!]
\begin{minipage}{8in}
\includegraphics[height=2.14in]{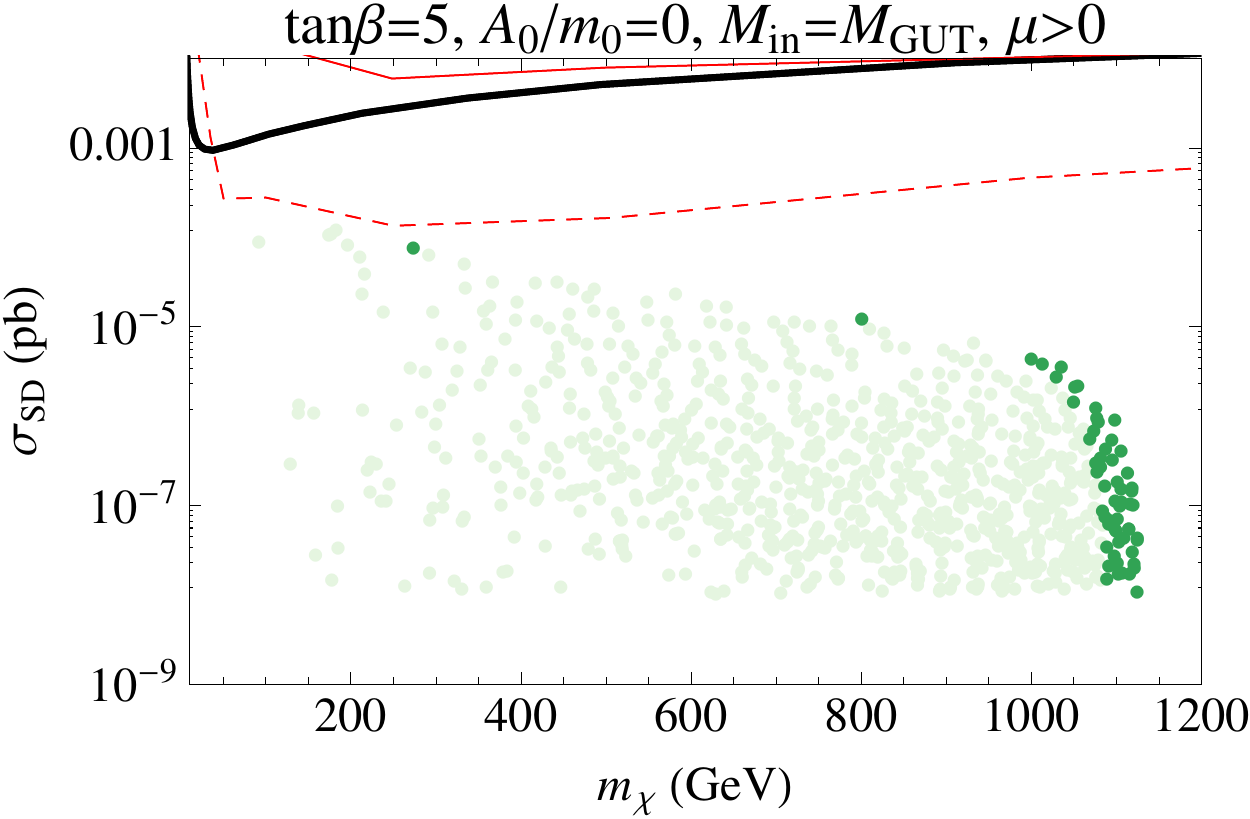}
\includegraphics[height=2.14in]{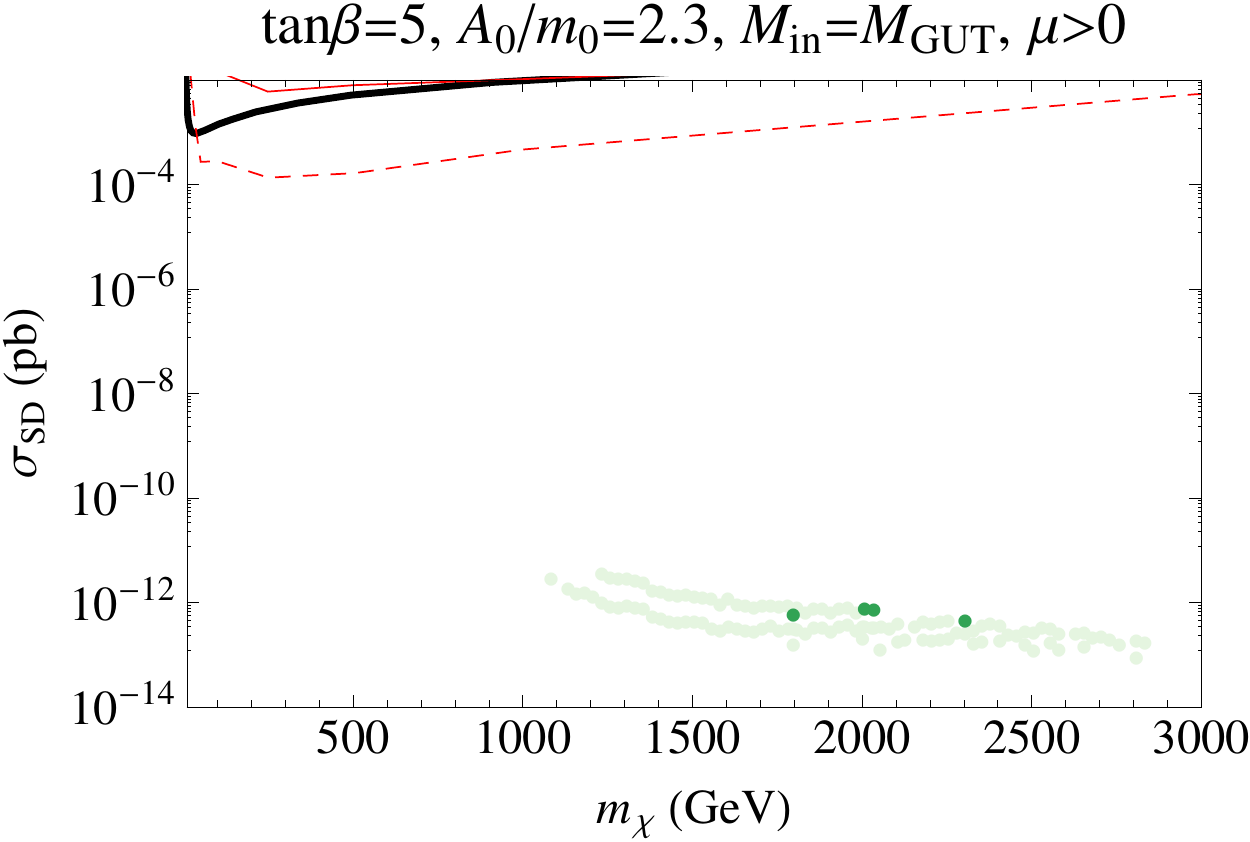}
\hfill
\end{minipage}
\begin{minipage}{8in}
\includegraphics[height=2.14in]{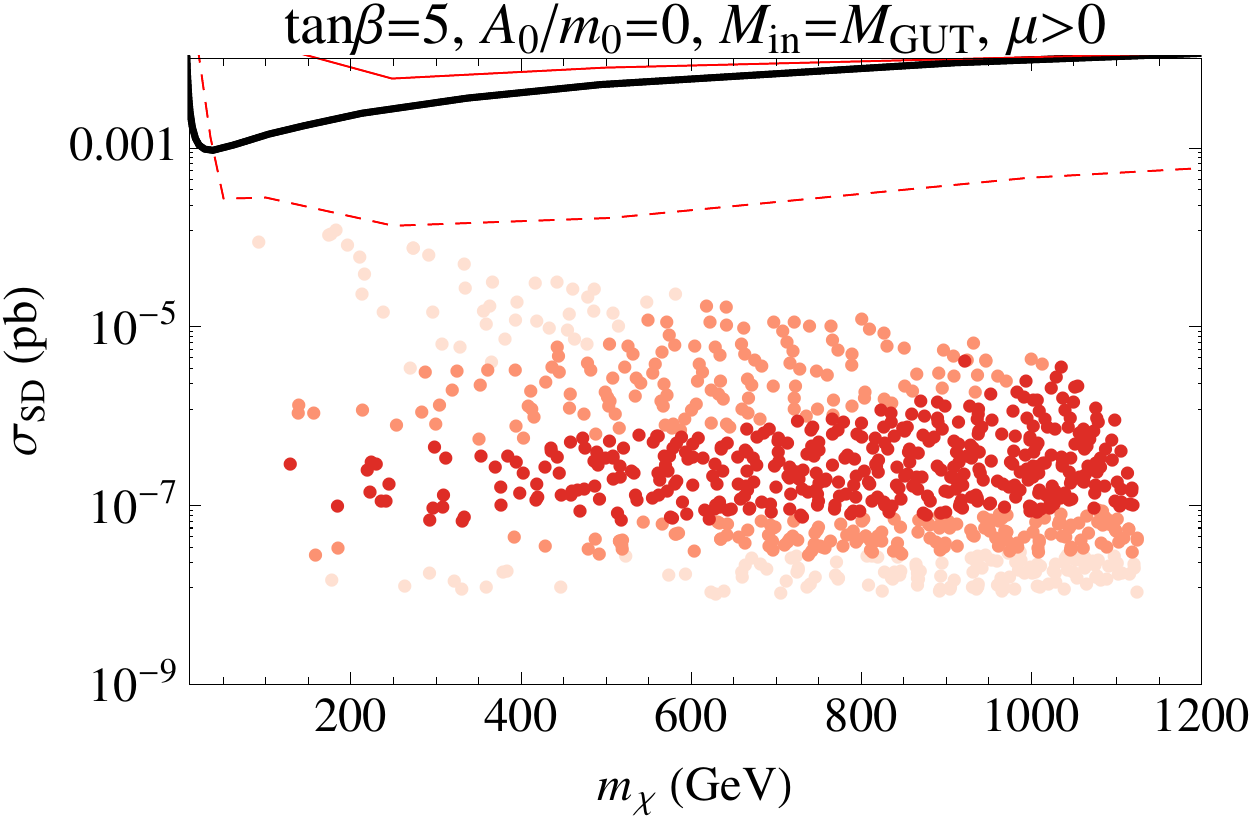}
\includegraphics[height=2.14in]{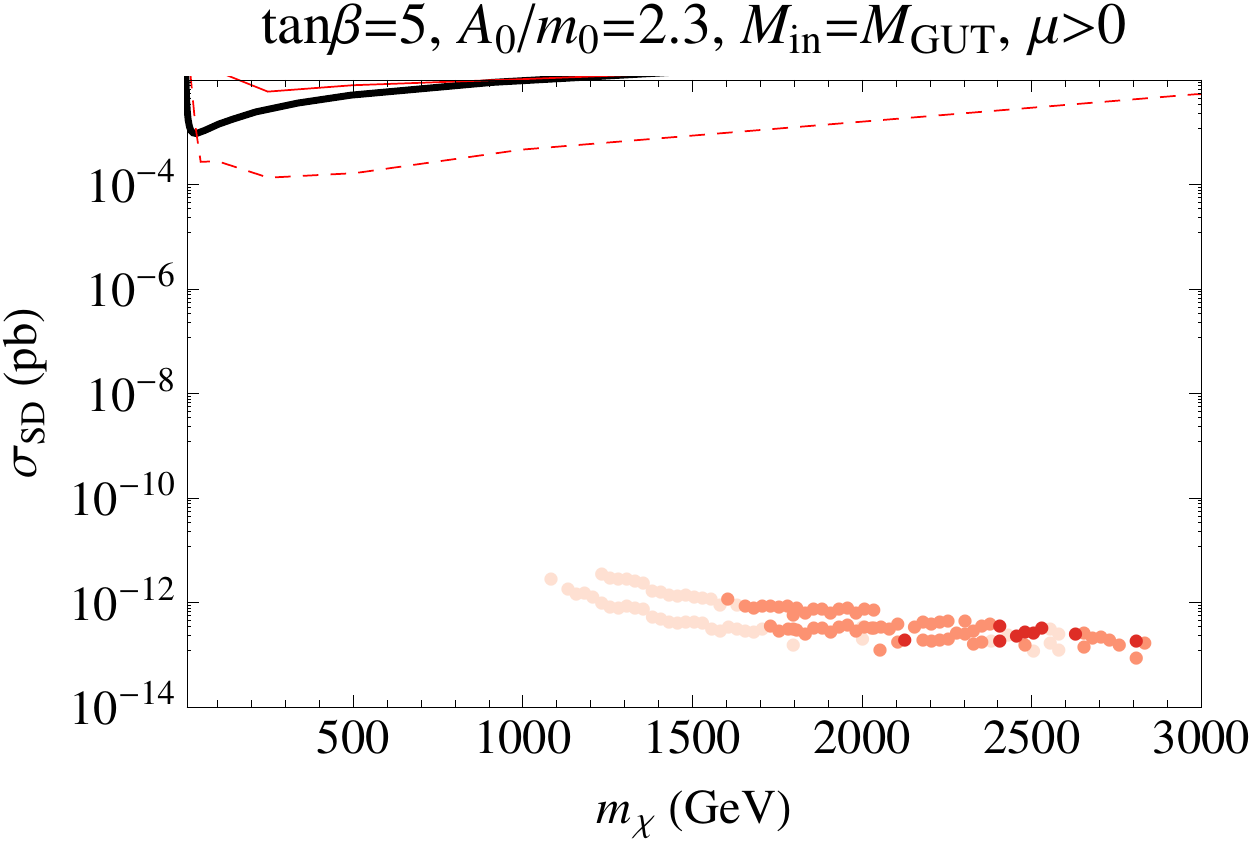}
\hfill
\end{minipage}
\caption{
{\it
As in Fig.~\protect\ref{fig:CMSSMSI}, but showing the spin-dependent cross section.
The solid curve is the current PICO bound~\protect\cite{pico}, and the red solid/dashed curves are the
IceCube bounds~\protect\cite{ice} assuming annihilations into ${\bar b}b/W^+W^-$, respectively. }}
\label{fig:CMSSMSD}
\end{figure}

Finally, in Fig.~\ref{fig:CMSSMSD} we show the spin-dependent cross
sections, $\sigma_{\rm SD}$, for the upper panels in
Fig.~\ref{fig:CMSSM} when $\mu > 0$. The points and shadings are
identical to those in the previous two figures. Here the thick black
solid curve is the upper limit from PICO \cite{pico} and the thin curves
are obtained from IceCube \cite{ice} limits based on annihilations into
$b {\bar b}$ pairs (solid) or $W^+ W^-$ pairs (dashed). For the
focus-point models, annihilations proceed primarily into electroweak
gauge bosons, or $hZ$ final states with some non-negligible
contributions from $t \bar{t}$, for which the $W^+ W^-$ may be
applicable. Models with $A_0 = 0$ lie just below the current bounds
again because of the highly-mixed nature of the LSP,
whilst the models with $A_0 = 2.3 m_0$ predict cross section far below
these bounds.

\subsection{mSUGRA}

In \cite{ELOS}, mSUGRA models were considered with $A_0/m_0 = 3-\sqrt{3}$ (the Polonyi \cite{Polonyi} value)
and $A_0/m_0 = 2$ for comparison.
The computed value of $\tan \beta$ is generally $\gtrsim 10$.
As a result,  proton lifetime limits for these cases would be short,
in violation of the experimental bounds unless a non-minimal version of SU(5) is adopted.

In the case of the Polonyi model, viable regions of the parameter space contain a gravitino LSP,
which would give negligible signals in direct detection experiments.
However, for the larger value $A_0 = 2 m_0$ there are some regions of parameter
space with a bino LSP with the relic density held in check by stau coannihilations, but
$\tan \beta \gtrsim 40$ for $m_h > 124$ GeV and there are strong constraints from
B-physics observables in that case. This model is highly constrained and we do not discuss it further
here, though we return later to mSUGRA models with universality imposed below the GUT scale.

\subsection{subGUT}

\begin{figure}[htb!]
\begin{minipage}{8in}
\includegraphics[height=3.3in]{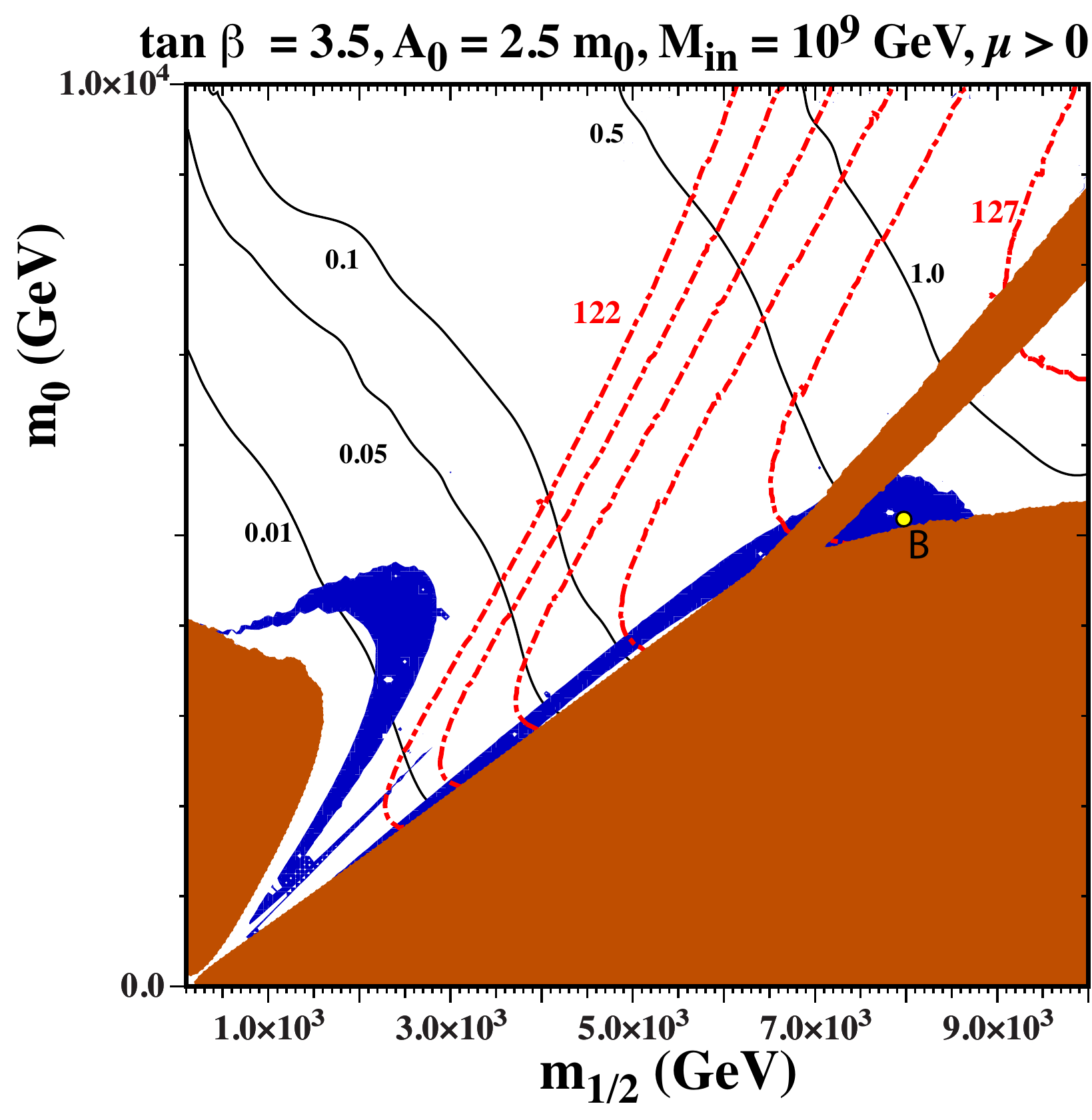}
\includegraphics[height=3.3in]{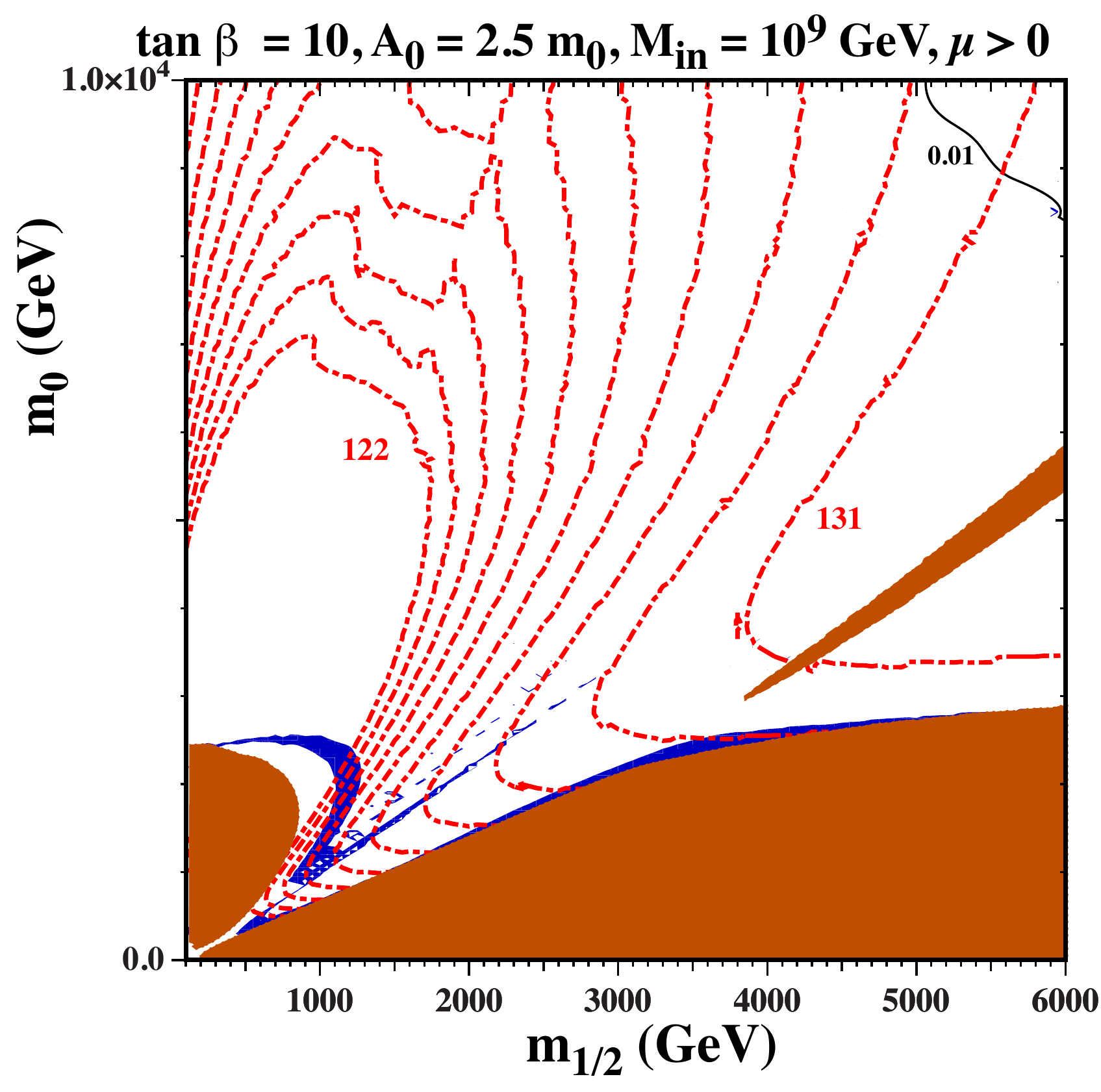}
\end{minipage}
\begin{minipage}{8in}
\includegraphics[height=3.3in]{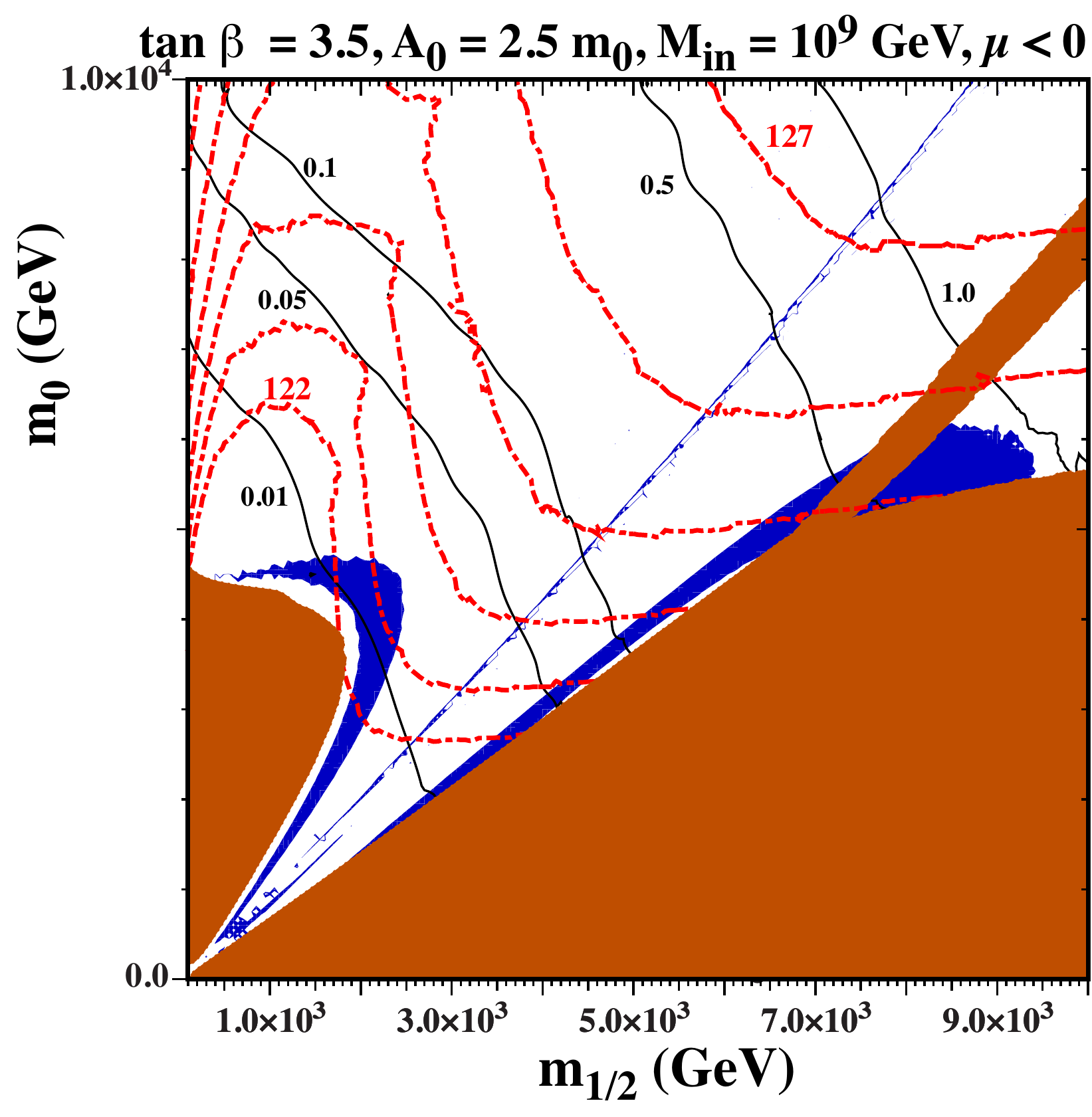}
\hspace*{-0.05in}
\includegraphics[height=3.3in]{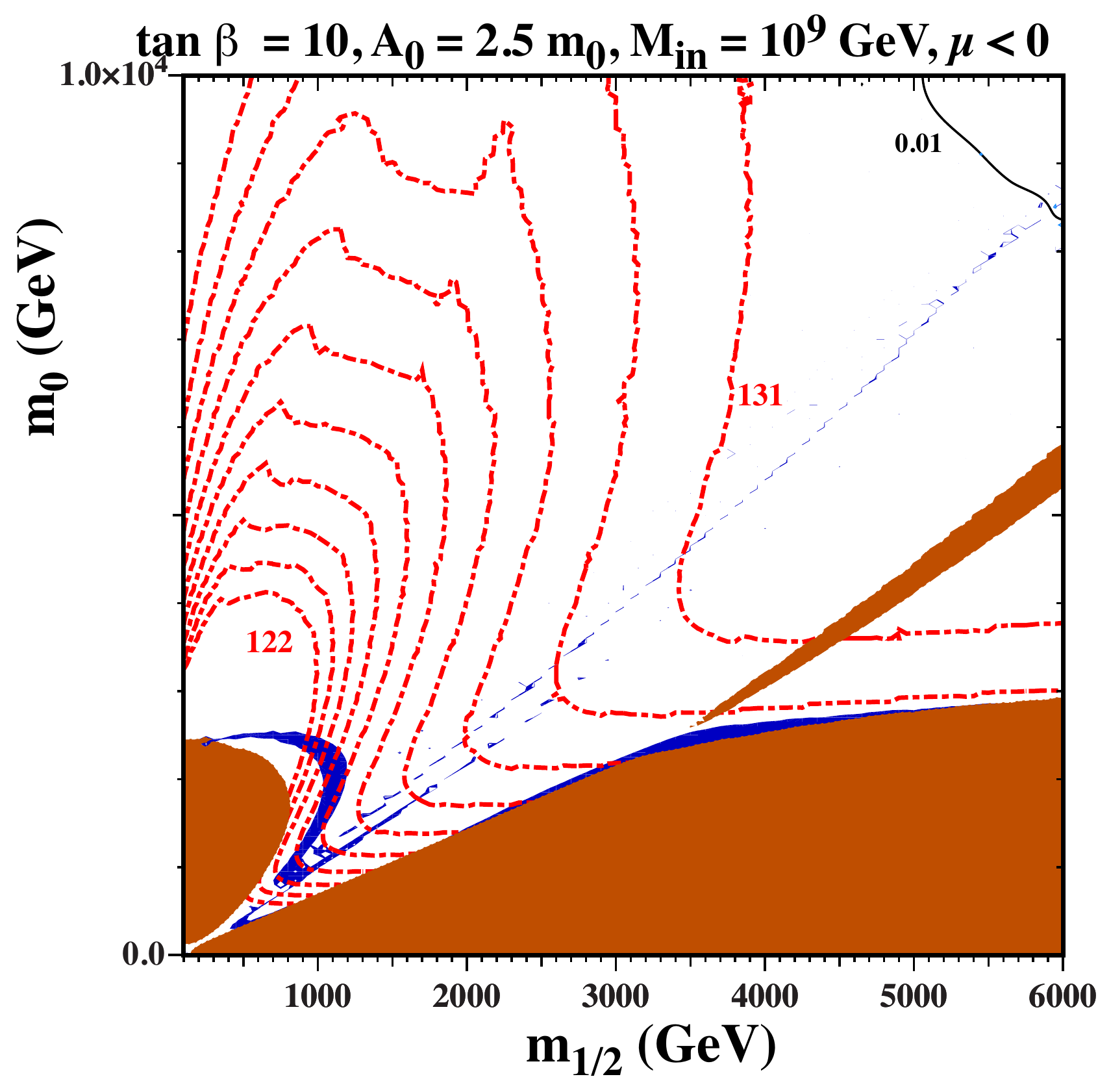}
\end{minipage}
\caption{
{\it
The subGUT CMSSM  $(m_{1/2}, m_0)$ planes for
$A_0 = 2.5 \, m_0$ with $M_{in} = 10^9$ GeV.
The left (right) panels have $\tan \beta = 3.5 (10)$.
The upper (lower) panels have $\mu > 0$ ($\mu < 0$).
The shadings and contour types are as in Fig.~\protect{\ref{fig:CMSSM}}.
The point labelled B refers to the point tested for phase dependence in the
Appendix.
} }
\label{fig:subGUT}
\end{figure}

In the left panels of Fig.~\ref{fig:subGUT}, we show examples of
$(m_{1/2}, m_0)$ planes with $\tan \beta = 3.5$, $A_0 = 2.5 \, m_0$,
$M_{in} = 10^9$~GeV and $\mu > 0$ (upper panel) and $\mu < 0$ (lower
panel). In both planes, one finds three distinct brown shaded regions where the
LSP is no longer neutral and/or uncolored. At the left, at low $m_{1/2}$
and $m_0 < 4$ TeV the lighter stop is the LSP, at low $m_0$ for all
$m_{1/2}$ the lighter stau is the LSP, and along a diagonal strip that
rises from the stau LSP region the lighter chargino is the LSP.
When $\mu$  is approximately equal to the bino mass, $M_1$, two of the neutralino mass eigenstates are
strongly mixed bino-Higgsino states. In this case, 1-loop corrections
to these masses can differ significantly from the correction to the second Higgsino (which
is nearly identical to the correction to the lighter chargino) and cause the chargino to
become the LSP.
To the left
of this region, the LSP is the bino and the relic density gets too
large. To the right of this region, the LSP is the Higgsino and, because
$m_{1/2}$ and $\mu$ are large here, the relic density again gets too large.

There are also three distinct regions in these panels of Fig.~\ref{fig:subGUT}
where the relic density is consistent with the Planck constraint.
Somewhat offset from the stop LSP region, we see a curved band which is produced
by stop coannihilation. This region is much broader than in the typical CMSSM case, due to the increased degeneracy of the SUSY particles. Then, just above the stau LSP region we see the familiar
stau coannihilation strip. At this value of $M_{in}$ it extends to far greater values
of $m_{1/2}$ than it would in the CMSSM with $M_{in} = M_{GUT}$: this is generally possible for
sufficiently small $M_{in}$, and is again due to the degeneracy of the SUSY particles.
We note that, to the right of the chargino LSP region, stau coannihilation occurs between the stau
and a Higgsino LSP instead of the more usual bino.
Finally, between the stau and stop
strips, we see a narrow funnel region where $2 m_\chi \approx m_{A,H}$. For $\mu > 0$, the funnel
extends to $m_{1/2}, m_0 \lesssim 3$ TeV, whilst for $\mu < 0$,
it extends past the end of the plot to $m_0 > 10$ TeV.
Looking now at the red dashed contours of $m_h$, we see that
in the stop coannihilation strip and the funnel region the Higgs mass is somewhat too small: $m_h < 123$ GeV for $\mu> 0$,
whereas the uncertainty from {\tt FeynHiggs} is $\sim 1.5$~GeV. This is to be expected,
since the stop masses are light in this region and so the corrections to the Higgs mass are small. On the other hand, the Higgs mass
exceeds 124 GeV for $4~{\rm TeV} < m_{1/2}$, and hence much of the stau coannihilation region is acceptable.
For $\mu < 0$ (lower left panel of Fig.~\ref{fig:subGUT}), the values of $m_h$ are somewhat higher, and parts
of the stop coannihilation strip may be acceptable.

As in the previous subsections, the solid black lines are contours of the
proton decay lifetime. For $4~{\rm TeV} < m_{1/2}$ (where the Higgs mass is acceptable),
the lifetime exceeds the experimental bound of $5 \times 10^{33}$~yrs.
As discussed earlier, in order to obtain a sufficiently long lifetime,
we are forced to relatively small values of $\tan \beta$.
For  $\tan \beta$ larger than the value $3.5$ shown here, the lifetime
along the stau strip drops below the experimental bound, as seen in the
right panels of Fig.~\ref{fig:subGUT} where $\tan \beta = 10$ is chosen,
and one would need to abandon minimal SU(5), as the proton lifetime is less than
$10^{33}$~yrs over much of the plane.

Qualitatively, we see similar features
for the LSP and relic density in the right panels.
Smaller $\tan \beta$ ($< 3.5$) is possible, but one needs to go to higher values of $m_{1/2}$ to ensure
a sufficiently heavy Higgs boson. For smaller $A_0/m_0$, the extent of the stau strip
is reduced, making it difficult to obtain both a heavy enough Higgs and a long proton lifetime.
This reduction in the stau coannihilation strip is due to a reduction in the Higgsino masses as $A_0$ is reduced.
The stau coannihilation band in this figure is actually assisted by several other supersymmetric particles with masses
similar to the stau, including the charged and neutral Higgsinos.  The Higgsino masses are roughly set by
$m_0$ ($\mu$ is set by EWSB conditions which is related to the stop mass and hence to $m_0$).
Since the bino and stau masses continue to grow as $m_{1/2}$ is increased,
eventually the Higgsino becomes the LSP with no potential coannihilation partners.
At this point, the stau coannihilation band disappears and the Higgsino becomes the LSP.
Lowering $A_0$ reduces the value of $m_0$ for which the Higgsino becomes the LSP,
and so reduces the size of the stau coannihilation strip.  However, if we rely on non-minimal SU(5) to
lengthen the lifetime of the proton, we can go to larger values of $\tan\beta$, as seen in the right
panels of Fig.~\ref{fig:subGUT}. This allows one to obtain simultaneously a large enough Higgs mass and a small enough relic density.

\begin{figure}[htb!]
\begin{minipage}{8in}
\includegraphics[height=2.24in]{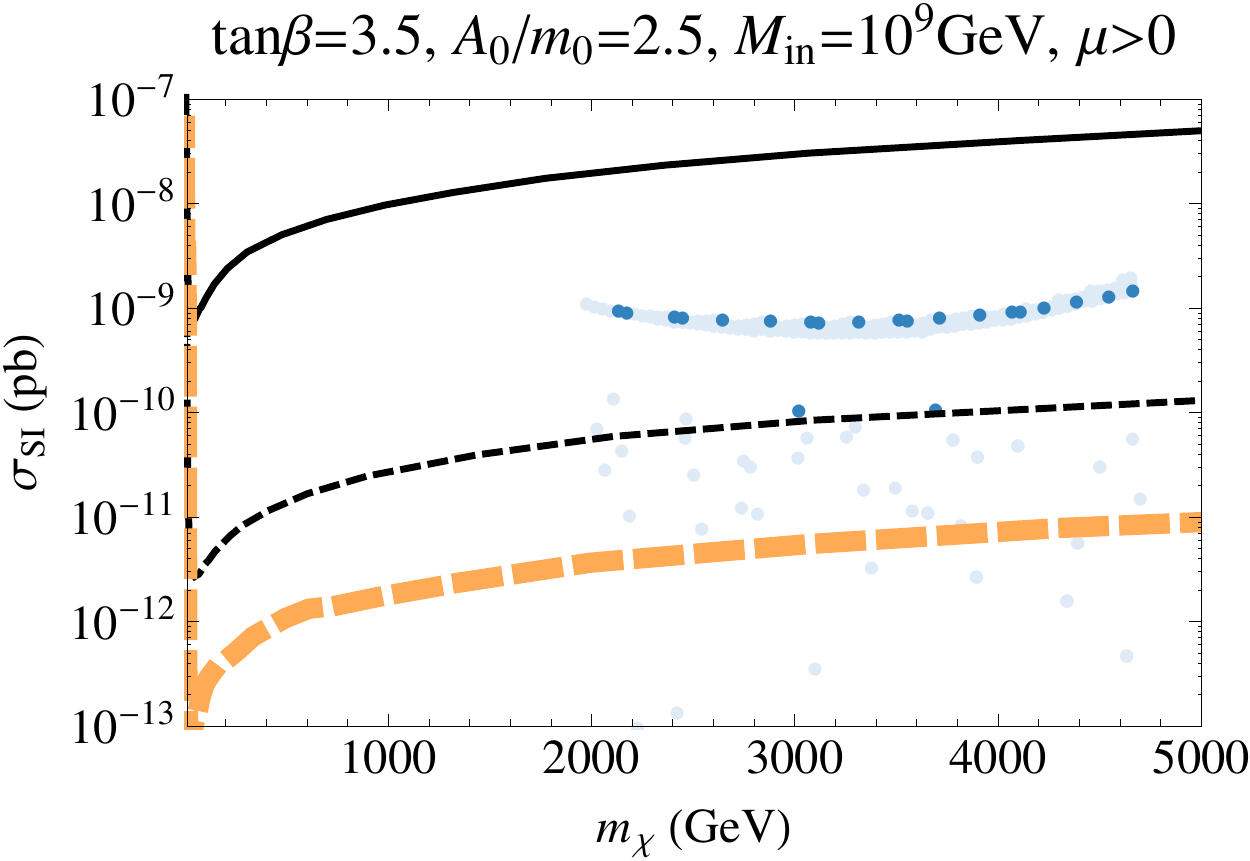}
\includegraphics[height=2.24in]{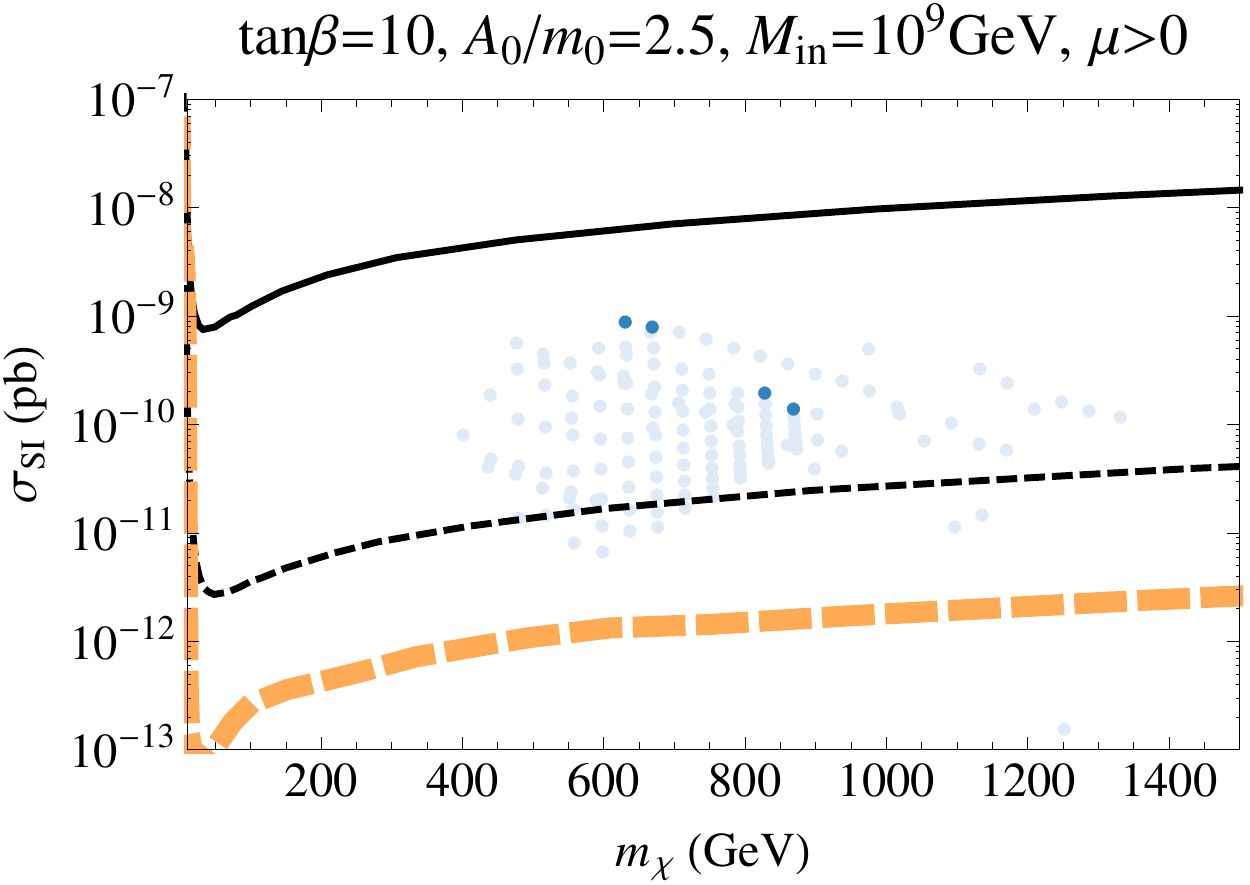}
\hfill
\end{minipage}
\begin{minipage}{8in}
\includegraphics[height=2.24in]{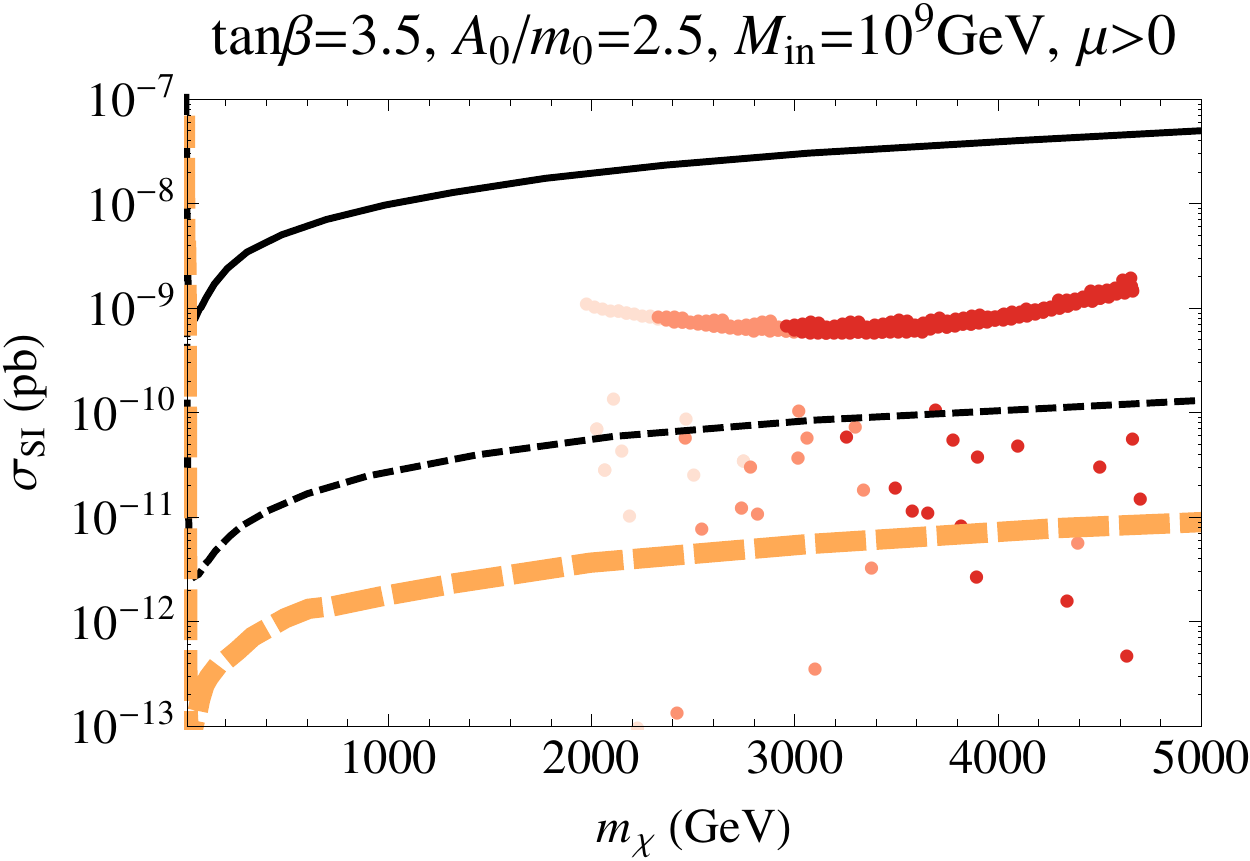}
\includegraphics[height=2.24in]{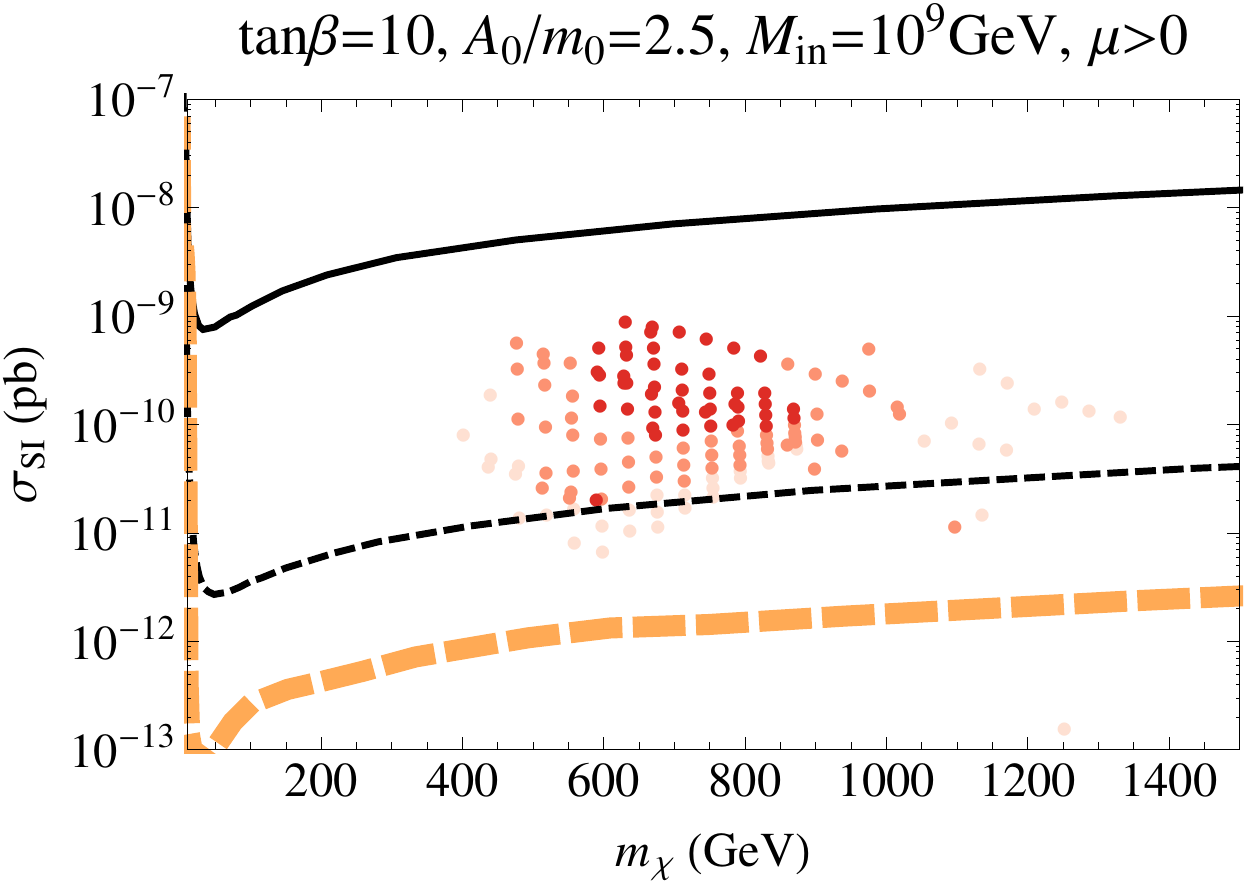}
\hfill
\end{minipage}
\caption{
{\it
As in Fig.~\protect{\ref{fig:CMSSMSI}} for the subGUT case with  $A_0 = 2.5 m_0$,
$\mu > 0$, $M_{in} = 10^9$ GeV and $\tan \beta = 3.5$ (left),
$\tan \beta = 10$ (right),  shown in the upper panels of Fig.~\protect{\ref{fig:subGUT}}.
 }}
\label{fig:subGUTSI}
\end{figure}

\begin{figure}[htb!]
\begin{minipage}{8in}
\includegraphics[height=2.22in]{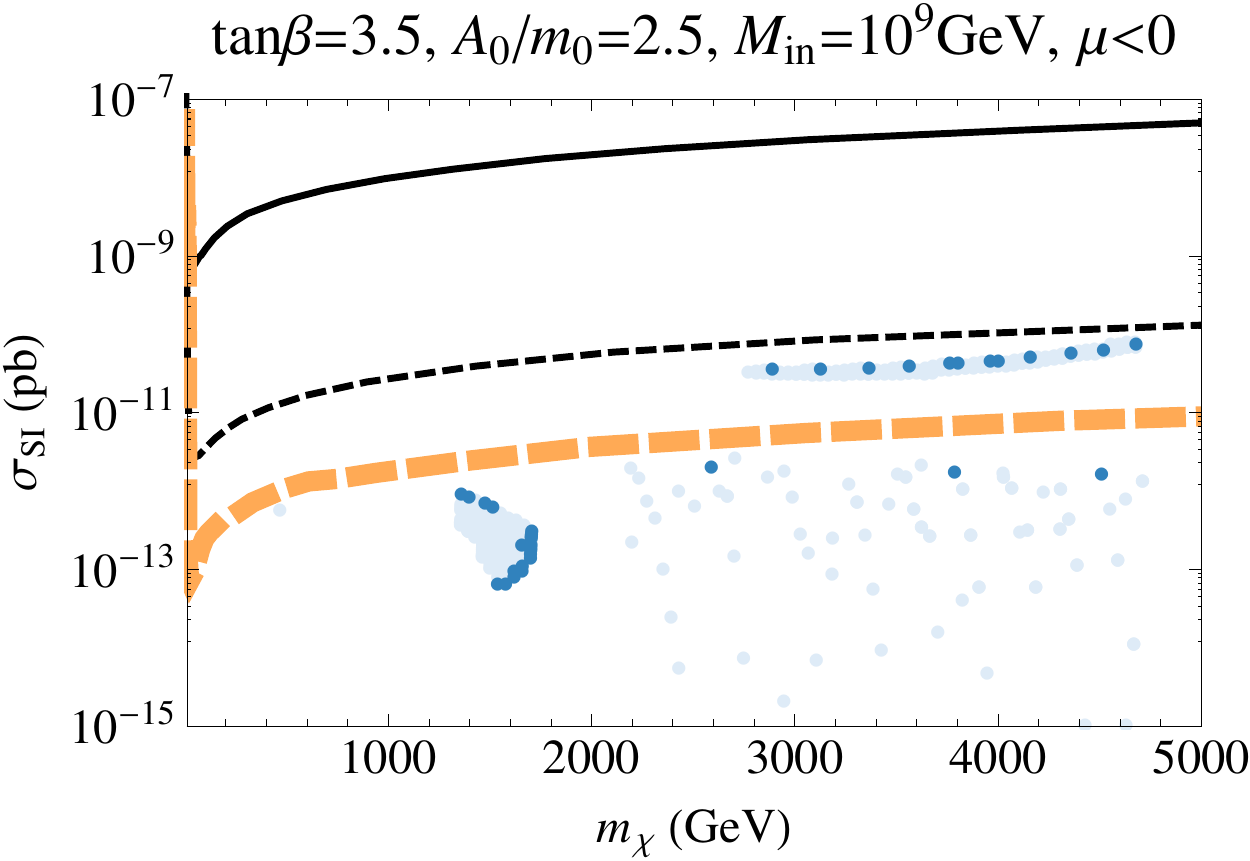}
\includegraphics[height=2.22in]{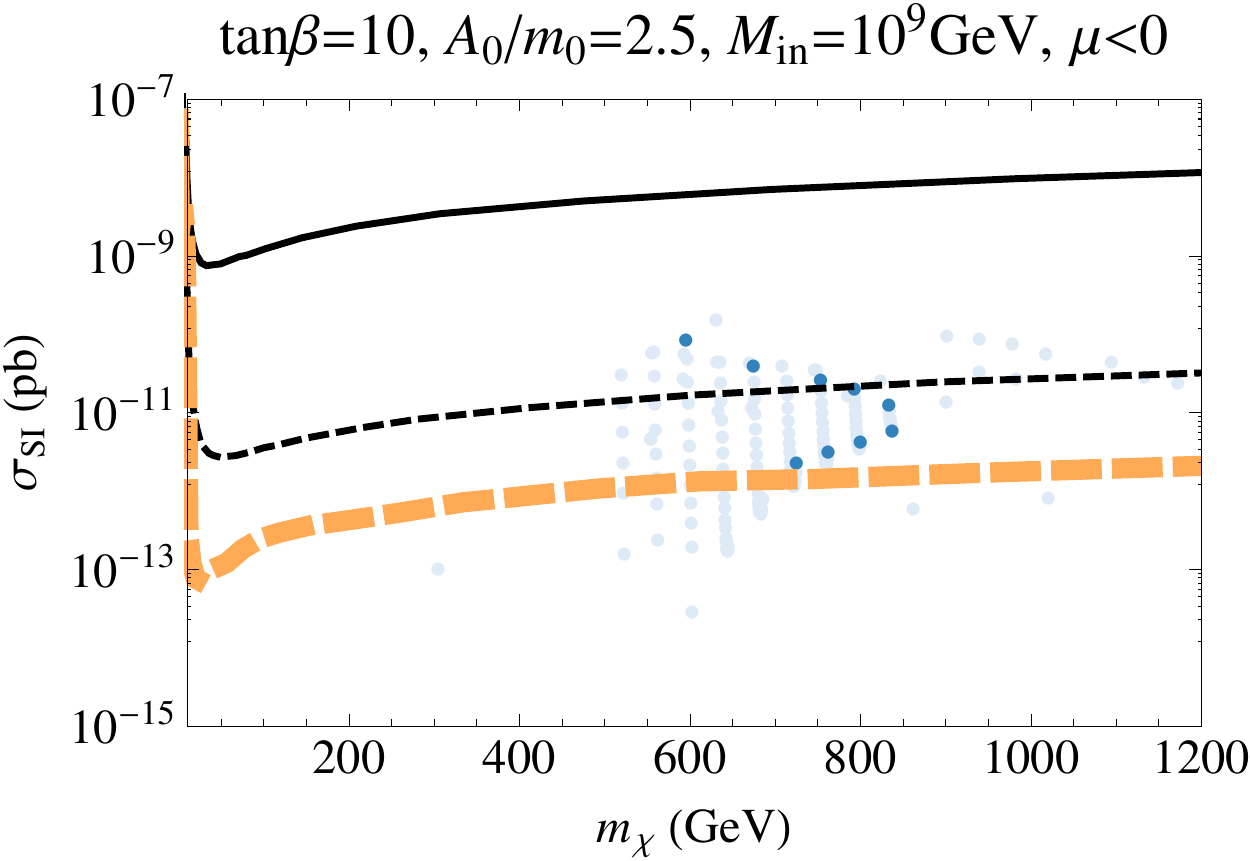}
\hfill
\end{minipage}
\begin{minipage}{8in}
\includegraphics[height=2.22in]{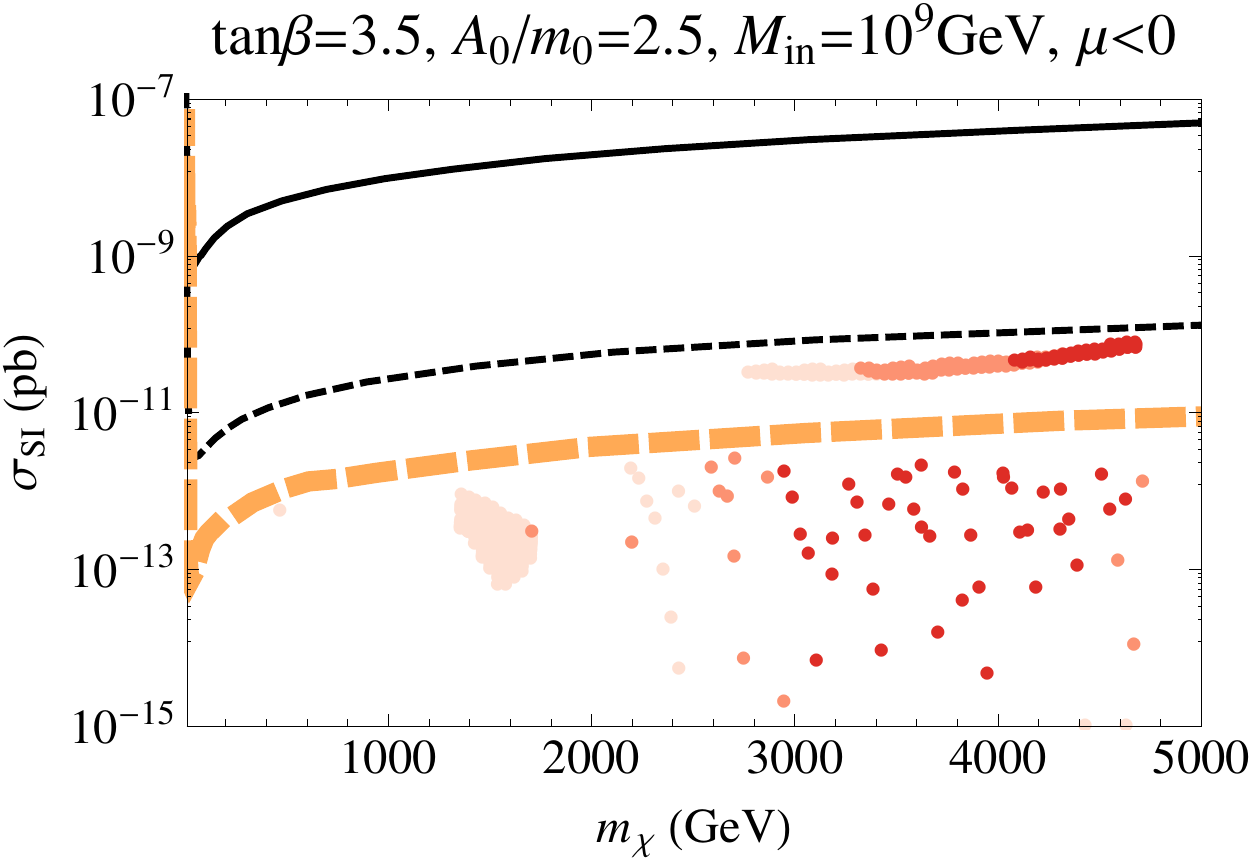}
\includegraphics[height=2.22in]{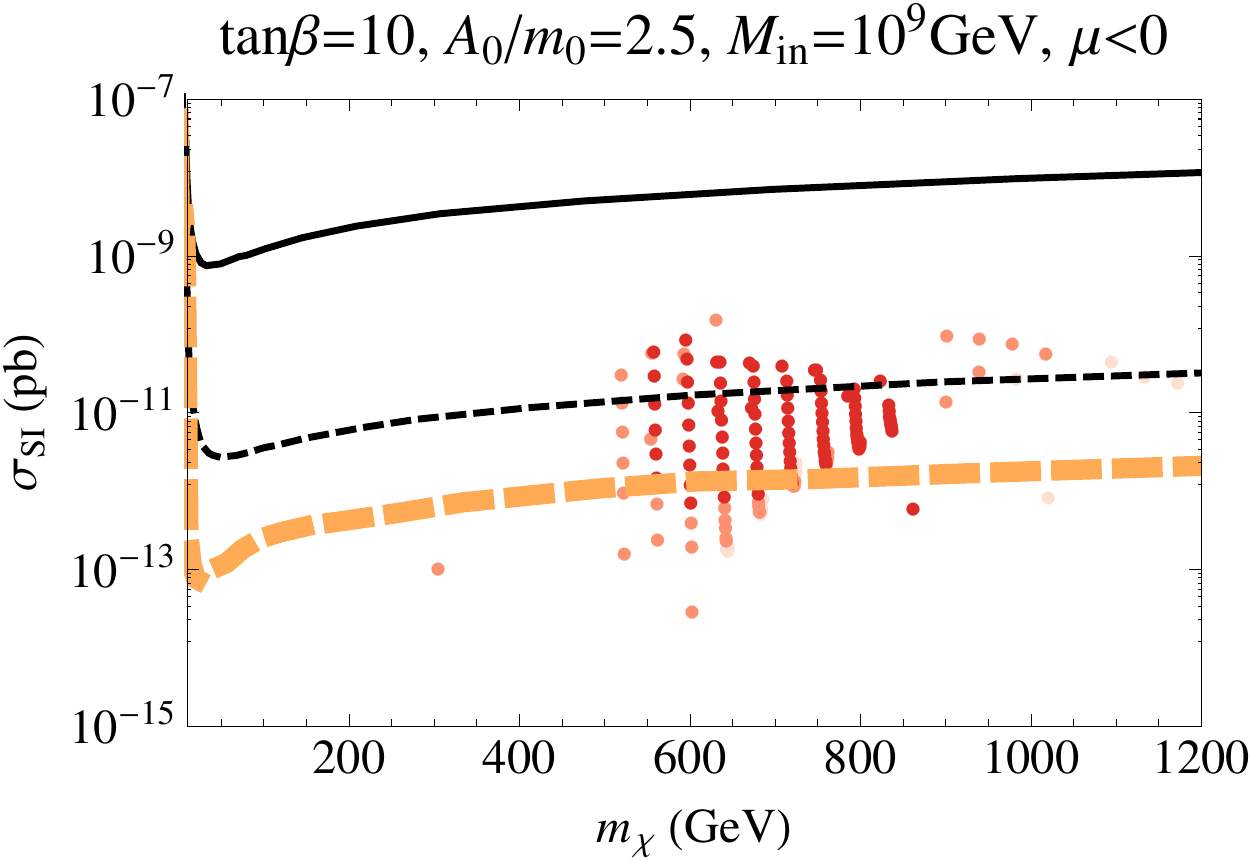}
\hfill
\end{minipage}
\caption{
{\it
As in Fig.~\protect{\ref{fig:CMSSMSI}} for the subGUT case with  $A_0 = 2.5 m_0$,
$\mu < 0$, $M_{in} = 10^9$ GeV and $\tan \beta = 3.5$ (left),
$\tan \beta = 10$ (right), shown in the lower panels of Fig.~\protect{\ref{fig:subGUT}}.
 }}
\label{fig:subGUTSIn}
\end{figure}

Results for the SI cross section for the subGUT CMSSM cases displayed in Fig.~\ref{fig:subGUT}
are shown in Figs.~\ref{fig:subGUTSI} (for $\mu > 0$) and \ref{fig:subGUTSIn} (for $\mu < 0$).
We see in the upper panels of Fig.~\ref{fig:subGUTSI} that the SI cross sections for models
with a relic density compatible with the Planck range (darker blue points) are generally below the current upper limit
but within reach of the LZ experiment.
The dark shaded points originate from what appears to be the stau coannihilation strip.
As noted above, obtaining the correct relic density at such large LSP masses requires
additional coannihilation and mass degeneracies. In this case, the Higgsinos also have masses
comparable to the bino mass. Along this horizontal strip of points in the left panel of Fig.~\ref{fig:subGUTSI},
the mass difference ($\mu - M_1$) is relatively constant and hence from Eq.~(\ref{eq:siappbino}),
when $\tan \beta$ is small, we obtain a cross section which is relatively constant as well.
The same is true for most of the under-dense (paler blue) points
for $\tan \beta = 10$ (upper right panel), but under-dense points for $\tan \beta = 3.5$ may have SI
cross sections below the LZ sensitivity though above the neutrino background level. The lower panels of
Fig.~\ref{fig:subGUTSI} show that many of the points within reach of the LZ experiment have values of
$m_h$ close to the experimental value. Fig.~\ref{fig:subGUTSIn} (for $\mu < 0$) exhibits somewhat lower
values of the SI cross section, in general.
Note that the cluster of points with $m_\chi \sim 1500$ GeV correspond to
bino LSPs in the stop coannihilation region. These points do not appear when $\mu > 0$ as that
region has $m_h < 122$ GeV and hence not included in our scan for elastic cross sections.
Consequently, most models with a relic density compatible
with the Planck range are beyond the LZ sensitivity, and some of the $\tan \beta = 3.5$ points are below the neutrino
background level. The same is true {\it a fortiori} for the points with under-dense relic neutralinos.
Finally we note in passing, that the 3 nearly horizontal points below the neutrino background
(for $\mu < 0$) originate from the funnel region (there are similar points when $\mu > 0$ but
more difficult to discern in the figure).

\begin{figure}[htb!]
\begin{minipage}{8in}
\includegraphics[height=3.2in]{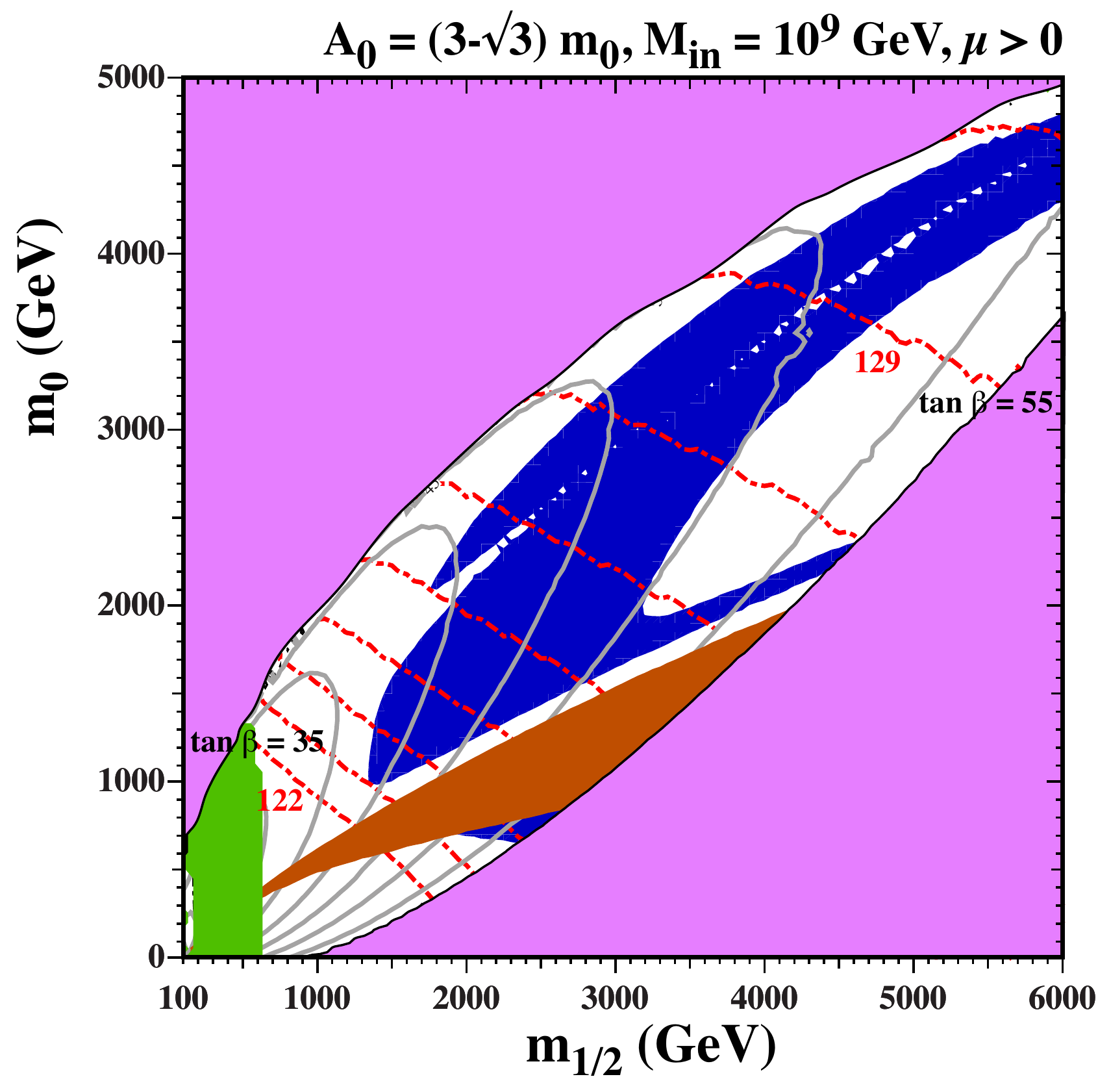}
\hspace*{-0.17in}
\includegraphics[height=3.2in]{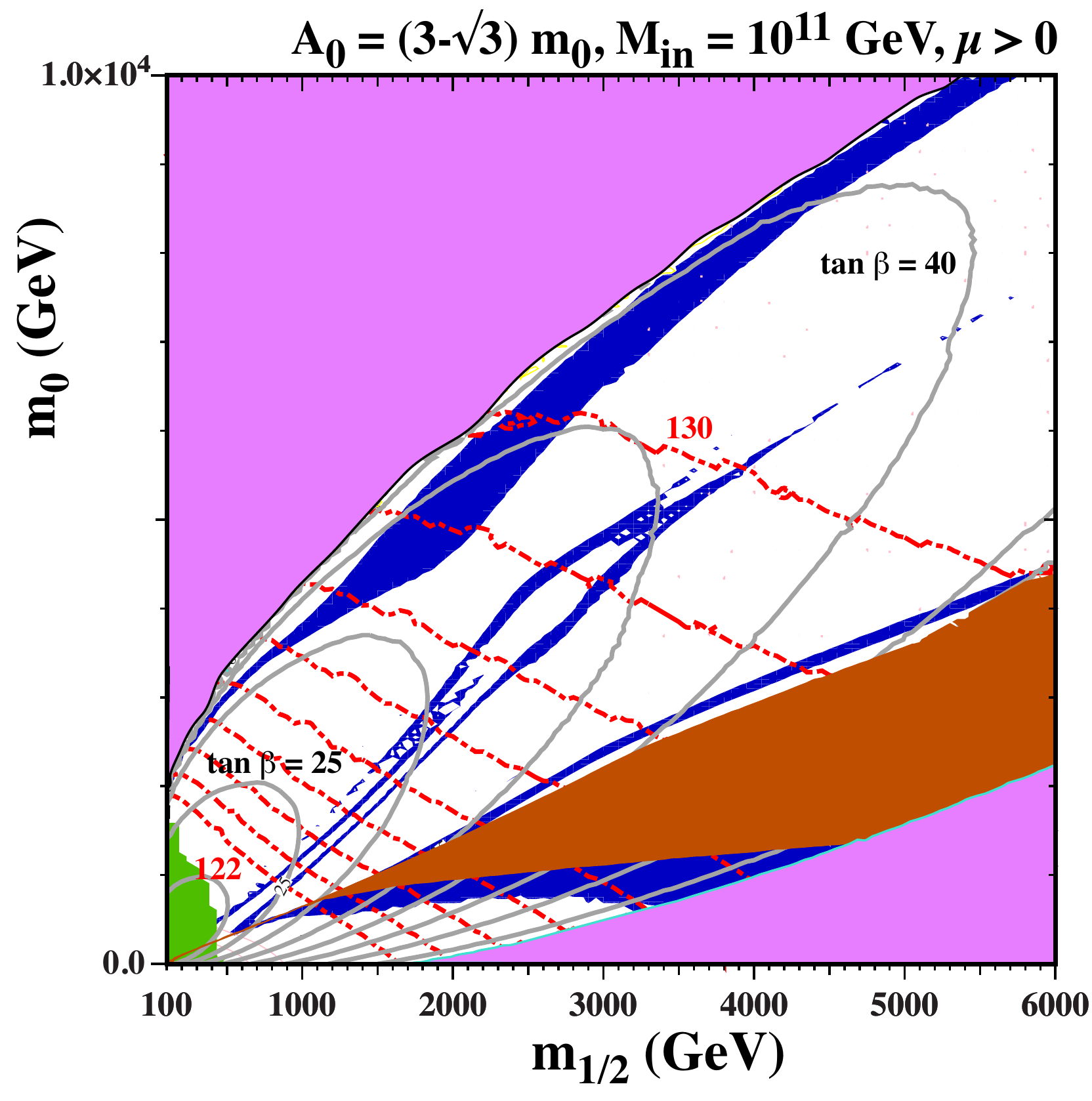}
\hfill
\end{minipage}
\caption{
{\it
As in Fig.~\protect{\ref{fig:CMSSM}} for the subGUT mSUGRA case with  $A_0 = (3-\sqrt{3}) m_0$,
and $M_{in} = 10^9$ GeV (left) and  $M_{in} = 10^{11}$ GeV (right). In addition to the shadings described for
Fig.~\protect{\ref{fig:CMSSM}}, the green shaded region is excluded by $b \to s \gamma$, the gray lines
show contours of $\tan \beta$ in increments of 5 as labelled.
}}
\label{fig:subGUTmsugra}
\end{figure}

Turning now to subGUT mSUGRA models, we consider only the Polonyi model, i.e., $A_0=(3-\sqrt{3})m_0$,
with $\mu > 0$.
The left panel of Fig.~\ref{fig:subGUTmsugra} is for $M_{in}=10^9$~GeV.
The mauve shaded regions in the upper left
and lower right parts of the plane are where electroweak symmetry breaking fails
($\mu^2 < 0$ in the upper left and a diverging Yukawa coupling due to an excessive value for $\tan \beta$ in the lower right), and in the
central brown shaded region the stau is the LSP. Above this region, various processes contribute to bringing
the relic density into the Planck range. Over much of this plane, the LSP is mostly Higgsino which is nearly degenerate with the next lightest superparticle (NLSP) which is a chargino in this case as well as the 2nd Higgsino.
  In the blue shaded area above the stau LSP region,
 in addition to neutralino coannihilations,  stau coannihilation also enhances the cross section in this strip.
In the wide blue shaded region above the stau strip (recall that we are here showing regions where the relic
density lies between .06 and 0.2)  the Higgs funnel (lower part of this strip and  conventional focus-point region (upper part) have merged.
Below the stau LSP region, the gravitino becomes the LSP and can be dark matter. The values of
$\tan\beta$ are quite large for all points in the left panel of Fig.~\ref{fig:subGUTmsugra}.  For this reason, the proton lifetime in
minimal SU(5) models is much too short, and some non-minimal model must be considered~\footnote{We note also
that the $b \to s \gamma$ constraint excludes a region at small $m_{1/2}$ and $m_0$ (shaded green), and that the
$B_s \to \mu^+ \mu^-$ constraints is also relevant in much of the allowed region of the plane.}.
The right side of Fig.~\ref{fig:subGUTmsugra} is for $M_{in}=10^{11}$~GeV, and shares the qualitative features
of the electroweak symmetry breaking and stau LSP constraints. The values of $\tan \beta$ are somewhat
smaller than in the $M_{in}=10^9$~GeV case, but still much too large to obtain a sufficiently long proton lifetime in minimal SU(5) models.
The dark matter constraint is satisfied in a focus-point strip close to the electroweak symmetry breaking boundary,
which has now demerged from the funnel and stau coannihilation strip. The rapid-annihilation funnel is now
clearly visible as a separate well-defined region.
In the stau strip and in the funnel, the LSP is once again a bino, though the masses of the Higgsinos are
not much larger. There is also
a gravitino dark matter region below the stau LSP region.

\begin{figure}[htb!]
\begin{minipage}{8in}
\includegraphics[height=2.24in]{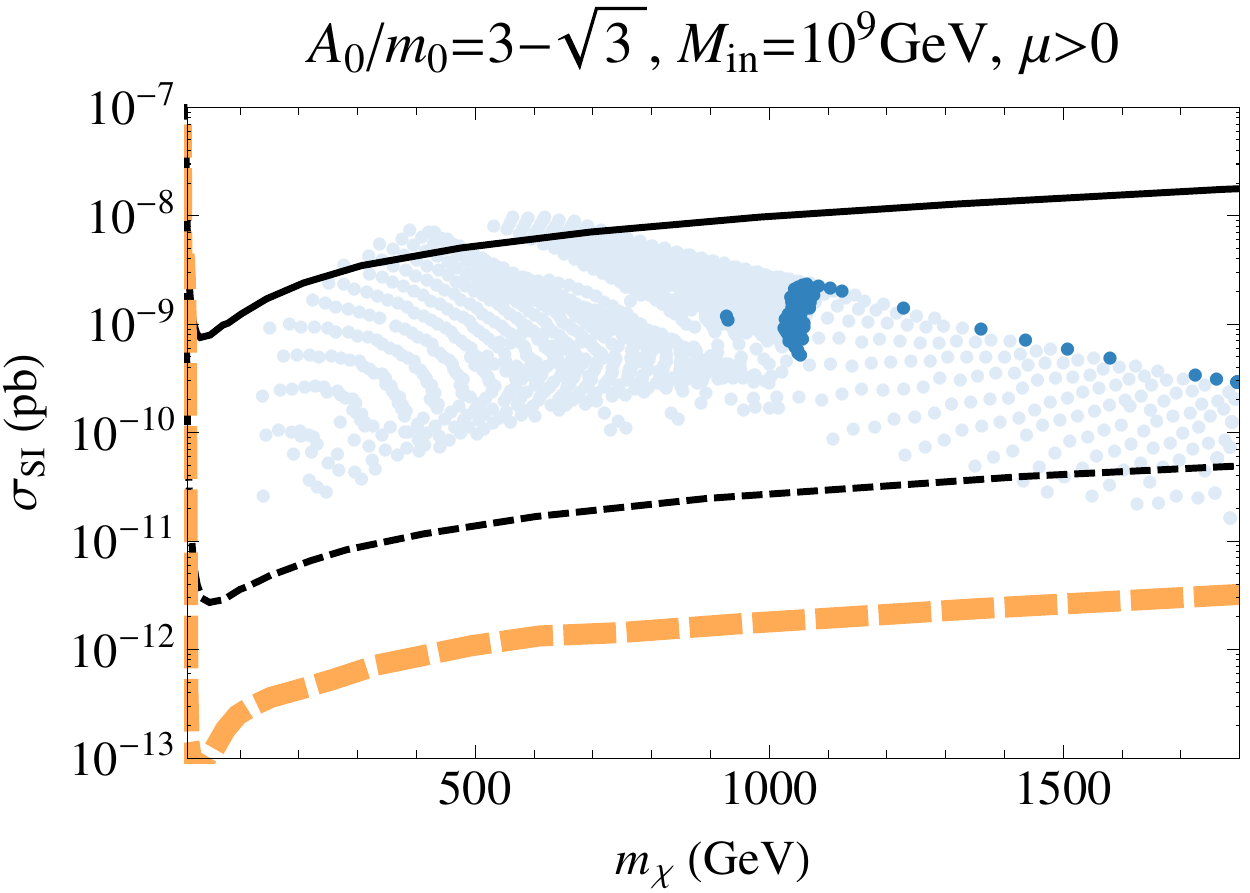}
\includegraphics[height=2.24in]{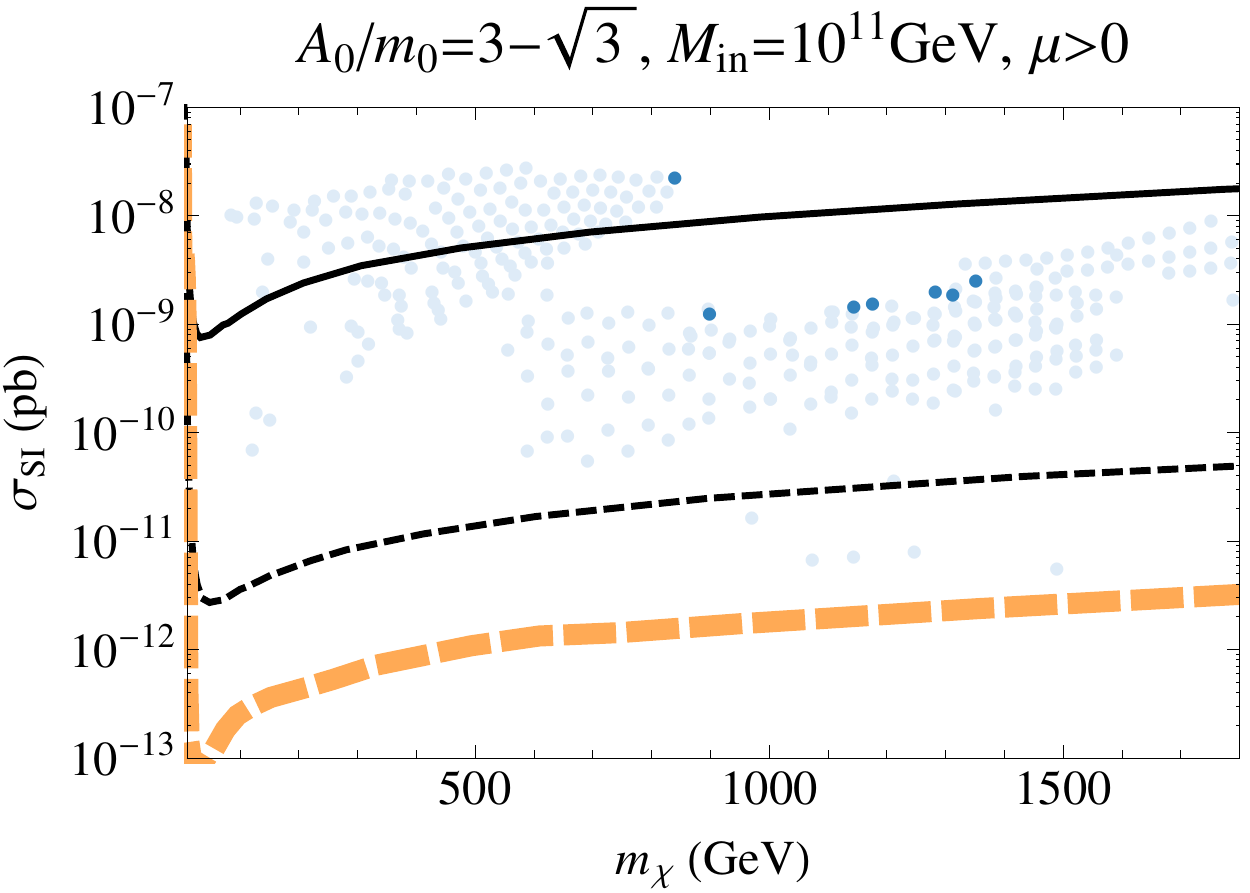}
\hfill
\end{minipage}
\begin{minipage}{8in}
\includegraphics[height=2.24in]{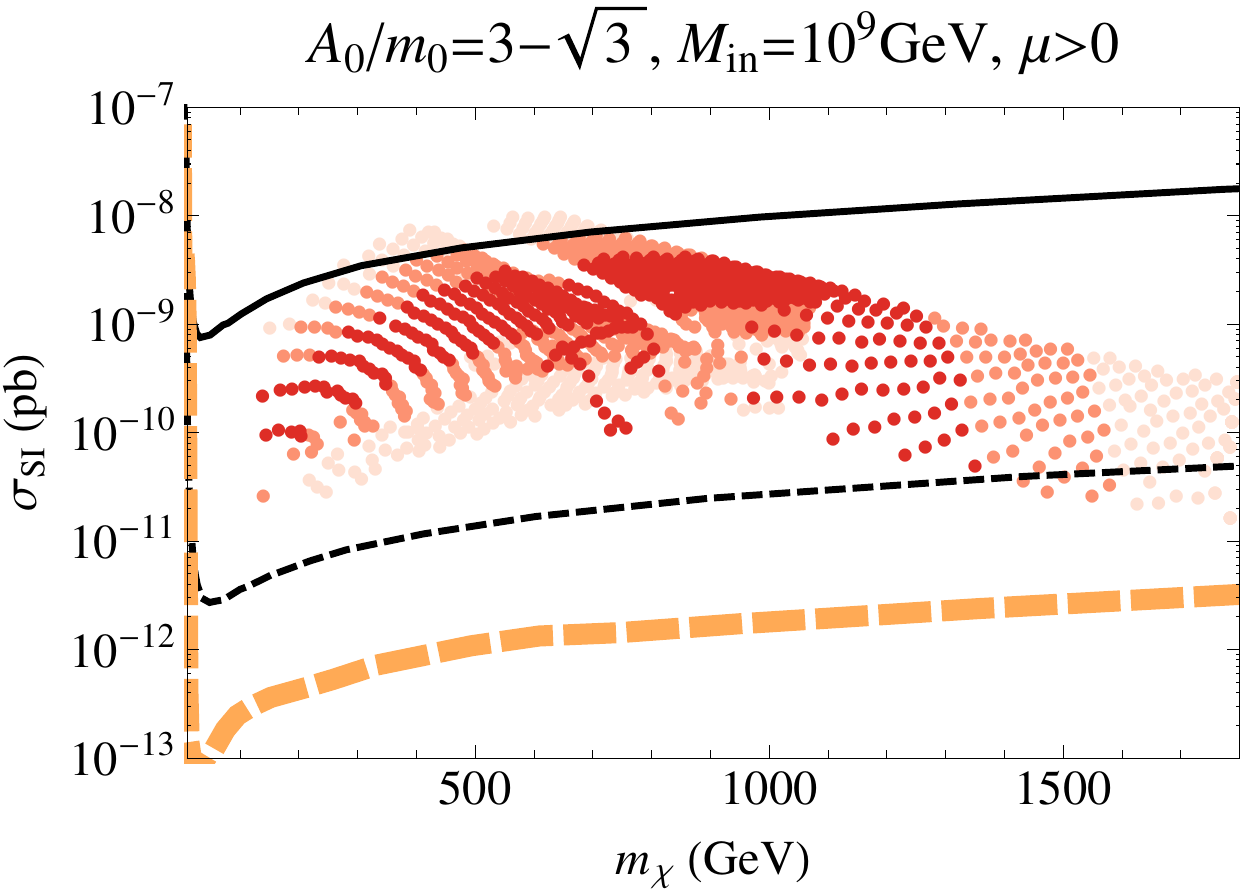}
\includegraphics[height=2.24in]{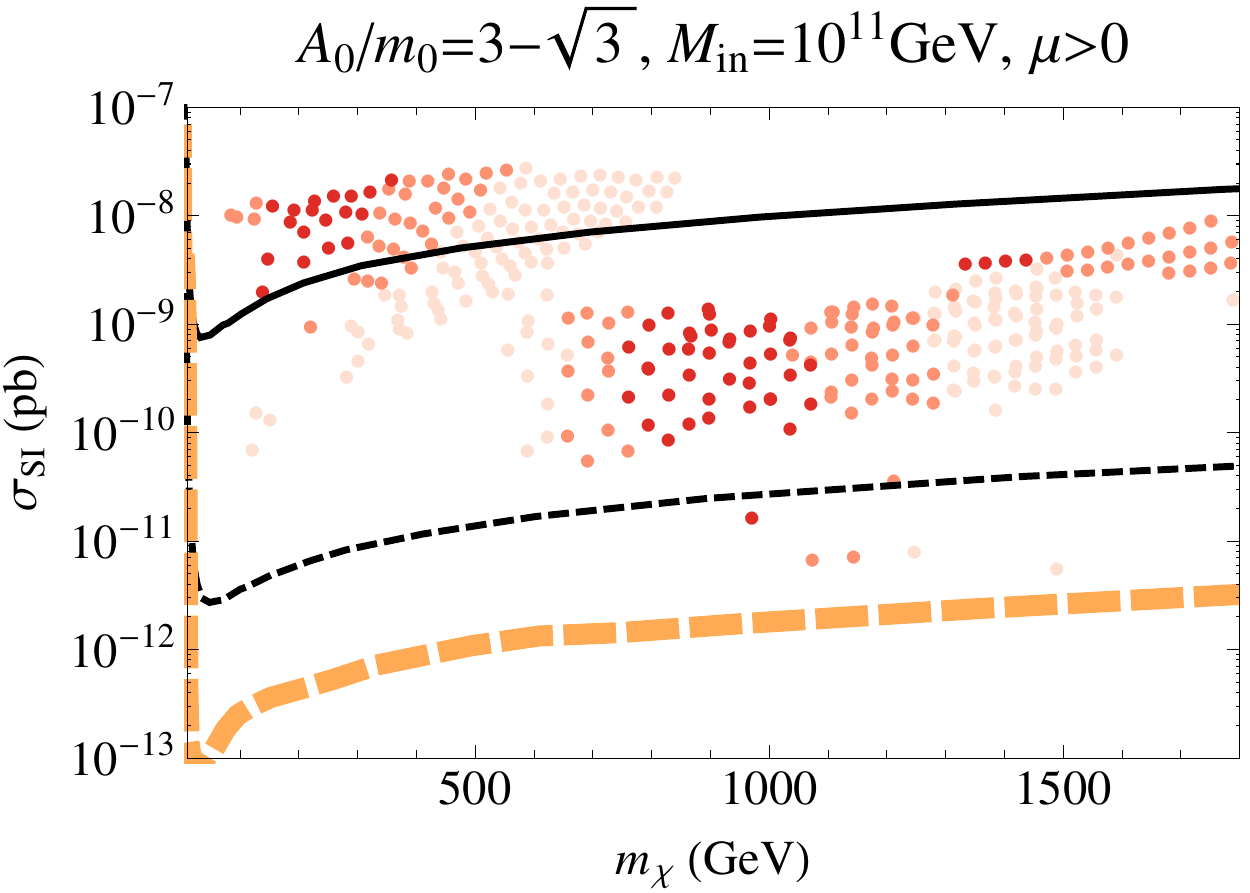}
\hfill
\end{minipage}

\caption{
{\it
As in Fig.~\protect{\ref{fig:CMSSMSI}} for the subGUT mSUGRA case with  $A_0 = (3 - \sqrt{3}) m_0$, and $M_{in} = 10^9$ GeV (left panels) and $M_{in} = 10^{11}$ GeV (right panels)
shown in Fig.~\protect{\ref{fig:subGUTmsugra}}.
 }}
\label{fig:subGUTmsugraSI}
\end{figure}

Fig.~\ref{fig:subGUTmsugraSI} displays results for the SI cross section in
these subGUT mSUGRA models. We see in the upper panels that the SI
cross section is generally between the current LUX upper limit and the prospective
LZ sensitivity, though some models (particularly for $M_{in}=10^{11}$~GeV) have
cross sections above the LUX limit and a few under-dense models have SI cross sections
below the LZ sensitivity. For $M_{in} = 10^9$ GeV, we clearly see the pile of points
with the Planck relic density at $m_\chi \approx 1100$ GeV corresponding to a Higgsino LSP
near the broad intersection of the stau strip and focus point swath. The dark blue points
in this figure continue to higher Higgsino masses along the stau coannihilation strip.
Very low mass points (all lightly shaded) correspond to regions in Fig. \ref{fig:subGUTmsugra}a
that are to the left of the blue shaded region.  In the white region to the left, the relic density is
small, and in the white region to the right (between the stau strip and funnel) the relic density is too high.
For $M_{in} = 10^{11}$ GeV, we see two very distinct regions in Fig. \ref{fig:subGUTmsugraSI}.
The region with lower masses ($m_\chi \lesssim 800$ GeV and cross section between
$10^{-9}$ and $10^{-8}$ pb) originate from the focus point region.
The remainder of the points come from either the funnel or the stau strip
and can be more easily distinguished by the lower panels showing the Higgs mass ranges.
The relative paucity of dark shaded blue points stems from the fact that the true Planck strips are quite
thin in this case. Note also that there is no pile up of points at 1100 GeV as the LSP
is most bino rather than Higgsino at the higher value of $M_{in}$.
We see in the lower panels of this figure that models with
$m_h \in [124, 126]$~GeV lie in the intersection region for $M_{in} = 10^9$ GeV,
and as noted above the dark brown shaded points for $M_{in} = 10^{11}$ GeV
at low masses come from the focus point where as we now see that the middle group
around $m_\chi \sim 1000$ GeV originate in the funnel, and the group at larger masses
lie in the stau strip.
 All of the dark shaded points lie within reach of the LZ experiment (though some
$M_{in}=10^{11}$~GeV models are excluded already by the LUX upper limit).

\begin{figure}[htb!]
\begin{minipage}{8in}
\includegraphics[height=2.2in]{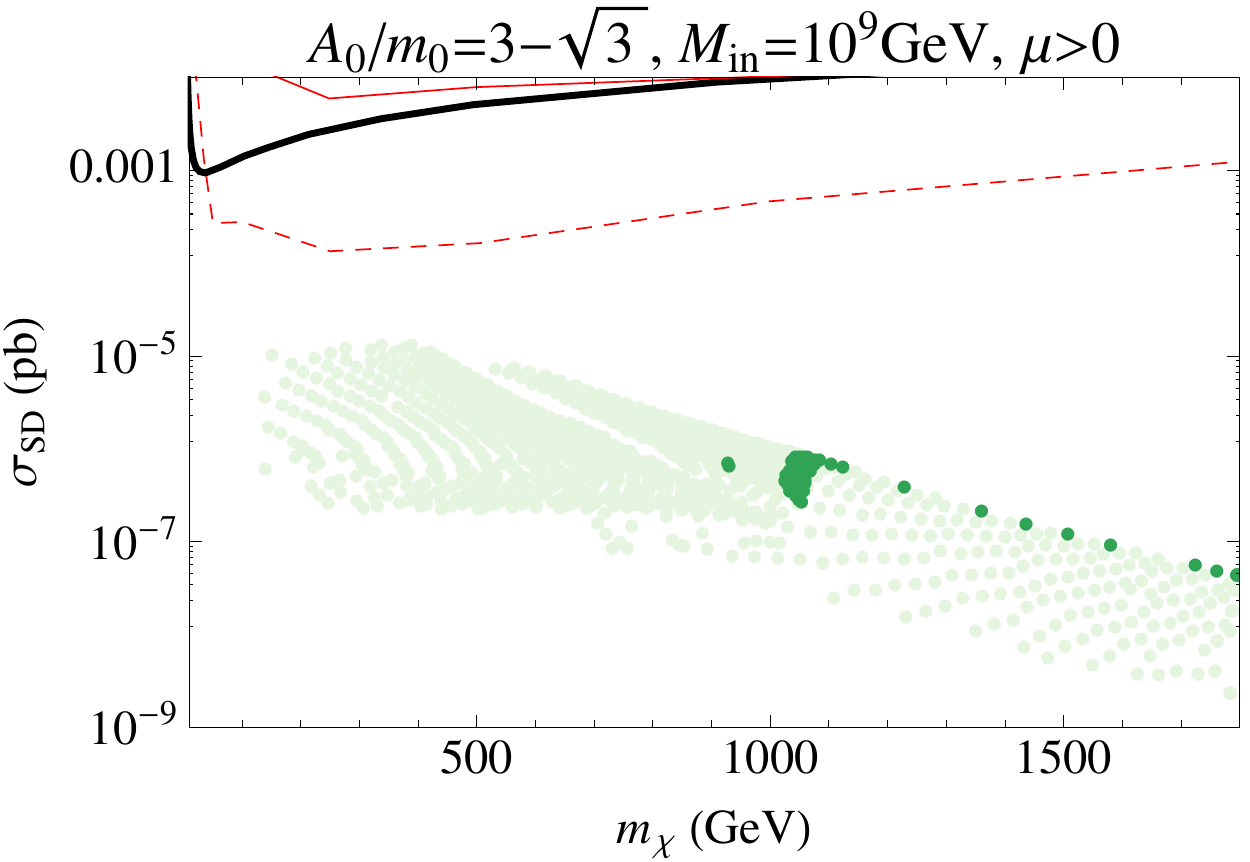}
\includegraphics[height=2.2in]{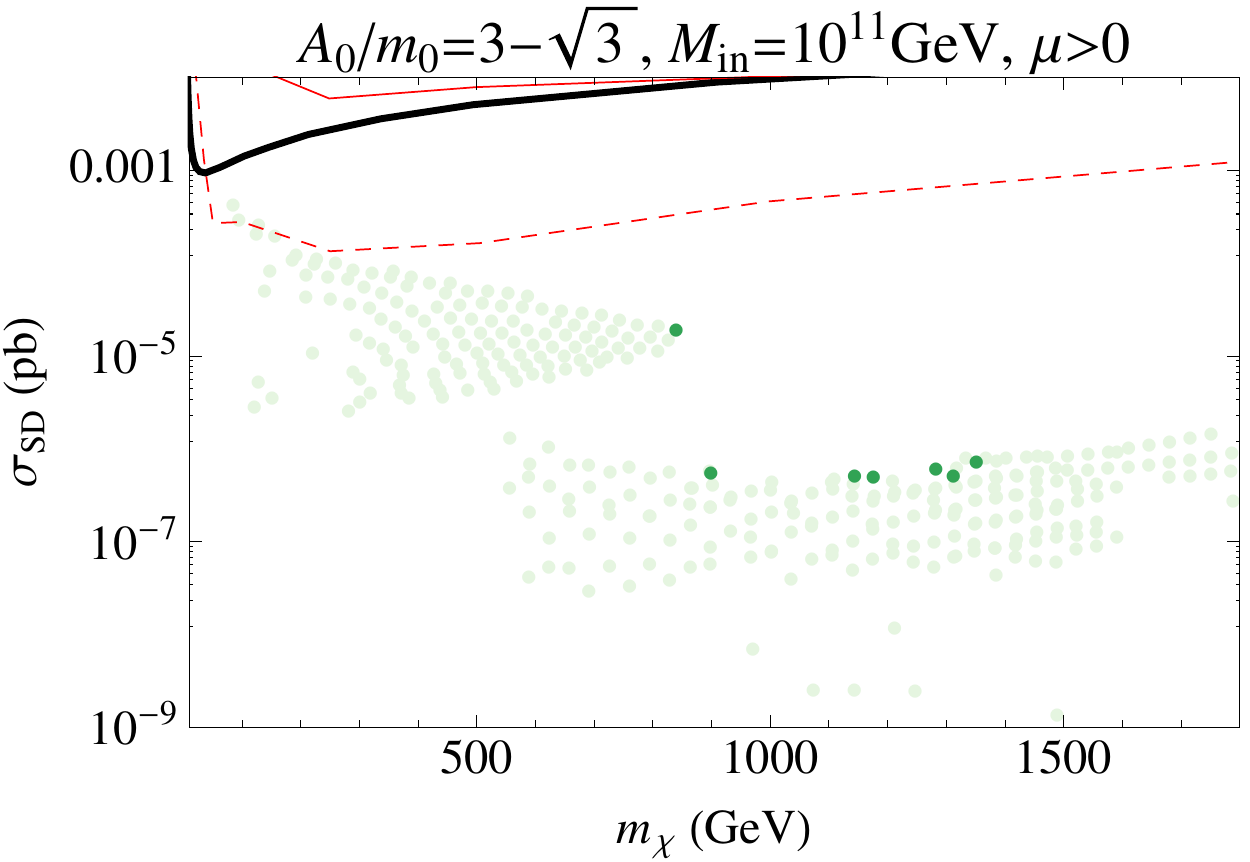}
\hfill
\end{minipage}
\begin{minipage}{8in}
\includegraphics[height=2.2in]{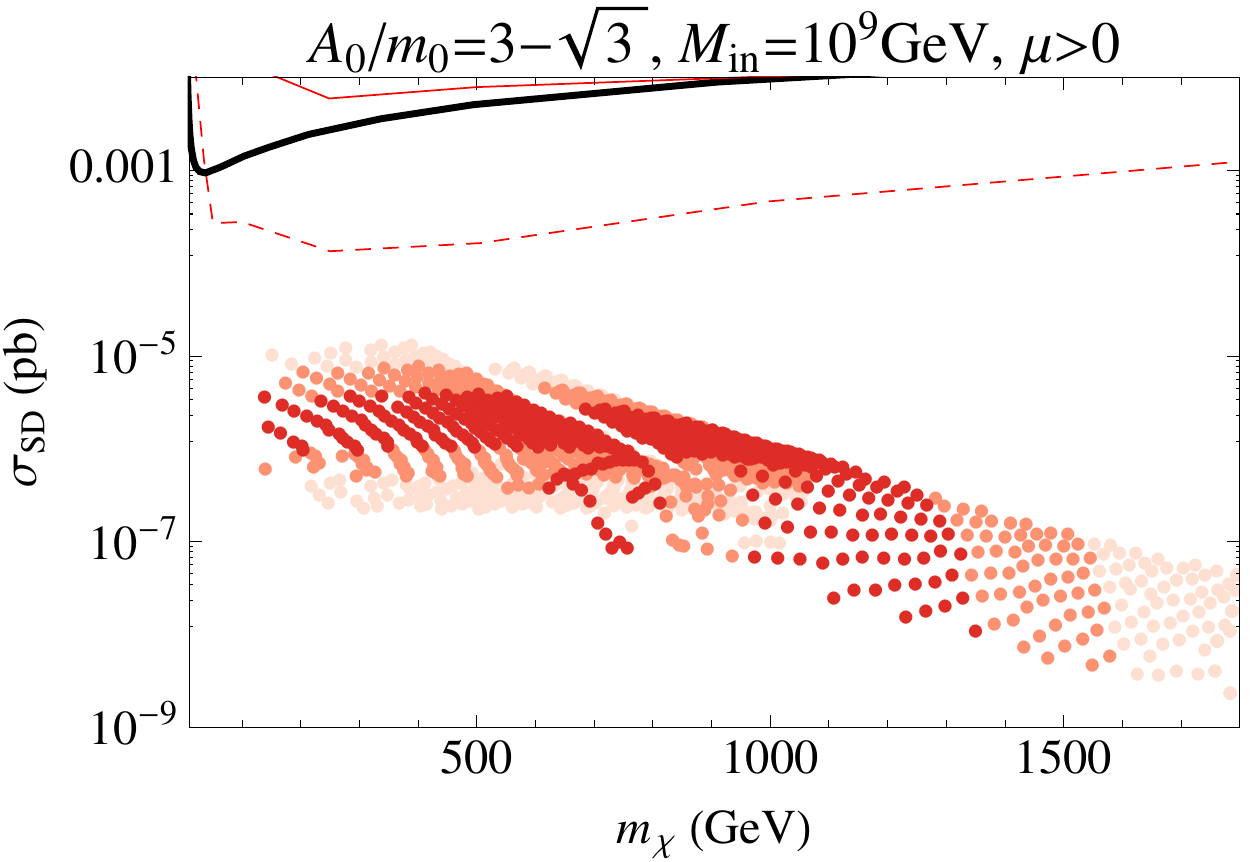}
\includegraphics[height=2.2in]{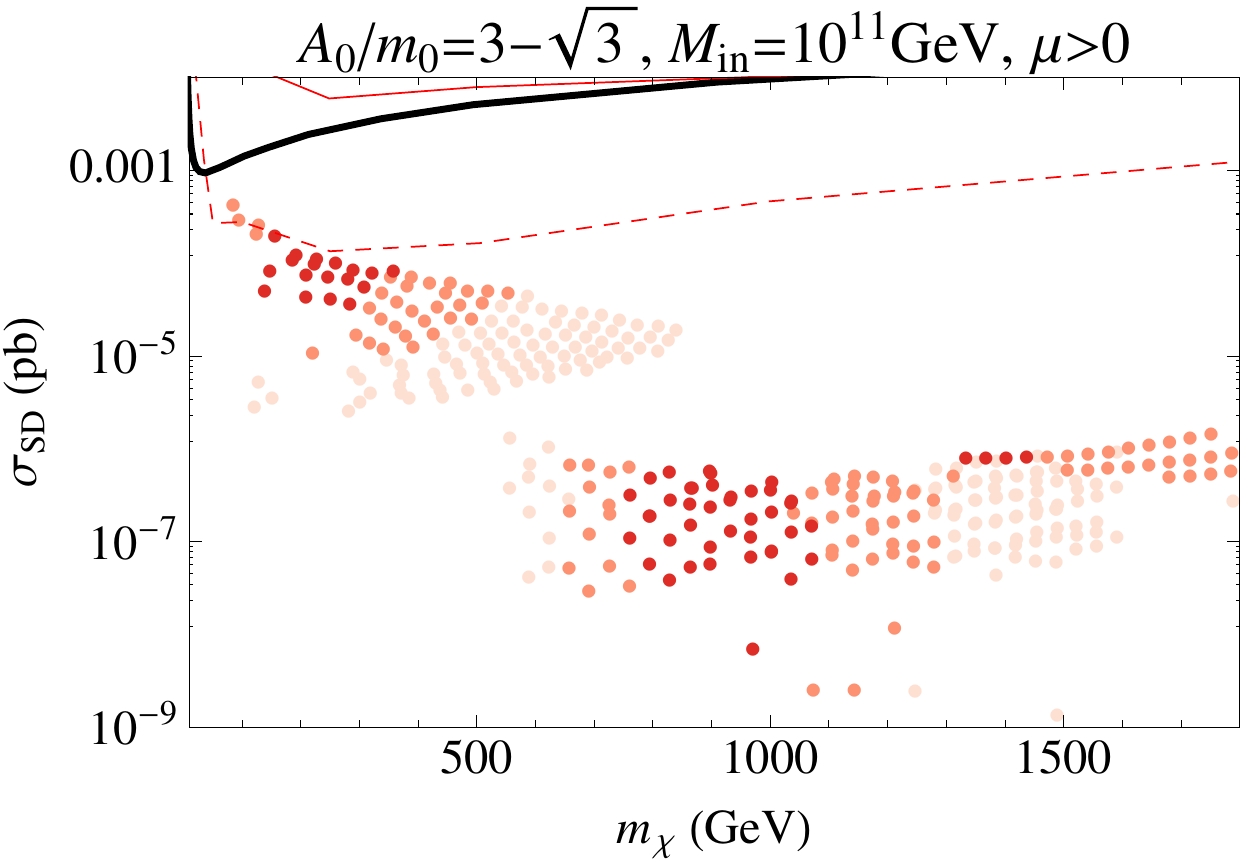}
\hfill
\end{minipage}
\caption{
{\it
As in Fig.~\protect{\ref{fig:CMSSMSD}} for the subGUT mSUGRA case with  $A_0 = (3 - \sqrt{3}) m_0$, and
$M_{in} = 10^9$ GeV (left panels) and $M_{in} = 10^{11}$ GeV (right panels)
shown in Fig.~\protect{\ref{fig:subGUTmsugra}}.
 }}
\label{fig:subGUTmsugraSD}
\end{figure}

We conclude this subsection by showing results for the SD cross sections in
these subGUT mSUGRA models in Fig.~\ref{fig:subGUTmsugraSD}. We see in
the upper panels that the SD cross sections are generally smaller than the PICO bound~\cite{pico},
and also below the IceCube upper limits~\cite{ice} for both ${\bar b} b$ and $W^+ W^-$ final
states (which are likely to be more similar to the model final states). There is a handful of
$M_{in} = 10^{11}$ GeV models (most of them under-dense) whose predictions lie close to the IceCube $W^+ W^-$ limit,
but most model predictions are significantly below it. We see in the lower right panel of
Fig.~\ref{fig:subGUTmsugraSD} that many of the models close to the IceCube $W^+ W^-$ limit
have {\tt FeynHiggs} $m_h$ values close to the experimental value. Comparing this figure with
Fig.~\ref{fig:subGUTmsugraSI}, it seems that there are better prospects for discovering SI
scattering in these subGUT mSUGRA scenarios. However, we recall that these models yield
proton lifetimes that are too short in minimal SU(5), pointing to the need for some non-minimal model.

\subsection{NUHM1}

As mentioned earlier, in the NUHM1 one has the freedom to treat $\mu$ as a free parameter,
and in the following we show $(m_{1/2}, m_0)$ planes for some representative choices
of $\mu$, $\tan \beta$ and $A_0$. In fact, as long as $\tan \beta$ is small enough to obtain
an acceptable proton lifetime, the qualitative behaviour of the parameter space is relatively
insensitive to $A_0$, though there is some dependence of the Higgs mass contours
on $A_0$, as could be expected.

\begin{figure}[htb!]
\begin{minipage}{8in}
\includegraphics[height=3.3in]{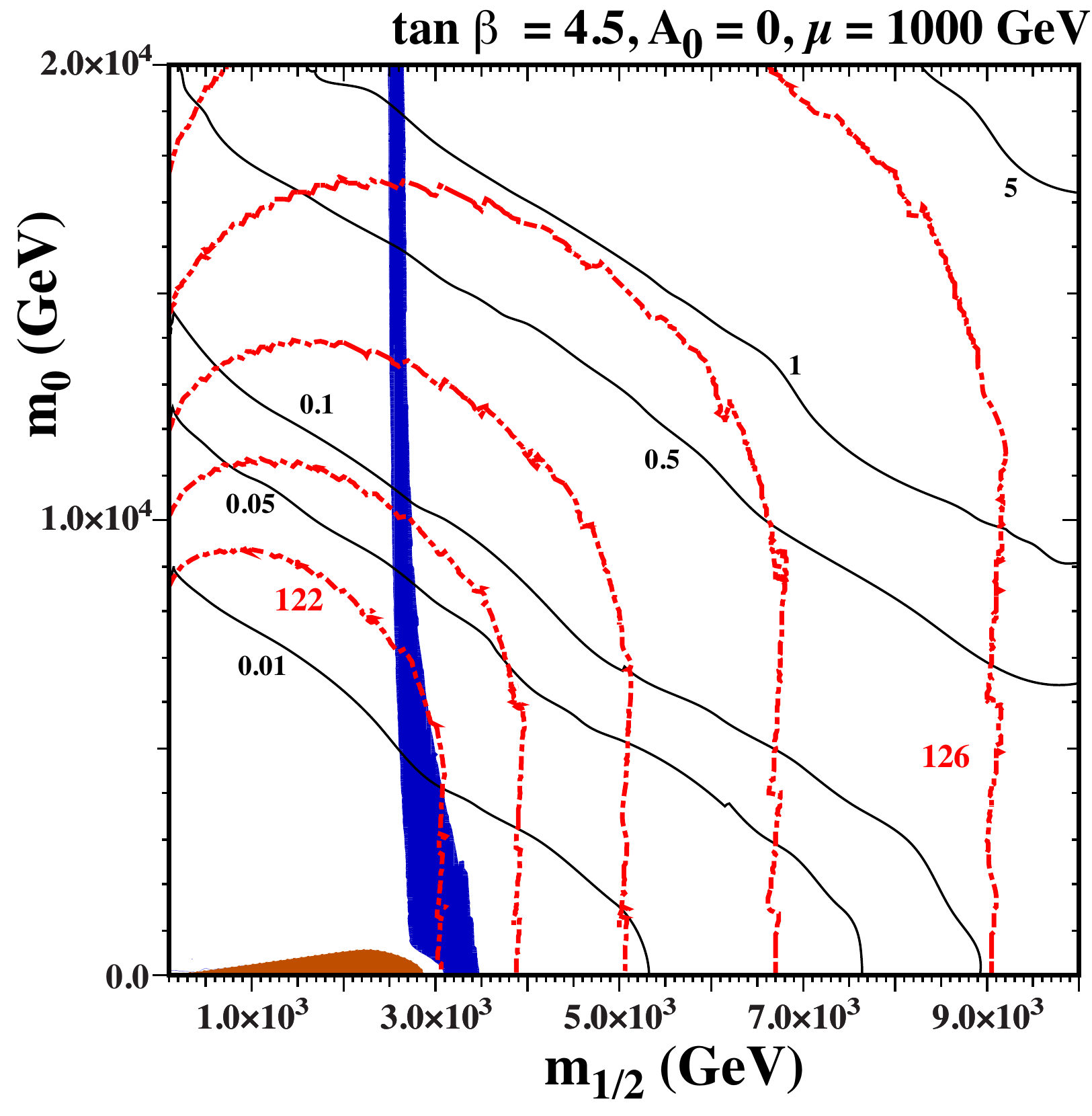}
\hspace*{-0.08in}
\includegraphics[height=3.3in]{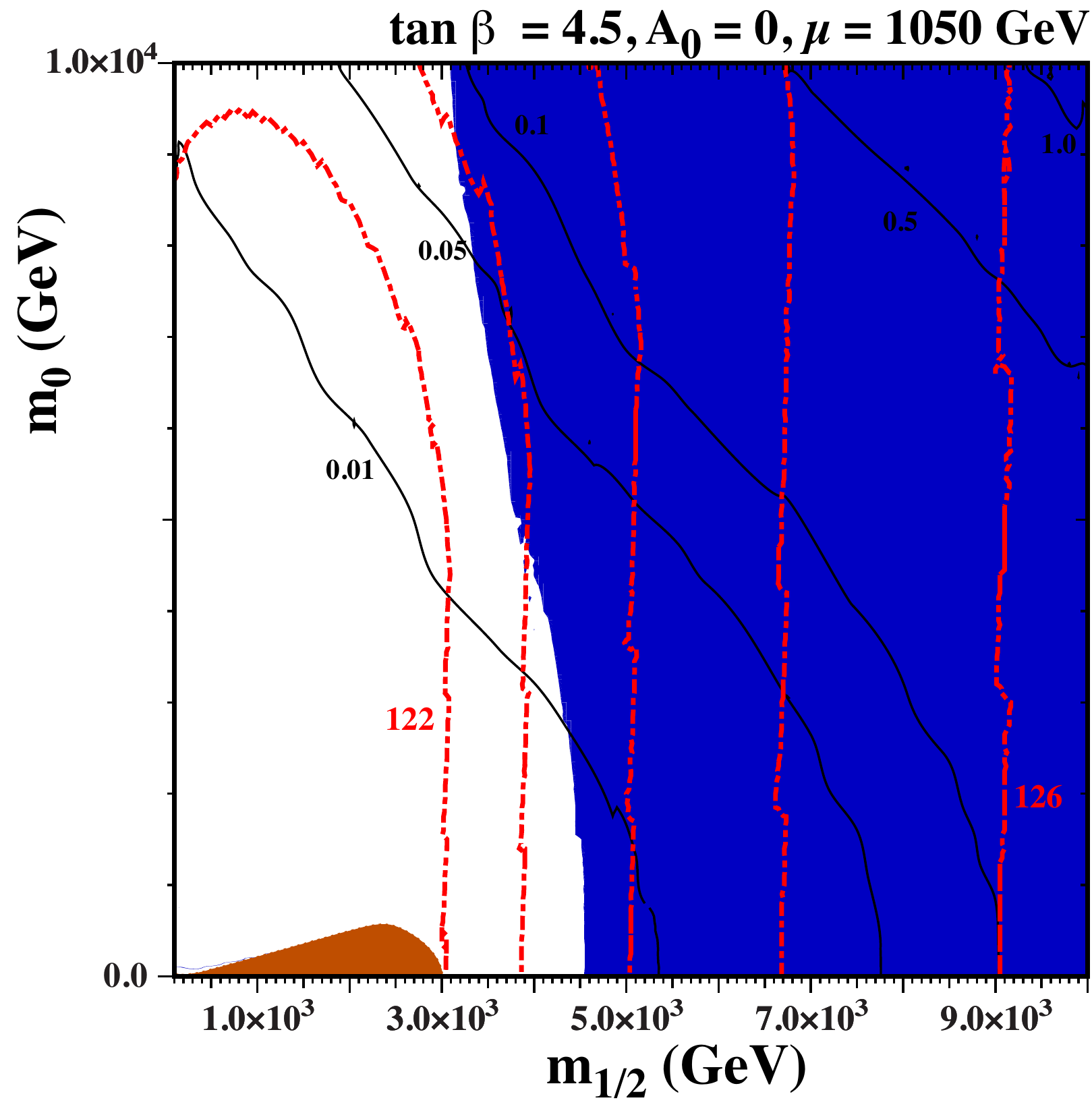}
\end{minipage}
\begin{minipage}{8in}
\includegraphics[height=3.3in]{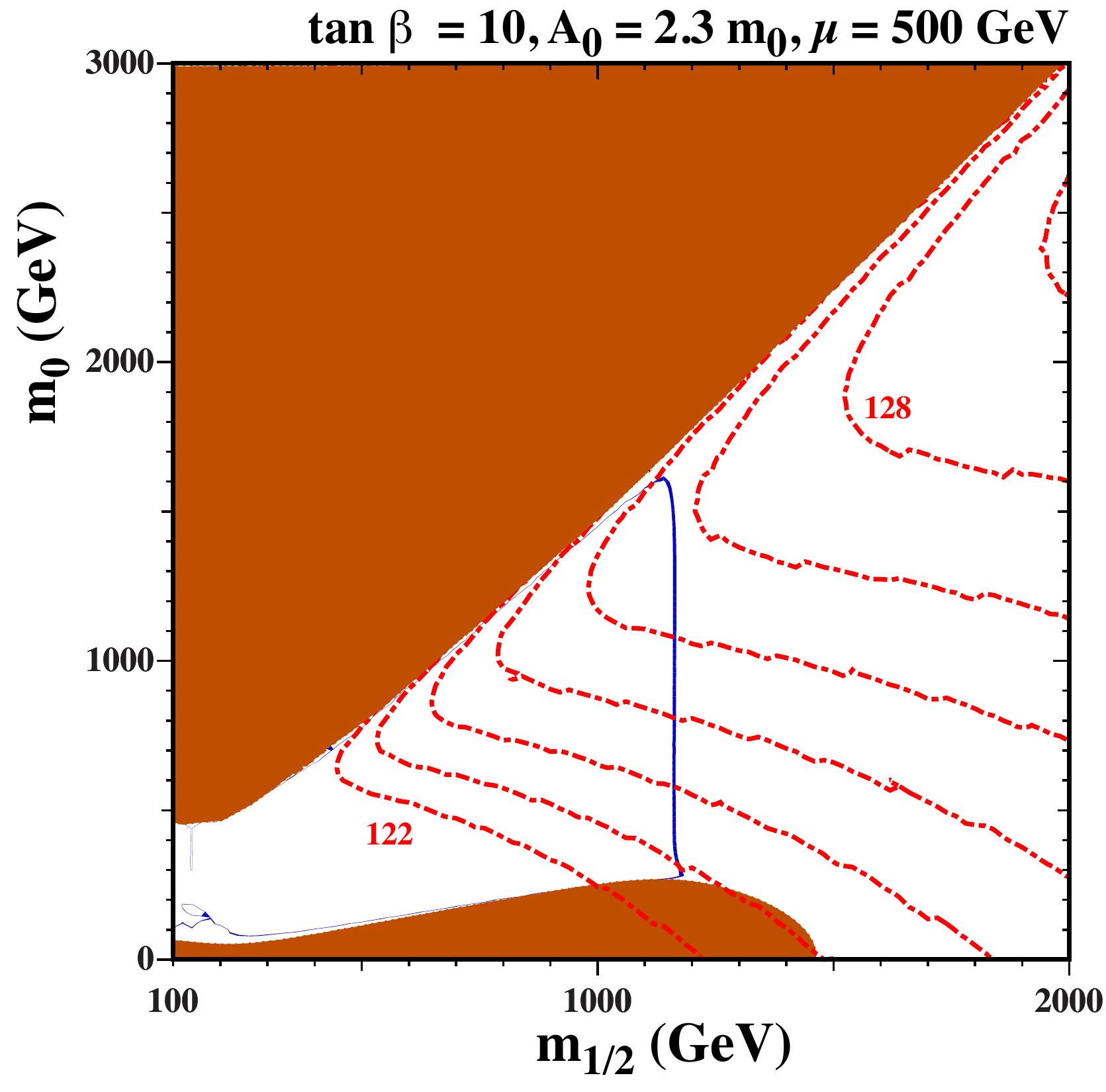}
\hspace*{-0.17in}
\includegraphics[height=3.3in]{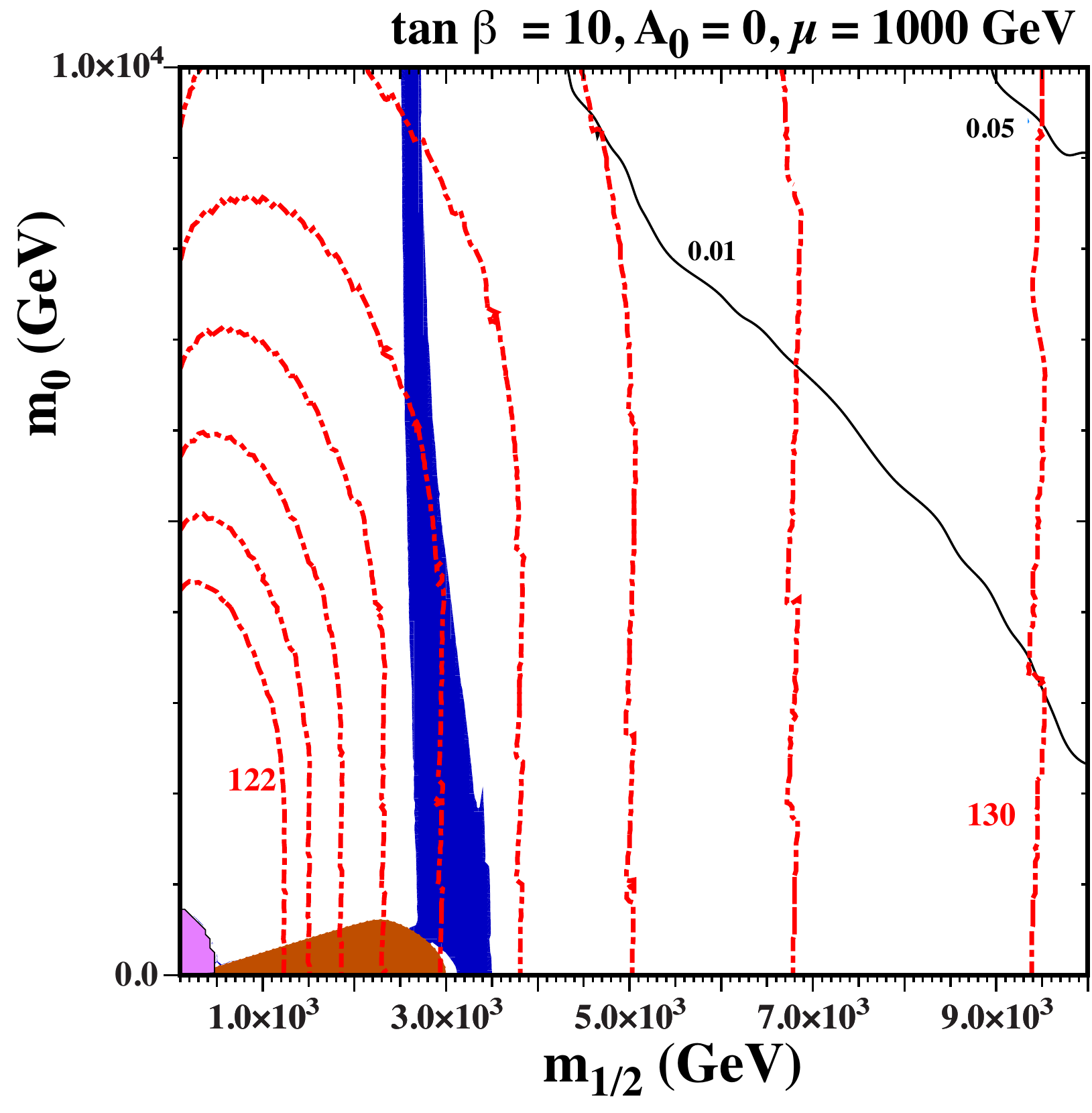}
\end{minipage}
\caption{
{\it
The NUHM1  $(m_{1/2}, m_0)$ planes for $\tan \beta = 4.5$ (upper) and $\tan \beta = 10$ (lower). We take
$\mu = 1000$ GeV in the upper left panel and 1050 GeV in the upper right panel, both with $A_0 = 0$.
In the lower panels, $\mu = 500$ GeV with $A_0 = 2.3 m_0$ (left) and $\mu = 1000$ GeV with $A_0 = 0$
(right). The shading and contour types are as in Fig.~\protect{\ref{fig:CMSSM}}.
}}
\label{fig:nuhm1}
\end{figure}

We show in the upper left panel of Fig.~\ref{fig:nuhm1} the $(m_{1/2}, m_0)$ plane for
$\tan \beta = 4.5$, $A_0 = 0$ and $\mu = 1000$~GeV, which exhibits a small stau LSP region
at low $m_0$ and $m_{1/2}$. Since $\mu$ is fixed, the composition of the LSP changes as $m_{1/2}$ is
increased. At small $m_{1/2}$ the LSP is mainly bino and the relic density is too high.
As $m_{1/2}$ is increased, the Higgino component increases and the relic density passes through the Planck
range across a relatively narrow, near-vertical transition strip.
(Note that, in all four panels of this figure, the blue region corresponds to just the 3$\sigma$
Planck range rather than the extended range used in previous figures.) At larger
$m_{1/2}$ the LSP is a Higgsino with a mass of about 1050 GeV which is slightly low for a
Higgsino LSP and, as a result, the relic density is somewhat too small when $m_{1/2} \gtrsim 3$ TeV. In this panel,
we see that we obtain an acceptable Higgs mass ($m_h > 124$ GeV)  when $m_0 \gtrsim 13$ TeV.
The proton lifetime is sufficiently large ($\tau_p \gtrsim 0.25 \times 10^{35}$ yrs) for this value of $m_0$.

In the upper right panel of Fig.~\ref{fig:nuhm1}, we have increased $\mu$ slightly to 1050 GeV.
The most striking feature is that the
dark matter region fills the right part of the plane: indeed, it
extends infinitely far to the right towards large gaugino masses. In this case, when
the gaugino mass is large, the LSP is a nearly pure Higgsino, as is the NLSP.
This near-degeneracy facilitates coannihilation that brings the relic density within the acceptable range,
with $\Omega_\chi h^2$ being determined predominantly by $\mu$~\cite{osi}.
The Higgsino mass in this case
is very close to 1100 GeV, which remains constant at large $m_{1/2}$. Thus there is
a very large (infinite) area where the relic density matches the Planck result.
At low $m_{1/2}$, the relic density is too large and drops monotonically  as the gaugino mass is increased
and asymptotes to the Planck density at very large $m_{1/2}$. 
For $m_0 \lesssim 10$ TeV, when $A_0 = 0$,
the Higgs mass contours are nearly vertical and the value of $\tan \beta = 4.5$ was
chosen to maximize the area with good relic density and Higgs masses.
The area between
$m_{1/2} = 5$ TeV and 9 TeV has $m_h$ between 124 and 126 GeV,
and increasing (decreasing) $\tan \beta$ by 0.5 would raise (lower) $m_h$ by roughly 1 GeV. Much of this
region has $\tau_p \gtrsim 0.05 \times 10^{35}$ yrs:
requiring $\tau_p > 5 \times 10^{33}$~yrs implies either $m_{1/2} \gtrsim 7.8$ TeV for small $m_0$ or
$m_0 \gtrsim 8$ TeV for $m_{1/2} \simeq 4$ TeV.
As $\mu$ is increased past 1050 GeV, the left edge of the blue shaded region moves quickly to the right 
and the relic density would be too large over much of the plane. The relic density would now asymptote to
a value in excess of the Planck density.
The Higgs mass is independent of $A_0/m_0$
for small $m_0$ but the Higgs mass contours bend to the left
as $A_0/m_0$ is increased, so that the Higgs mass becomes
large at larger $m_0$.

In the lower panels of Fig.~\ref{fig:nuhm1}, we have taken $\tan \beta = 10$.
In the left panel, $\mu = 500$ and the transition strip from bino to Higgsino dark matter is much narrower and
occurs at much lower $m_{1/2} \approx 1200$ GeV. Had we chosen $A_0 = 0$ as in the previous
plots, the Higgs mass would be far too small. This can be compensated in this panel by choosing larger $A_0$, and
we have chosen $A_0 = 2.3 m_0$ in this panel. As in the CMSSM, there is now a shaded region
where the LSP is a stop in the upper left of the panel. There is a barely visible stop coannihilation
strip that runs close to the stop LSP boundary, from the transition strip down to smaller $m_{1/2}$ and $m_0$.
There is also a narrow stau coannihilation strip running on top of the stau LSP region at low $m_0$.
In the right panel, we have again taken $A_0 = 0$ and increased $\mu$ to 1000 GeV.
The relic density region resembles that in the upper left panel of the same figure, though the Higgs masses
are now notably larger. The transition strip is now centered on $m_h = 126$ GeV, which is compatible
within the experimental measurement within the theoretical uncertainties.
We note also that the proton lifetime is far smaller in the lower panels due to the larger value of
$\tan \beta$. Indeed, in the lower left panel $\tau_p$ is always below 0.01 $\times 10^{35}$ yrs.

\begin{figure}[htb!]
\begin{minipage}{8in}
\includegraphics[height=2.2in]{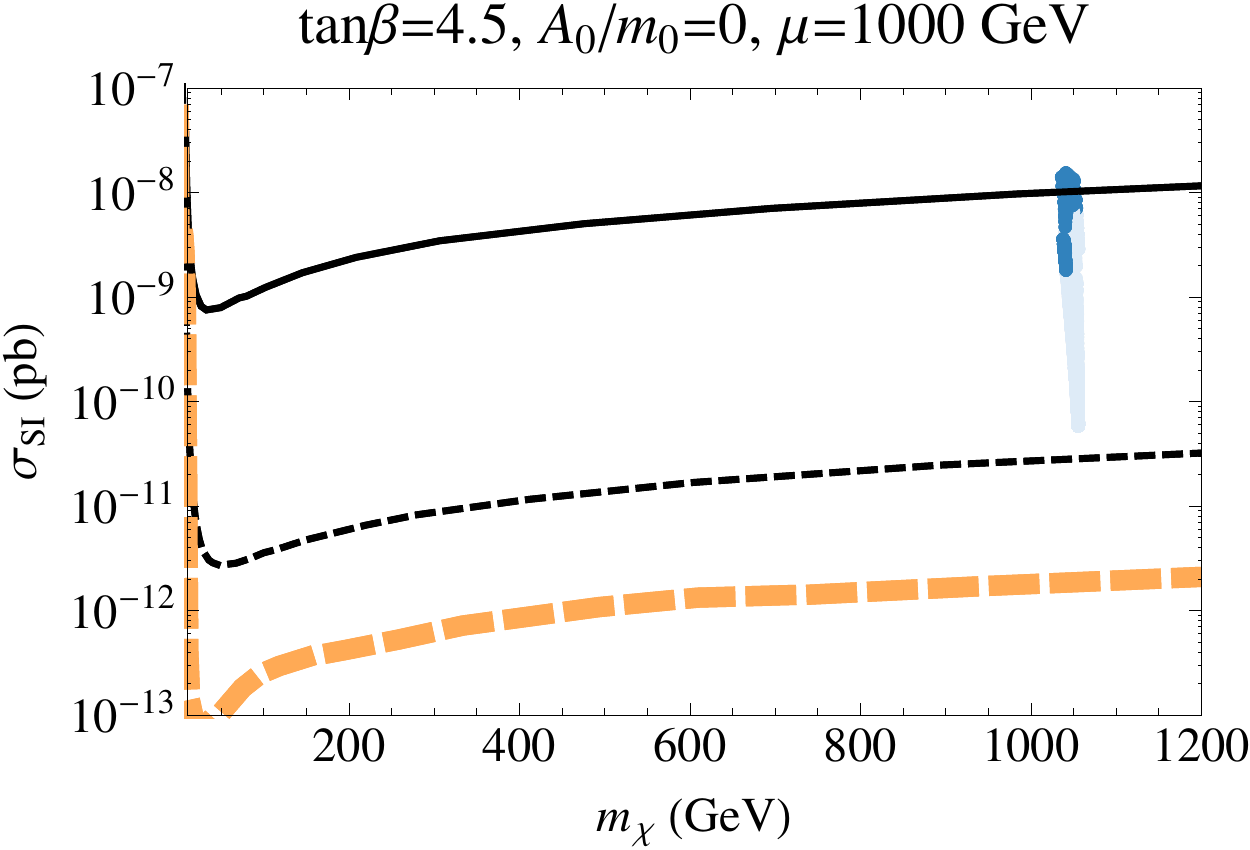}
\includegraphics[height=2.2in]{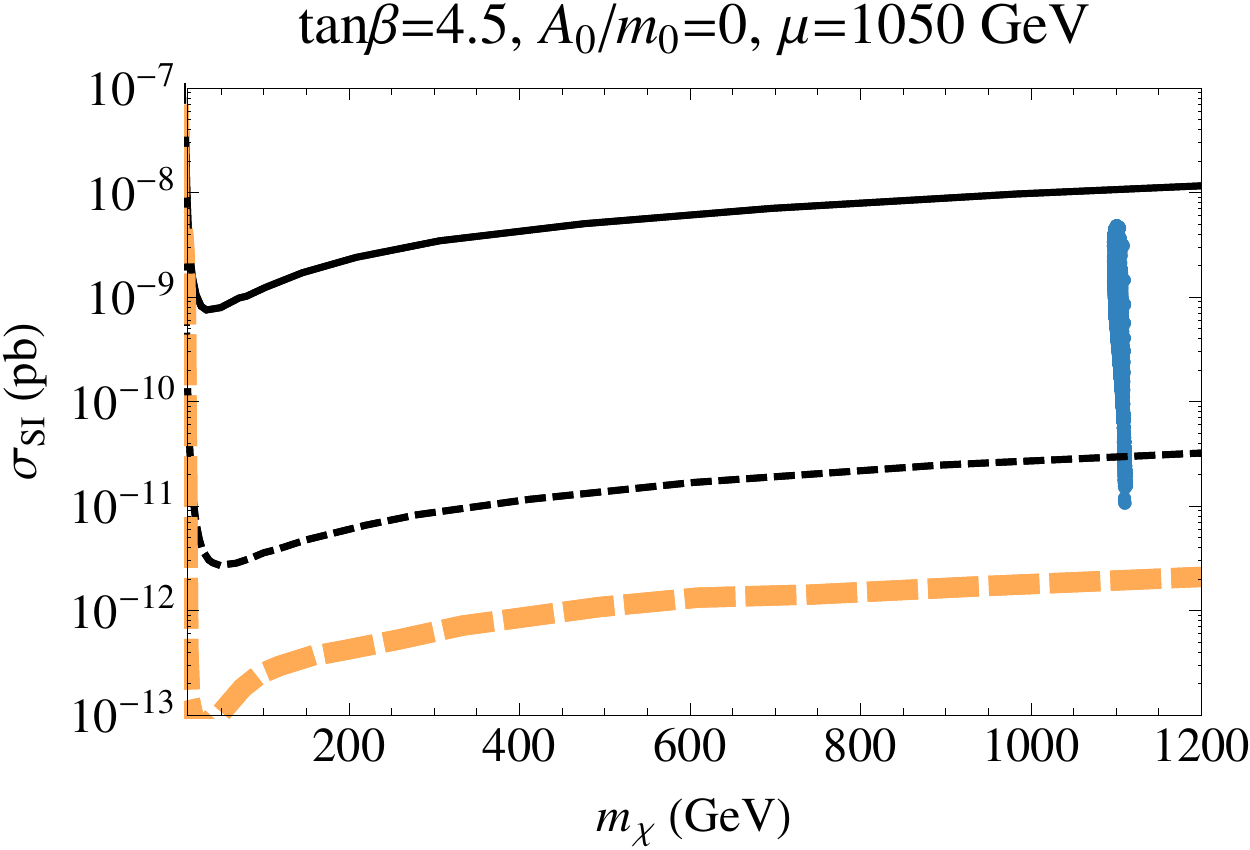}
\hfill
\end{minipage}
\begin{minipage}{8in}
\includegraphics[height=2.2in]{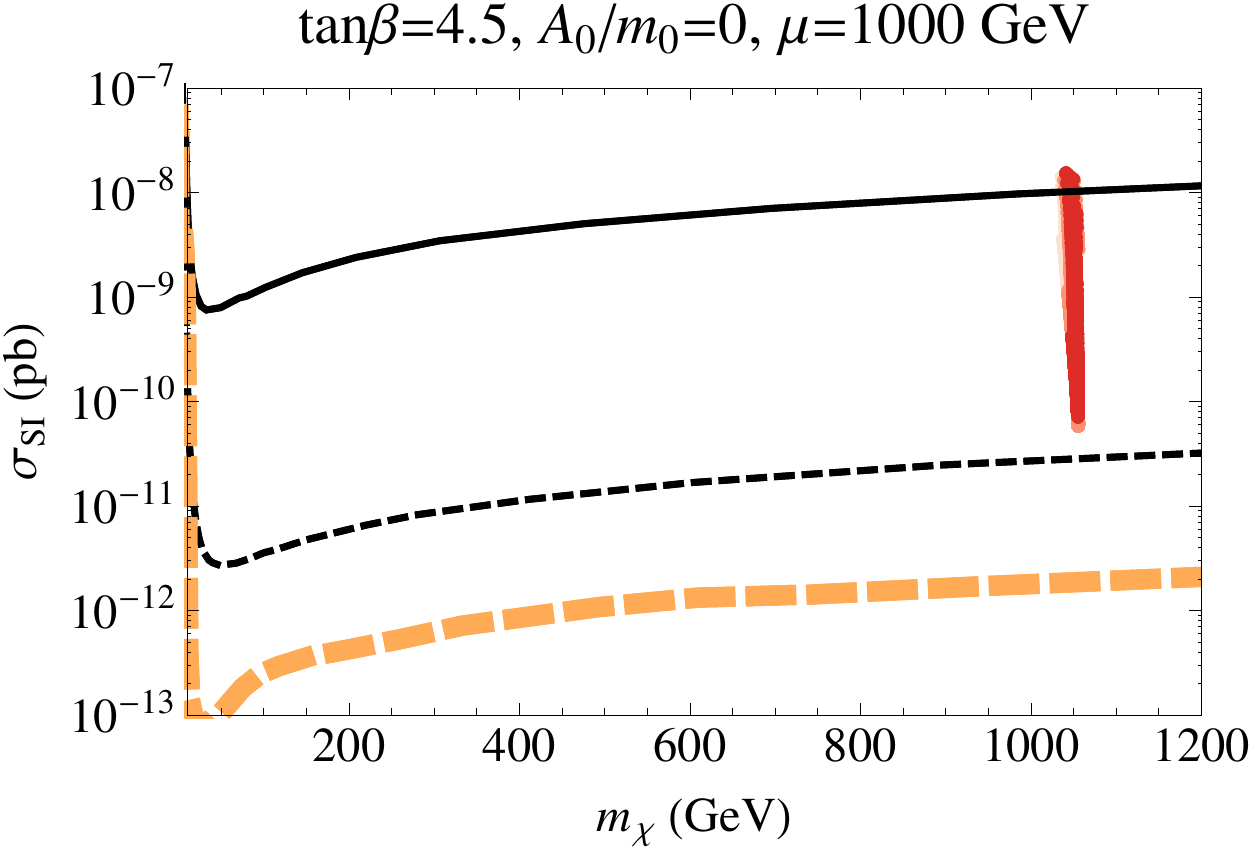}
\includegraphics[height=2.2in]{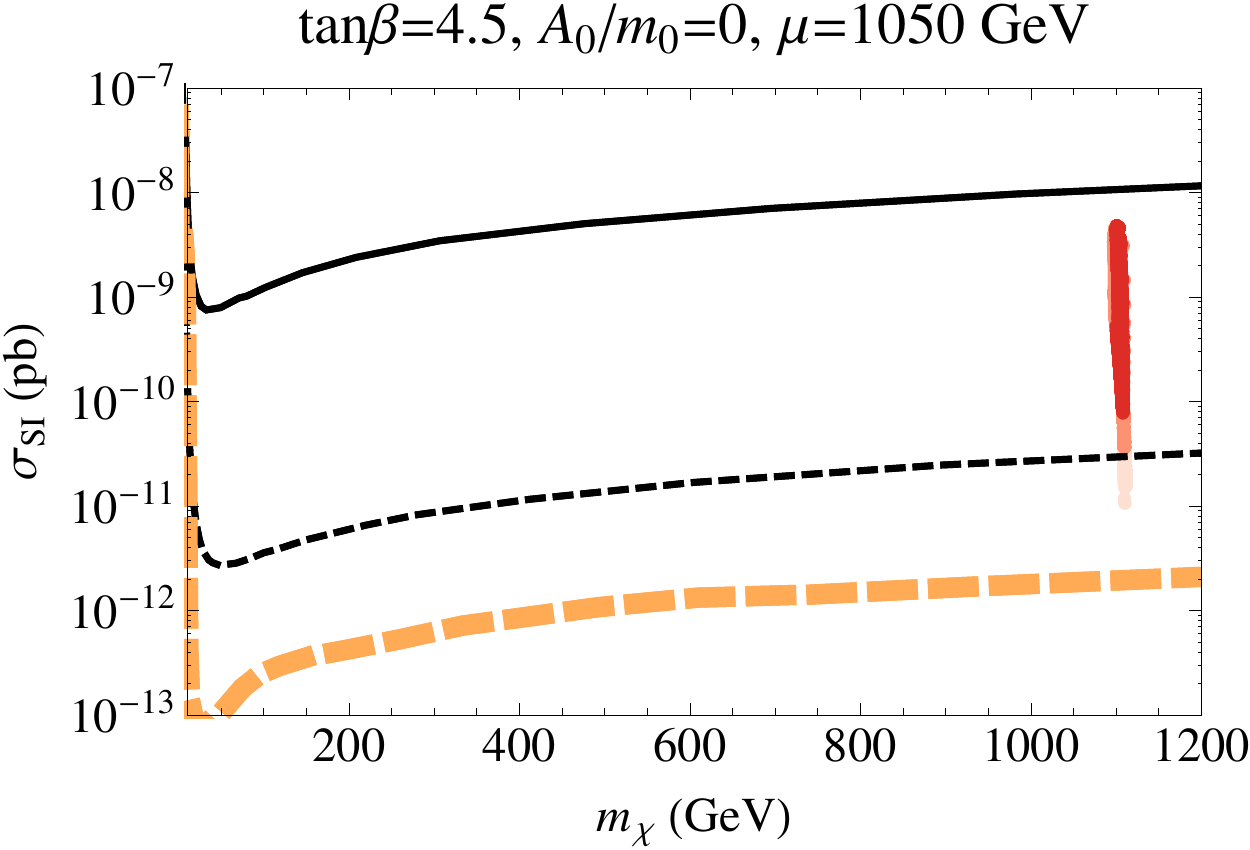}
\hfill
\end{minipage}

\caption{
{\it
As in Fig.~\protect{\ref{fig:CMSSMSI}} for the NUHM1 cases with  $\tan \beta =  4.5$ and
$A_0 = 0$ with $\mu = 1000$ GeV (left) and $\mu = 1050$ GeV (right).
 }}
\label{fig:NUHMSI1}
\end{figure}

\begin{figure}[htb!]
\begin{minipage}{8in}
\includegraphics[height=2.2in]{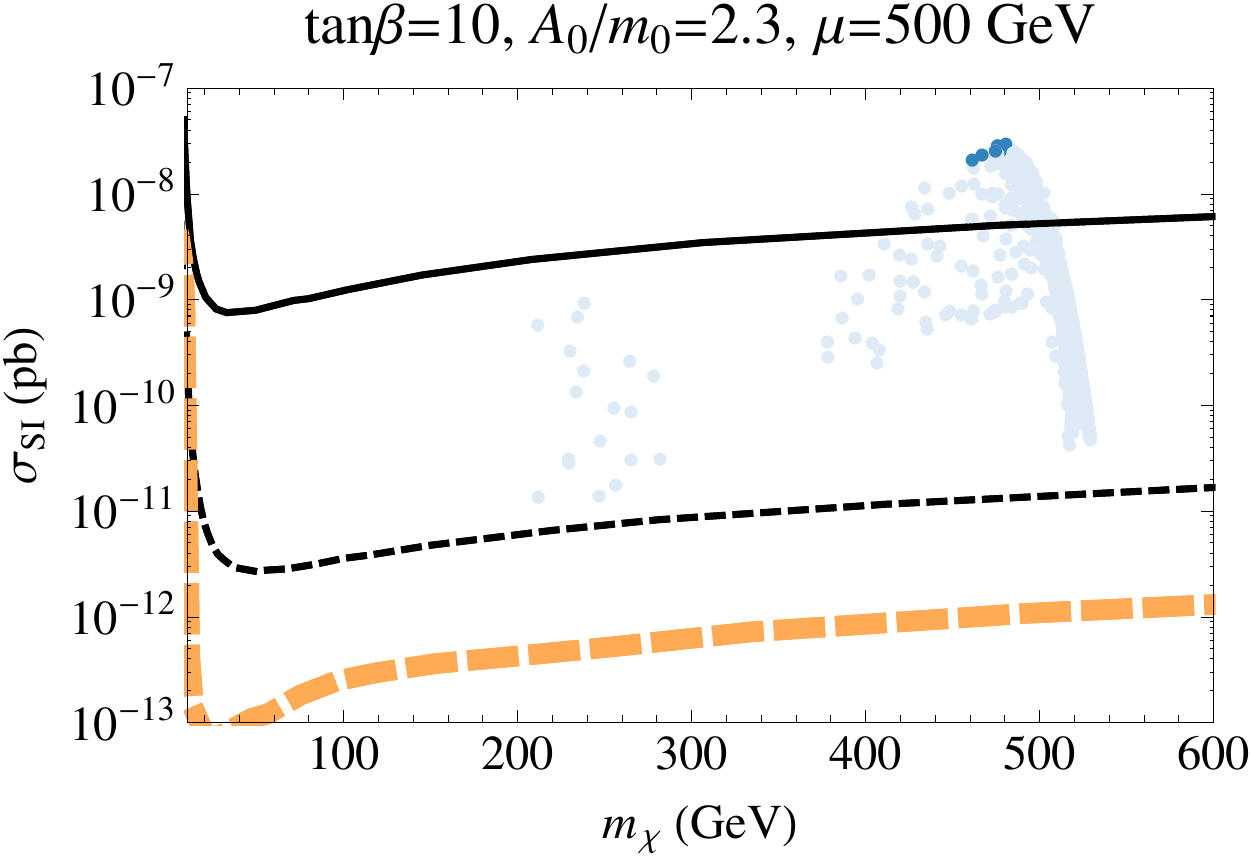}
\includegraphics[height=2.2in]{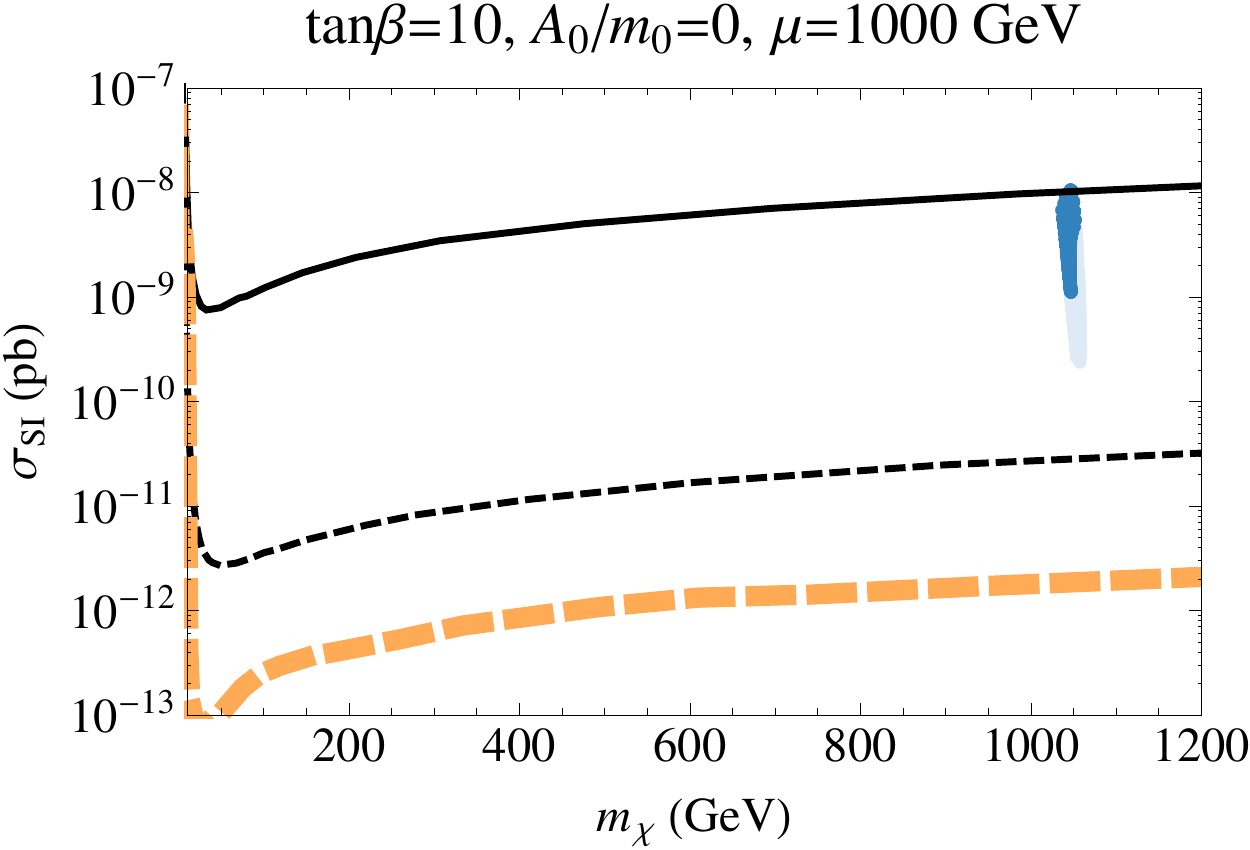}
\hfill
\end{minipage}
\begin{minipage}{8in}
\includegraphics[height=2.2in]{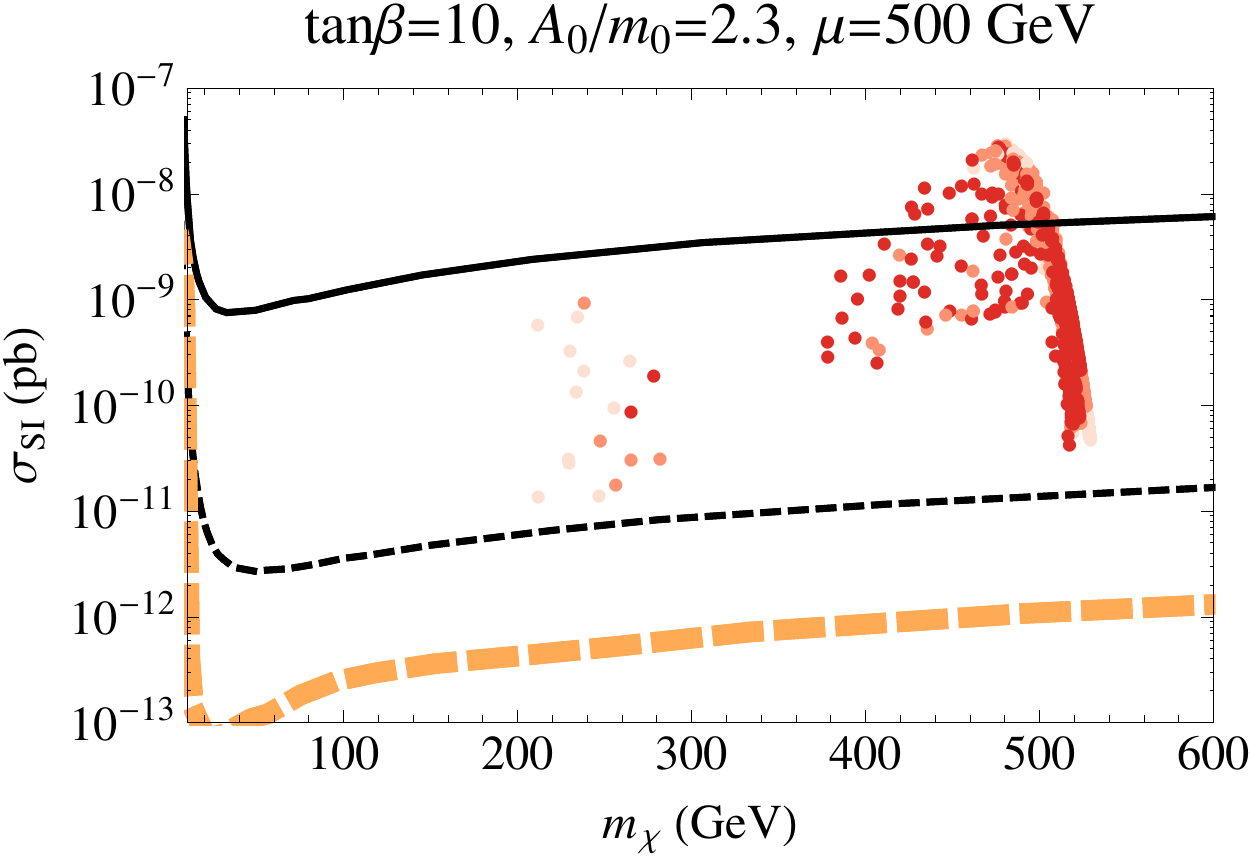}
\includegraphics[height=2.2in]{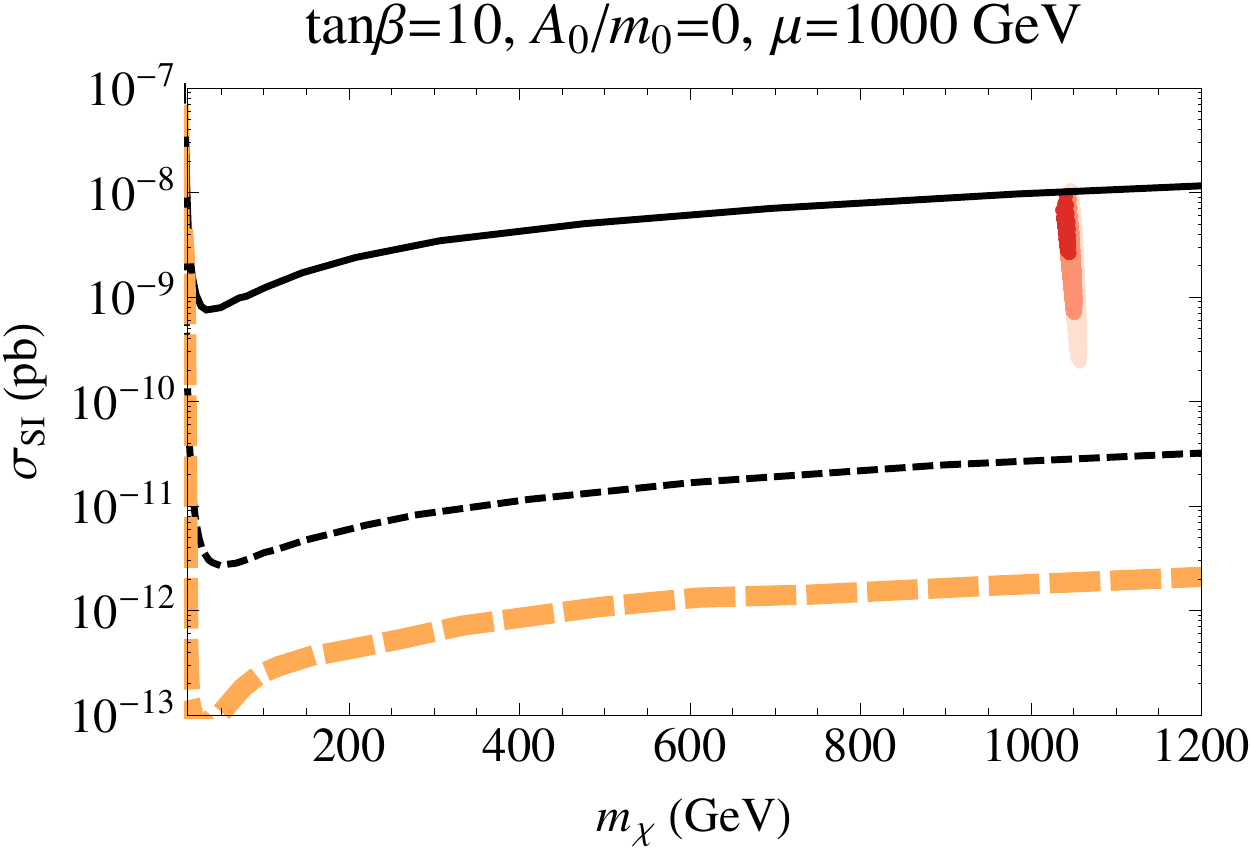}
\hfill
\end{minipage}

\caption{
{\it
As in Fig.~\protect{\ref{fig:CMSSMSI}} for the NUHM1 with  $\tan \beta = 10$ and
$A_0 = 2.3 m_0$ with $\mu = 500$ GeV (left) and $A_0 = 0$ with $\mu =1000$ GeV (right).
 }}
\label{fig:NUHMSI2}
\end{figure}

The elastic scattering cross sections for the four panels of Fig.~\ref{fig:nuhm1} are shown in
Figs.~\ref{fig:NUHMSI1} and \ref{fig:NUHMSI2}.
The left panels of Fig.~\ref{fig:NUHMSI1} correspond to the NUHM1 model with $\tan \beta = 4.5$, $A_0 = 0$, and $\mu = 1000$ GeV.  Viable points (with the correct relic density or less) have gaugino
masses of around 3 TeV (for the correct relic density) or greater (less than the Planck density).
In either case, the LSP mass is just over 1 TeV, which explains why all the points
line up vertically. Most of the points (though not all) lie below the current LUX limit and all of them
lie above the LZ projected reach.
Note that, in principle, this vertical strip could extend further down, into the neutrino background,
if we continued to sample points at higher $m_{1/2}$.  Our sampling of points includes only points with
$m_h$ between 122 and 128 GeV.
Concerning the right panels with $\mu = 1050$~GeV, we recall that much of the $(m_{1/2}, m_0)$ plane
contains a Higgsino LSP with the desired relic density. In that region, the mass
of the LSP is always very close to 1100 GeV and that fact is readily seen
in the right panels of Fig.~\ref{fig:NUHMSI1}, where all the points stack vertically at $m_\chi \simeq 1100$ GeV.
All of these points lie below the current LUX bound, but most of them  are within the projected reach of
LZ. As in the previous example, the vertical strip of points could go lower if we sampled to
larger $m_{1/2}$ where $m_h > 128$ GeV.
As seen in the lower right panel of this figure, the points
with $m_h$ between 124 and 126 GeV are all accessible to LZ.

In Fig.~\ref{fig:NUHMSI2}, we show the spin-independent cross sections for $\tan \beta = 10$.
In the left panels with $\mu = 500$ GeV, we again see a pile-up at a fixed LSP mass, $m_\chi \sim 500$~GeV,
corresponding in this case to the transition strip and the region to its right.
Since the strip is so narrow in this case, there are very few dark-shaded points, and these
have cross sections that exceed the LUX bound. There are also a few points at lower
$m_\chi$ that originate near the stop or stau coannihilation strips.  All of the points shown
lie within the LZ projected reach.  One also sees in the lower left panel that many of the points
have a Higgs mass in the 124--126 GeV range.
Indeed, for an underdense Higgsino-like LSP with $m_\chi \approx 500$ GeV (to the right of the transition strip in Fig.~\ref{fig:nuhm1}), there are also many points with $m_h>126$ GeV, though these points are eclipsed in Fig.~\ref{fig:NUHMSI2} by those with more favorable $m_h$.
In the right panels of  Fig.~\ref{fig:NUHMSI2}, for
$\mu = 1000$ GeV, all points in the transition strip and to the right of the strip
have a narrow range of LSP masses fixed by the value of $\mu$. These points lie just below
the current LUX bound.  Points that are most compatible with Planck dark matter abundance and have $m_h \approx 125$ GeV have the largest SI cross sections, within an order of magnitude of the current LUX limit.  All points considered here, with 122 GeV $<m_h<128$ GeV, should be accessible with LZ.

\section{Discussion}

Large parts of the CMSSM parameter space are excluded by the absence (so far) of proton decay,
if the CMSSM is embedded within the minimal SU(5) GUT. There are regions of parameter space with $\tan \beta \lesssim 5$
and $(m_{1/2}, m_0) \gtrsim$ several TeV that are still allowed, however. Typically, these models predict
a spin-independent dark matter-nucleon scattering cross section that falls below the current LUX upper limit
but could be accessible to the planned LZ experiment. In fact, it is the constraints coming from the Higgs boson mass which exclude the bulk of the model points which are beyond the reach of LZ. The prospects for direct detection
of spin-independent dark matter scattering are reduced for $\mu < 0$ and for $A_0 > 0$, and the
cross sections for spin-dependent dark matter scattering are generally substantially below the current
bounds from PICO and IceCube.

In view of the limited perspectives within the CMSSM, we have explored in this paper the prospects for probing
other MSSM scenarios via proton decay and dark matter detection. In one class of scenarios, called subGUT
models, universality of the soft supersymmetry-breaking masses is retained, but is imposed at some scale
$M_{in} < M_{GUT}$. Within the subGUT CMSSM there are more possibilities for bringing the supersymmetric
relic density within the range allowed by Planck and other experiments even if $m_{1/2}$ and $m_0$ are each
several TeV, thanks in particular to the more compressed spectrum and consequently the
greater possibilities for coannihilation processes that bring the dark matter density down into the allowed range.
However, small values of $\tan \beta \lesssim 5$ are preferred again, as in the GUT-scale CMSSM, unless
one adopts a non-minimal GUT structure. For $M_{in} = 10^9$~GeV and $\tan \beta = 3.5$ or 10, we find
spin-independent dark matter-nucleon scattering cross sections that are well within the range allowed by LUX, and the
spin-independent cross section may fall below the neutrino background level, particularly for $\mu < 0$. 

In mSUGRA models, the possibilities are very limited if $M_{in} = M_{GUT}$, but open up for $M_{in} < M_{GUT}$.
On the other hand, in mSUGRA models $\tan \beta$ is no longer a free parameter, and the electroweak vacuum conditions
typically require large values that give severe problems within the minimal SU(5) GUT framework. That said,
spin-independent dark matter scattering may again lie within reach of the LZ experiment, whereas spin-dependent
scattering cross sections generally lie below the PICO and IceCube upper limits.

In the NUHM1 one may regard the Higgs mixing parameter $\mu$ as an extra free
parameter compared to the CMSSM. This freedom opens up new possibilities for models that
respect the dark matter, Higgs mass, and proton
decay constraints. In particular, since varying $\mu$ varies the Higgsino component of the LSP, there is
the possibility of a `well-tempered' transition region as well as the more familiar stau and stop
coannihilation possibilities for bringing the relic neutralino density into (or below) the Planck range.
Moreover, for some $\mu$ values as seen in the upper right panel of Fig.~\ref{fig:nuhm1}, in particular,
the relic density may lie within the Planck range up to indefinitely high values of $m_{1/2}$ and $m_0$.
In this case, the proton lifetime may certainly be long enough to survive the present experimental lower
limit whereas, as seen in Fig.~\ref{fig:NUHMSI1}, the spin-independent dark matter scattering cross
section is likely to be within reach of the planned LZ experiment.

In conclusion, the examples studied in this paper show that there are certainly interesting possibilities for probing
supersymmetric models beyond the CMSSM via searches for proton decay and direct dark matter
scattering.

\section*{Acknowledgements}

The work of J.E. was supported in part by the London Centre for Terauniverse Studies
(LCTS), using funding from the European Research Council via the Advanced Investigator
Grant 267352 and from the UK STFC via the research grant ST/J002798/1. The work of F.L.
was also supported by the European Research Council Advanced Investigator
Grant 267352. The work of J.L.E., N.N. and K.A.O.
was supported in part by DOE grant DE-SC0011842 at the University of Minnesota.
The work of N.N. was also supported by Research
Fellowships of the Japan Society for the Promotion of Science for Young
Scientists.
The work of P.S.~was supported in part by NSF Grant No. PHY-1417367.  P.S.~would also like to thank CETUP* (Center for Theoretical Underground Physics and Related Areas) for its hospitality and partial support during the 2015 Summer Program.

\section*{Appendix}
\appendix

\section{Wilson Coefficients and RGEs for Proton Decay}
\label{sec:protondecayapp}

In this Appendix, we summarize the matching conditions and RGEs used in
the proton decay calculation discussed in Sec.~\ref{pdecay}. The Wilson
coefficients $C^{ijkl}_{5L}$ and $C^{ijkl}_{5R}$ in
Eq.~\eqref{eq:efflaggut} are given at the GUT scale by
\begin{align}
 C^{ijkl}_{5L}(M_{GUT})&
=\frac{1}{M_{H_C}}y_{u_i}
e^{i\varphi_i}\delta^{ij}V^*_{kl}y_{d_l}~,\nonumber \\
C^{ijkl}_{5R}(M_{GUT})
&=\frac{1}{M_{H_C}}y_{u_i}V_{ij}V^*_{kl}y_{d_l}
e^{-i\varphi_k}
~,
\label{eq:wilson5}
\end{align}
where $y_{u_i}$
and $y_{d_l}$ are up- and down-type Yukawa couplings at the GUT scale,
respectively, $V_{ij}$ is the Cabibbo-Kobayashi-Maskawa (CKM)
matrix, and $\varphi_i$ $(i=1,2,3)$ denote the extra phases appearing in the
GUT Yukawa couplings. They are taken such that they satisfy
$\sum_{i}\varphi_i = 0$, and thus there are two independent degrees of
freedom~\cite{Ellis:1979hy}. For the definition of the GUT Yukawa
couplings, we follow the convention of~\cite{evno}. These unknown phases
cause uncertainty in our calculation, whose significance is estimated
below.

These Wilson coefficients are run down to the SUSY breaking scale
using RGEs. Since the theory is supersymmetric, the RGEs of the
Wilson coefficients are readily obtained from the anomalous dimensions
of the fields in the corresponding effective operators, thanks to the
non-renormalization property \cite{Grisaru:1979wc} of the holomorphic
operators. Hence we have
\begin{align}
 \frac{d}{d\ln Q} C^{ijkl}_{5L} &=
\frac{1}{16\pi^2}\biggl[-\frac{2}{5}g_1^2 -6g_2^2 -8g_3^2
+y_{u_i}^2 +y_{d_i}^2 + y_{u_j}^2 +y_{d_j}^2 + y_{u_k}^2 + y_{d_k}^2 +
 y_{e_l}^2
\biggr] C^{ijkl}_{5L}~, \nonumber \\
 \frac{d}{d\ln Q} C^{ijkl}_{5R} &=
\frac{1}{16\pi^2}\biggl[-\frac{12}{5}g_1^2 -8g_3^2
+2y_{u_i}^2 + 2y_{e_j}^2 + 2y_{u_k}^2 + 2y_{d_l}^2
\biggr] C^{ijkl}_{5R}~,
\end{align}
where $Q$ denotes the renormalization scale.

The matching conditions at the sfermion mass scale are
\begin{align}
 C_i^{\widetilde{H}}
&=\frac{y_ty_\tau}{(4\pi)^2}
F(\mu, m_{\widetilde{t}_R}^2,m_{\tau_R}^2) C^{*331i}_{5R}~, \nonumber \\
 C^{\widetilde{W}}_{jk} &=
\frac{\alpha_2}{4\pi}\left[
F(M_2, m_{\widetilde{Q}_1}^2,
 m_{\widetilde{Q}_j}^2) +F(M_2, m_{\widetilde{Q}_j}^2,
 m_{\widetilde{L}_k}^2)\right]C^{jj1k}_{5L} ~, \nonumber \\
 \overline{C}^{\widetilde{W}}_{jk} &=
-\frac{3}{2}
\frac{\alpha_2}{4\pi}\left[
F(M_2, m_{\widetilde{Q}_j}^2,
 m_{\widetilde{Q}_j}^2) +F(M_2, m_{\widetilde{Q}_1}^2,
 m_{\widetilde{L}_k}^2)\right] C^{jj1k}_{5L}~,
\end{align}
where $m_{\widetilde{t}_R}$, $m_{\widetilde{\tau}_R}$,
$m_{\widetilde{Q}_j}$, and $m_{\widetilde{L}_k}$
are the masses of the right-handed stop, the right-handed stau, left-handed
squarks, and left-handed sleptons, respectively, $\alpha_i \equiv
g_i^2/(4\pi)$, and
 \begin{align}
F(M, m_1^2, m_2^2) &\equiv
\frac{M}{m_1^2-m_2^2}
\biggl[
\frac{m_1^2}{m_1^2-M^2}\ln \biggl(\frac{m_1^2}{M^2}\biggr)
-\frac{m_2^2}{m_2^2-M^2}\ln \biggl(\frac{m_2^2}{M^2}\biggr)
\biggr]~.
\label{eq:funceq}
\end{align}

Between the SUSY-breaking scale and the electroweak scale, the RGEs for the
Wilson coefficients are given by \cite{Alonso:2014zka}
\begin{align}
 \frac{d}{d\ln Q} C^{\widetilde{H}}_{i}
&= \biggl[\frac{\alpha_1}{4\pi}\biggl(-\frac{11}{10}\biggr)
+\frac{\alpha_2}{4\pi}\biggl(-\frac{9}{2}\biggr)
+\frac{\alpha_3}{4\pi}(-4) +\frac{1}{2}\frac{f_{t}^2}{16\pi^2}
\biggr]C^{\widetilde{H}}_{i}
 ~,\nonumber \\
 \frac{d}{d\ln Q} C^{\widetilde{W}}_{jk}
&= \biggl[\frac{\alpha_1}{4\pi}\biggl(-\frac{1}{5}\biggr)
+\frac{\alpha_2}{4\pi}(-3)
+\frac{\alpha_3}{4\pi}(-4) +\frac{f_{u_j}^2}{16\pi^2}
\biggr]C^{\widetilde{W}}_{jk}
+\frac{\alpha_2}{4\pi}(-4)[
2C^{\widetilde{W}}_{jk} + \overline{C}^{\widetilde{W}}_{jk}]
 ~,\nonumber \\
 \frac{d}{d\ln Q} \overline{C}^{\widetilde{W}}_{jk}
&= \biggl[\frac{\alpha_1}{4\pi}\biggl(-\frac{1}{5}\biggr)
+\frac{\alpha_2}{4\pi}(-3)
+\frac{\alpha_3}{4\pi}(-4) +\frac{f_{u_j}^2}{16\pi^2}
\biggr]\overline{C}^{\widetilde{W}}_{jk}
+\frac{\alpha_2}{4\pi}(-4)[
2C^{\widetilde{W}}_{jk} + \overline{C}^{\widetilde{W}}_{jk}] ~,
\end{align}
where $f_{u_j}$ denote the SM up-type Yukawa couplings. At the
electroweak scale, the effective operators are matched onto the effective
interactions that induce the $p\to K^+\bar{\nu}_k$ decay mode. The
interactions are written as
\begin{align}
 {\cal L}(p\to K^+\bar{\nu}_i^{})
=&C_{RL}(usd\nu_i)\bigl[\epsilon_{abc}(u_R^as_R^b)(d_L^c\nu_i^{})\bigr]
+C_{RL}(uds\nu_i)\bigl[\epsilon_{abc}(u_R^ad_R^b)(s_L^c\nu_i^{})\bigr]
\nonumber \\
+&C_{LL}(usd\nu_i)\bigl[\epsilon_{abc}(u_L^as_L^b)(d_L^c\nu_i^{})\bigr]
+C_{LL}(uds\nu_i)\bigl[\epsilon_{abc}(u_L^ad_L^b)(s_L^c\nu_i^{})\bigr]
~,
\end{align}
and we have
\begin{align}
 C_{RL}(usd\nu_\tau)&=-V_{td}C^{\widetilde{H}}_{2}(m_Z)~,\nonumber \\
 C_{RL}(uds\nu_\tau)&=-V_{ts}C^{\widetilde{H}}_{1}(m_Z)~,\nonumber \\
 C_{LL}(usd\nu_k)&=\sum_{j=2,3}V_{j1}V_{j2}
C^{\widetilde{W}}_{jk}(m_Z)~,\nonumber \\
 C_{LL}(uds\nu_k)&=\sum_{j=2,3}V_{j1}V_{j2}
C^{\widetilde{W}}_{jk}(m_Z)~.
\end{align}
We note that $\overline{C}^{\widetilde{W}}_{jk}$ does not contribute to the
electroweak matching conditions: it is relevant only to the RGEs.

The above coefficients are then run down to the hadronic scale $Q_{\text{had}}=
2$~GeV, where the matrix elements of the effective operators are
evaluated. The QCD contributions to the RGE for this step are calculated at
two-loop level in Ref.~\cite{Nihei:1994tx}. For a generic coefficient $C$, the two-loop RGE is
\begin{equation}
  \frac{d}{d \ln Q}C =
-\biggl[
4\frac{\alpha_s}{4\pi}+\biggl(\frac{14}{3}+\frac{4}{9}N_f
+\Delta\biggr)\frac{\alpha_s^2}{(4\pi)^2}
\biggr]C ~,
\end{equation}
where $\alpha_s$ is the strong coupling constant, $N_f$ is the number of
quark flavors, and $\Delta=0$ ($\Delta=-10/3$) for $C_{LL}$
($C_{RL}$). The analytical solutions of the RGEs are given in
Refs.~\cite{Nagata:2013sba, Nihei:1994tx}.

For the hadron matrix elements of the effective operators, we use the results
given by the lattice QCD simulation in~\cite{Aoki:2013yxa}. Using
these results, we obtain finally the partial decay width of the $p\to K^+ \bar{\nu}_i$
mode:
\begin{equation}
 \Gamma(p\to K^+\bar{\nu}_i)
=\frac{m_p}{32\pi}\biggl(1-\frac{m_K^2}{m_p^2}\biggr)^2
\vert {\cal A}(p\to K^+\bar{\nu}_i)\vert^2~,
\end{equation}
where $m_p$ and $m_K$ are the masses of the proton and kaon,
respectively.  The
amplitude ${\cal A}(p\to K^+\bar{\nu}_i)$ is the sum of the Wilson
coefficients multiplied by the corresponding hadron matrix elements:
\begin{align}
 {\cal A}(p\to K^+\bar{\nu}_e)&=
C_{LL}(usd\nu_e)\langle K^+\vert (us)_Ld_L\vert p\rangle
+C_{LL}(uds\nu_e)\langle K^+\vert (ud)_Ls_L\vert p\rangle ~,
\nonumber \\
 {\cal A}(p\to K^+\bar{\nu}_\mu)&=
C_{LL}(usd\nu_\mu)\langle K^+\vert (us)_Ld_L\vert p\rangle
+C_{LL}(uds\nu_\mu)\langle K^+\vert (ud)_Ls_L\vert p\rangle ~,
\nonumber \\
 {\cal A}(p\to K^+\bar{\nu}_\tau)&=
C_{RL}(usd\nu_\tau)\langle K^+\vert (us)_Rd_L\vert p\rangle
+
C_{RL}(uds\nu_\tau)\langle K^+\vert (ud)_Rs_L\vert p\rangle
\nonumber \\
&+
C_{LL}(usd\nu_\tau)\langle K^+\vert (us)_Ld_L\vert p\rangle
+C_{LL}(uds\nu_\tau)\langle K^+\vert (ud)_Ls_L\vert p\rangle
~.
\end{align}
The following are the numerical values of the hadron matrix elements
at the scale of $Q_{\text{had}}=2$~GeV found in~\cite{Aoki:2013yxa}:
\begin{align}
\langle K^+\vert (us)_L ^{}d_L^{}\vert p\rangle &= 0.036(12)(7)
~~\text{GeV}^2~,
 \nonumber\\
\langle K^+\vert (ud)_L ^{}s_L^{}\vert p\rangle &= 0.111(22)(16)
~~\text{GeV}^2~,
\nonumber \\
\langle K^+\vert (us)_R ^{}d_L^{}\vert p\rangle &= -0.054(11)(9)
~~\text{GeV}^2~,
\nonumber \\
\langle K^+\vert (ud)_R ^{}s_L^{}\vert p\rangle &= -0.093(24)(18)
~~\text{GeV}^2 ~.
\end{align}
The first and second parentheses represent statistical and systematic
errors, respectively.

\begin{figure}[t]
\centering
\includegraphics[height=3in]{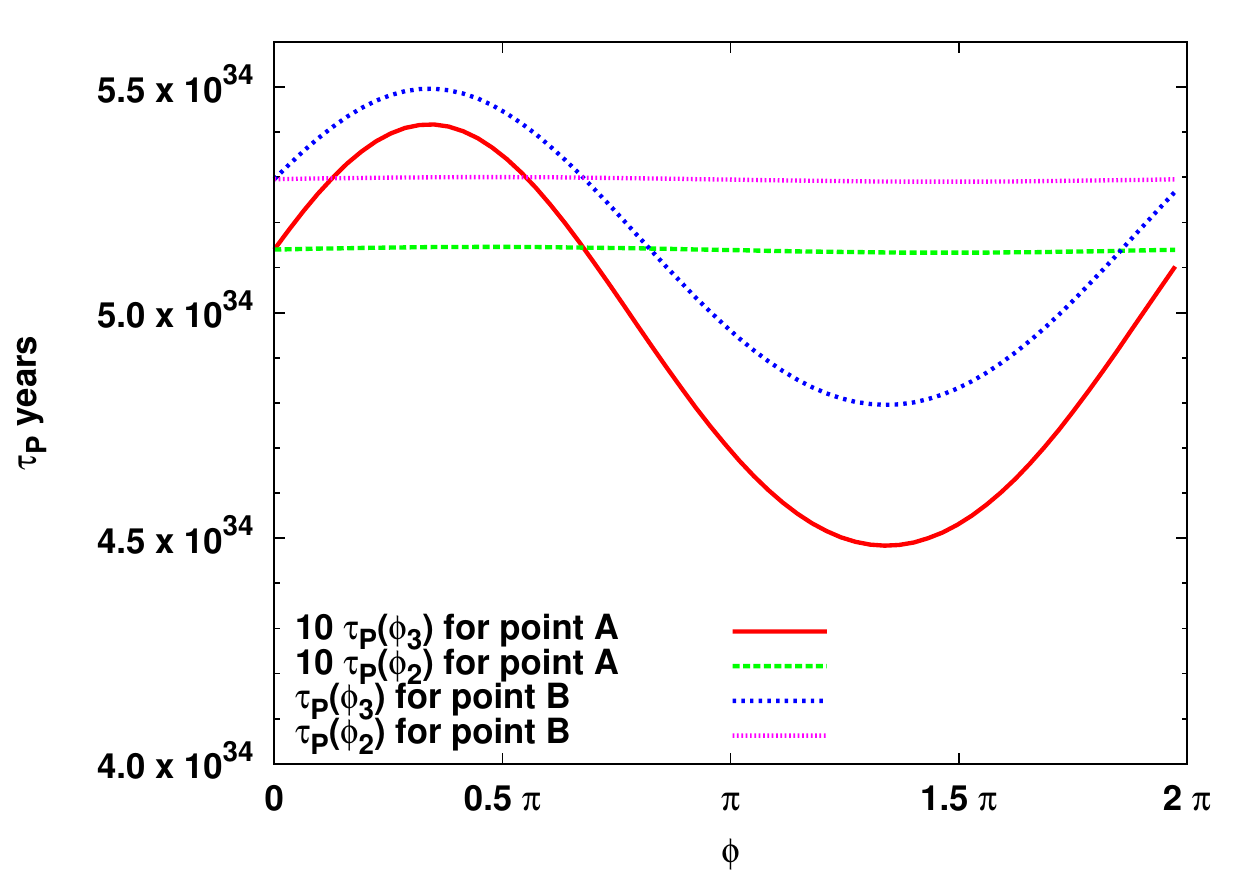}
\
\caption{
{\it
The variation of the proton lifetime due
 to the phases of the Wilson coefficients for point A ($m_{1/2}= 2.2$ TeV, $m_0=10$ TeV, $A_0=0$, $\tan\beta=5$ and $M_{in}=M_{GUT}$) and point B ($m_{1/2}=8$ TeV, $m_0=5.2$ TeV, $A_0=2.5m_0$, $\tan\beta=3.5$, and $M_{in}=10^9$ GeV). }}
\label{fig:phases}
\end{figure}

As mentioned above, new phases appearing in the GUT Yukawa couplings
yield uncertainty in the calculation \cite{Goto:1998qg}, though we have
neglected the effects of these phases in the main text. To justify our
neglect of these parameters, here we will show the variation in the
proton lifetime when the phases of the Wilson coefficients are included.
We select two points for analysis and present their phase dependence in
Fig.~\ref{fig:phases}.  Point A is for $m_{1/2}= 2.2$ TeV, $m_0=10$ TeV,
$A_0=0$, $\tan\beta=5$ and $M_{in}=M_{GUT}$ and has been labelled in
Fig.~\ref{fig:CMSSM}.  Point B is for $m_{1/2}=8$ TeV, $m_0=5.2$ TeV,
$A_0=2.5m_0$, $\tan\beta=3.5$, and $M_{in}=10^9$ GeV and has been
labelled in Fig.~\ref{fig:subGUT}. The variation arises due to
cancellations between the wino and Higgsino contribution to the proton
decay. Diagrams representing these processes are found in
Fig.~\ref{fig:1loop}.  As is clearly seen from Fig.~\ref{fig:phases},
the variation is small in comparison to the
uncertainty coming from the Yukawa couplings. Thus, we ignore these
effects in the main text.


\end{document}